\documentclass[a4paper,11pt,english]{article}

\usepackage{amsfonts,amssymb,amsthm}    
\usepackage[affil-it,auth-sc]{authblk}          
            
\usepackage{bigints}
\usepackage{bm}                         
\usepackage{braket}              
\usepackage[latin1]{inputenc}   
\usepackage{caption}
\usepackage{enumerate}
\usepackage{enumitem}           
\usepackage[left=27mm,right=27mm,top=27mm,bottom=29mm]{geometry}           
\usepackage{latexsym}            
\usepackage{mathrsfs}           
\usepackage{mathtools}                  
\usepackage{mparhack,fixltx2e,relsize}     
\usepackage[only,llbracket,rrbracket]{stmaryrd} 
\usepackage{subfig}                      
\usepackage{cite}
\usepackage{upgreek}
\usepackage[english]{varioref}   
\usepackage[english]{babel}                   
\usepackage{graphicx}
\usepackage{color}
\usepackage{caption}
\usepackage[titletoc,toc,title]{appendix}
\usepackage{hyperref}
\usepackage{bookmark}

\DeclareMathOperator{\Tan}{Tan}

\hypersetup{
  breaklinks=true,
  bookmarksnumbered,
  bookmarksopen=true, 
  bookmarksopenlevel=2, 
  pdftitle={Steady-state propagation of a mode II crack in couple stress elasticity},
  pdfauthor={Panos A. Gourgiotis, Andrea Piccolroaz},
  pdfkeywords={dynamic fracture, couple-stress, microstructure, mode-II crack, Wiener-Hopf} }

\begin{document}
\selectlanguage{english}

\title{Steady-State Propagation of a Mode II Crack in Couple Stress Elasticity}

\author[1]{P.A. Gourgiotis\footnote{Corresponding author:\,e-mail:\,p.gourgiotis@unitn.it; phone:\,+39\,0461\,282594.}}
\author[2]{A. Piccolroaz}
\affil[1,2]{Department of Civil, Environmental \& Mechanical Engineering, University of Trento, \newline via Mesiano 77, Trento, Italy}

\date{}
\maketitle

\begin{abstract}
\noindent
The present work deals with the problem of a semi-infinite crack steadily propagating in an elastic body subject to plane-strain shear loading. It is assumed that the mechanical response of the body is governed by the theory of couple-stress elasticity including also micro-rotational inertial effects. This theory introduces characteristic material lengths in order to describe the pertinent scale effects that emerge from the underlying microstructure and has proved to be very effective for modeling complex microstructured materials. It is assumed that the crack propagates at a constant sub-Rayleigh speed. An exact full field solution is then obtained based on integral transforms and the Wiener-Hopf technique. Numerical results are presented illustrating the dependence of the stress intensity factor and the energy release rate upon the propagation velocity and the characteristic material lengths in couple-stress elasticity. The present analysis confirms and extends previous results within the context of couple-
stress elasticity concerning stationary cracks by including inertial and micro-inertial effects. 
\end{abstract}


\newpage

\tableofcontents

\newpage

\section{Introduction}
\noindent It is well known that classical continuum theories possess no intrinsic length scale and thus fail to predict the scale effects observed experimentally in problems with geometric lengths comparable to the lengths of the material microstructure (e.g. Fleck and Hutchinson, 1997). This discrepancy between the classical theoretical predictions and experimental results is found more pronounced for materials with a coarse-grain structure. In fact, the macroscopical behavior of most microstructured materials with non-homogeneous microstructure, like ceramics, composites, cellular materials, foams, masonry, bone tissues, glassy and semi-crystalline polymers, is strongly influenced by the microstructural characteristic lengths, especially in the presence of large stress (or strain) gradients (Maranganti and Sharma, 2007a; 2007b). 

On the other hand, generalized continuum theories intend to capture the effects of microstructure by enriching the classical continuum with additional material characteristic length scales, and, thus, extending the range of applicability 
of the 'continuum' concept in an effort to bridge the gap between classical continuum theories and atomic-lattice theories. A thorough review of generalized continuum theories can be found in Maugin (2010).

One of the most effective generalized continuum theories has proved to be the theory of couple-stress elasticity, also known as Cosserat theory with constrained rotations. This theory is the simplest gradient theory in which couple-stresses make their appearance. In particular, the couple-stress theory assumes an augmented form of the Euler-Cauchy principle with a non-vanishing couple traction, and a strain-energy density that depends upon both the strain and the gradient of rotation. Such assumptions are appropriate for materials with granular structure, where the interaction between adjacent elements may introduce internal moments. In this way, characteristic material lengths may appear representing the material microstructure. The presence of these material lengths implies that the couple-stress theory encompasses the analytical possibility of size effects, which are absent in the classical theory. The fundamental concepts of the couple-stress theory were first introduced by Cauchy (1851), Voigt (1887) 
and the Cosserat brothers (1909), but the subject was generalized and reached maturity only in the 1960s through the works of Toupin (1962), Mindlin and Tiersten (1962), and Koiter (1964). 

Early applications of couple-stress elasticity dealt mainly with stress-concentration problems concerning holes and inclusions. In recent years, the couple-stress theory and related gradient theories attracted a renewed and growing interest in dealing with problems of complex microstructured materials. This is due to the inability of the classical theory to predict the observed scale effects, and also due to the increasing demands for manufacturing devices at very small scales. This approach and related extensions have been recently employed successfully to model size effects in microstructured materials in, among other areas, fracture (see e.g. Huang et al. 1997; Chen et al. 1998; Huang et al., 1999; Georgiadis, 2003; Radi and Gei, 2004; Gourgiotis and Georgiadis 2007; Radi, 2008; Aravas and Giannakopoulos, 2009; Gourgiotis and 
Georgiadis, 2011; Piccolroaz et al., 2012; Sciarra and Vidoli, 2012b; Antipov, 2012), plasticity (see e.g. Fleck et al., 1994; Gao et al., 1999; Hwang et al., 2002; Dal Corso and Willis, 2011), and wave propagation (see e.g. Vardoulakis and Georgiadis, 1997; Georgiadis and Velgaki, 2003; Georgiadis et al., 2004; Engelbrecht et al., 2005; Polyzos and Fotiadis, 2012; Gourgiotis et al., 2013).

For materials with microstructure, the characteristic material length mentioned before may be on the same order as the length of the microstructure. For instance, Chen et al. (1998) developed a continuum model for cellular materials and found that the continuum description of these materials obey a gradient elasticity theory of the couple-stress type. 
In the latter study, the intrinsic material length was naturally identified with the cell size. Mora and Waas (2000) performing experiments in honeycomb materials estimated that the value of the characteristic material length $\ell$ in couple-stress elasticity is $10d \sim 15d$, where $d$ is the diameter of the cell of the honeycomb. Also, Chang et al. (2003) associated the microstructural material constants of the couple-stress theory with the particle size and the inter-particle stiffness in a granular material. In addition, couple-stress theory was successfully utilized in the past to model some materials with microstructure like foams (Lakes, 1983) and porous solids (Lakes, 1993). Generally, the couple-stress theory is intended to model situations where a material with microstructure is deformed in very small volumes, such as in the immediate vicinity of crack tips, notches, small holes and inclusions, and in micrometer indentations. A recent study by Bigoni and Drugan (2007) provides an interesting account of the determination of couple-stress moduli via homogenization of heterogeneous materials.

Regarding now solutions closely related to our problem, we note that due to the complexity of the equations of couple-stress theories, very few dynamic crack problems have been considered in the literature. In particular, Itou (1972) studied the plane-strain time-harmonic problem of a mode I finite-length crack in a Cosserat medium with no micro-inertia employing the method of dual integral equations. In addition, Itou (1981) evaluated numerically the stress intensity factor for the problem of a propagating Yoffe crack in the context of standard couple-stress elasticity, analyzing the influence of the crack tip speed on the asymptotics of the stress field. Han et al. (1990) investigated the dynamic propagation of a finite-length crack under mode-I loading in a micropolar elastic solid by numerically solving a system of singular integral equations. They provided solutions for dynamic stress intensity and couple stress intensity factors by using the obtained values of the strengths of the square root singularities in macro-rotation and the gradient of micro-rotation at the crack tips. Later, Georgiadis (2003) using an analytic function method formulated the time-harmonic problem of semi-infinite crack under anti-plane strain conditions in a material exhibiting gradient effects. Recently, Mishuris et al. (2012) using the Wiener-Hopf technique solved the problem of a semi-infinite mode III crack steadily propagating in an elastic solid using the theory of couple-stress elasticity including micro-rotational effects. They showed that the process zone near the crack tip is strongly influenced by the microstructural parameters, such as the characteristic material lengths, the micro-inertia, and the maximum crack speed. Finally, Itou (2013) solved the problem of two collinear finite-length cracks subjected to time-harmonic stress waves impinging normal to the crack faces by using the dynamic couple-stress theory with no micro-inertial effects. The Schmidt method was utilized to satisfy the boundary conditions along the crack faces. He presented asymptotic results for the stress and the couple-stress intensity factors and examined the influence on the solution of the ratio of the characteristic material length to the pertinent geometrical lengths of the problem.

In the present work, the dynamic theory of couple-stress elasticity is employed to deal with the problem of a mode II semi-infinite crack propagating steadily at a constant sub-Rayleigh speed. In this way, the static analysis presented in Gourgiotis et al. (2012) is extended to the case of steady-state propagation in order to study the effects of micro-inertia and crack-tip speed on the stress and deformation fields, as well as the variation of the fracture toughness due to the presence of the microstructure. A micro-inertia term is included in our study because previous experience with couple-stress analyses of surface waves and anti-plane crack problems showed that this term is important, especially at high frequencies (Georgiadis and Velgaki, 2003; Mishuris et al., 2012). In fact, including this term in the present problem gives dispersion curves that mostly resemble with the ones obtained by atomic-lattice considerations. It is worth noting that this is the first study in the literature concerning a crack 
propagation problem under plane-strain conditions in the context of couple-stress elasticity including also micro-inertial effects. 

The paper is organized as follows: In Section 2 the fully dynamical version of couple-stress elasticity theory with micro-inertia is presented. The micro-rotational inertia term is included in our formulation by considering the appropriate expression for the kinetic-energy density of the material particle. In the next Section, the basic equations in plane-strain are provided and the influence of couple-stresses on the propagation of Rayleigh surface waves is studied. Unlike the conventional elastic theories, this reasonably simple gradient theory can indeed predict the dispersive character of Rayleigh waves in a medium with microstructure. The steady-state mode II crack propagation problem is formulated in Section 4. The analytical full-field solution is then addressed in Section 5 based on the Fourier transform and the Wiener-Hopf technique. In Section 6, we present numerical results for the stresses ahead of the crack-tip and the crack-face displacements. In addition, a closed-form expression for the stress intensity factor is provided. The dependence of these quantities upon the crack speed and the characteristic material lengths is also discussed in detail. Finally, we evaluate the dynamic energy release rate ($J$-integral) for the steady-state propagating mode II crack. To this purpose, a new extended definition of the dynamic $J$-integral is given in the context of couple-stress elasticity. It is shown that the $J$-integral can be determined through the use of distribution theory using only asymptotic results and is strongly influenced by the microstructure of the material.

\section{Fundamentals of dynamic couple-stress elasticity}
\label{sec2}
\noindent
In this Section, we present the basic elastodynamic equations of couple-stress elasticity. Our derivation of basic results relies on the momentum balance laws. Interesting presentations of the theory can also be found in the works by Mindlin and Tiersten (1962), Koiter (1964), and Muki and Sternberg (1965). However, the latter formulations do not include inertia and micro-inertia effects since they are of quasi-static character. 

As mentioned before, couple-stress elasticity assumes that: (i) each material particle has three degrees of freedom, (ii) an augmented form of the Euler-Cauchy principle with a non-vanishing couple traction prevails, and (iii) the strain-energy density depends upon both strain and the gradient of rotation. In addition, the kinetic-energy density $T$, within a geometrically linear theory, takes the following form (Nowacki, 1986)
  \begin{equation}
    \label{k.energy}
      T = \frac{\rho}{2} \dot{u}_{q} \dot{u}_{q} + \frac{I}{2} \dot{\omega}_{q} \dot{\omega}_{q},
  \end{equation}
where $\rho$ is the mass density, $I$ is the micro-rotational inertia, $u_q$ is the displacement vector, $\omega_q=1/2 e_{qpk} \partial{_p} u_{k}$ is the rotation vector, 
$\partial{_p()}=\partial()/\partial{x_p}$, the superposed dot denotes time derivative, and the Latin indices span the range (1,2,3) (indicial notation and summation convention is used throughout). The second term in the RHS of Eq.\eqref{k.energy}, involving the spin, represents the micro-rotational inertia of the continuum. This term, which is not encountered within classical continuum mechanics, reflects the more detailed description of motion in the present theory.

For a body with bounding surface $S$ and volume $\mathcal{V}$, the balance laws for the linear and angular momentum read
\begin{equation}
    \label{LinMom}
		  \int_{S} T^{(n)}_{q}\,dS + \int_{\mathcal{V}}F_{q}\,d\mathcal{V} =  \int_{\mathcal{V}} \rho\ddot{u}_q\,d\mathcal{V},
 \end{equation}
\begin{equation}
    \label{AngMom}
		  \int_{S} (e_{qpk}x{_p}T^{(n)}_{k}+M^{(n)}_{q})\,dS + \int_{\mathcal{V}} (e_{qpk}x{_p}F_{k}+C_{q})\,d\mathcal{V} = \int_{\mathcal{V}} (e_{qpk}x{_p}\rho\ddot{u}_k+I\ddot{\omega}_q)\,d\mathcal{V},
 \end{equation}
where $ e_{qpk} $ is the Levi-Civita alternating symbol, $T^{(n)}_q$ is the surface force per unit area, $M^{(n)}_q$ is the surface moment per unit area, $F_{q}$ is the body force per unit volume, $C_{q}$ is the body moment per unit volume, and $x{_p}$ designate the components of the position vector.

Next, pertinent force-stress and couple-stress tensors are introduced by considering the equilibrium of the elementary material tetrahedron and enforcing (2) and (3), respectively. The force stress tensor $\sigma_{pq}$ and the couple-stress tensor $\mu_{pq}$ (both asymmetric) are then defined by
\begin{equation}
\label{tr-T}
T^{(n)}_{q} = \sigma_{pq}n_{p}, \quad M^{(n)}_{q} = \mu_{pq}n_{p},
 \end{equation}
where $n_{p}$ are the direction cosines of the outward unit vector $\bf{n}$, which is normal to the surface. In addition, just like the third Newton's law $ \bf{T}^{(\bf{n})} = -\bf{T}^{(\bf{-n})} $ is proved to hold by considering the equilibrium of a material \lq slice', it can also be proved that  $ \bf{M}^{(\bf{n})} = -\bf{M}^{(\bf{-n})}$. The couple-stresses $\mu_{pq}$ are expressed in dimensions of $[force][length]^{-1}$. Further, $\sigma_{pq}$ can be decomposed into its symmetric and anti-symmetric components as follows
\begin{equation}
    \label{eq6}
		   \sigma_{pq} = \tau_{pq}+\alpha_{pq},
 \end{equation}
with $\tau_{pq} = \tau_{qp}$ and $\alpha_{pq} = -\alpha_{qp}$, whereas it is advantageous to decompose $\mu_{pq}$ into its deviatoric and spherical parts in the following manner
\begin{equation}
    \label{eq7}
		   \mu_{pq} = m_{pq}+\frac{1}{3}\delta_{pq}\mu_{kk},  
 \end{equation}
where $\mu_{pq}^{(D)}=m_{pq}$, $\mu_{pq}^{(S)}=(1/3)\delta_{pq}\mu_{kk}$, and $\delta_{pq}$ is the Kronecker delta. Now, with the above definitions and the help of the Green-Gauss theorem, one may obtain the stress equations of motion. In particular, Eq.\eqref{LinMom} leads to the following force equation
\begin{equation}
    \label{eq10}
		    \partial{_p}{\sigma_{pq}} +F_{q} = \rho\ddot{u}_q,
 \end{equation}
which, by virtue of \eqref{eq6}, becomes
\begin{equation}
    \label{eq11}
		    \partial{_p}{\tau_{pq}} + \partial{_p}{\alpha_{pq}} + F_{q} = \rho\ddot{u}_q.
 \end{equation}
Further, Eq. \eqref{AngMom} in conjunction with \eqref{eq10} leads to the following moment equation
\begin{equation}
    \label{eq8}
		   \partial{_p}{\mu_{pq}} + e_{qpk}\sigma_{pk}+C_{q} = I\ddot{\omega}_q,
 \end{equation}
which can also be written as
\begin{equation}
    \label{eq9}
		   \frac{1}{2}e_{pqk}\partial{_l}{\mu_{lk}} + \alpha_{pq}+ \frac{1}{2}e_{pqk}C_{k} = \frac{I}{2}e_{pqk}\ddot{\omega}_k.
 \end{equation}
Combining Eqs. \eqref{eq7}, \eqref{eq11}, \eqref{eq9} and by taking into account that $\nabla\!\times\!\nabla\!\cdot\!\bm{\mu}^{(\textit{S})}=0$, yields the \textit{single} equation
\begin{equation}
    \label{eq13}
		    \partial{_p}{\tau_{pq}} + \frac{1}{2}e_{qpk}\partial{_p}\partial{_l}{m_{lk}} + \frac{1}{2}e_{qpk}\partial{_p}C_{k} + F_{q} = \rho\ddot{u}_q + \frac{I}{2}e_{qpk}\partial{_p}\ddot{\omega}_k.
 \end{equation}
which constitutes the final equation of motion.

For the kinematical description of the continuum, the following quantities are defined in the framework of the geometrically linear theory
\begin{equation}
\label{eq14}
\varepsilon_{pq}=\frac{1}{2}(\partial{_p}{u_{q}}+\partial{_q}{u_{p}}), \quad \omega_{pq}=\frac{1}{2}(\partial{_p}{u_{q}}-\partial{_q}{u_{p}}),
 \end{equation}
\begin{equation}
\label{eq16}
\omega_{q}=\frac{1}{2}e_{qpk}\partial{_p}u_{k}, \quad \kappa_{pq}=\partial{_p}\omega_{q},
 \end{equation}
where $\varepsilon_{pq}$ is the strain tensor, $\omega_{pq}$ is the rotation tensor, and $\kappa_{pq}$ is the curvature tensor (i.e. the gradient of rotation or the curl of the strain) expressed in dimensions of $[length]^{-1}$. Notice also that \eqref{eq16}$_2$ can alternatively be written as
\begin{equation}
    \label{eq18}
		     \kappa_{pq}=\frac{1}{2}e_{qlk}\partial{_p}\partial{_l}u_{k}=e_{qlk}\partial{_l}\varepsilon_{pk}.
 \end{equation}
Equation \eqref{eq18} expresses compatibility for curvature and strain fields. The compatibility equations for the strain components are the usual Saint Venant's compatibility equations. Further, the identity $\partial{_k}\kappa_{pq}=\partial{_k}\partial{_p}\omega_{q}=\partial{_p}\kappa_{kq}$ defines the compatibility equations for the curvature components. We notice also that $\kappa_{pp}=0$ and, therefore, the curvature tensor has only eight independent components.

The traction boundary conditions, at any point on a smooth boundary or section, consist of the following three reduced force-tractions and two tangential couple-tractions (Mindlin and Tiersten, 1962; Koiter, 1964)
\begin{equation}
\label{tr-P}
P_{q}^{(n)}=\sigma_{pq}n{_p}-\frac{1}{2}e_{qpk}n{_p}\partial{_k}m_{(nn)}, \quad R_{q}^{(n)}=m_{pq}n{_p}-m_{(nn)}n{_q},
\end{equation}
where $m_{(nn)}=n{_p}n{_q}m_{pq}$ is the normal component of the deviatoric couple-stress tensor $m_{pq}$. We remark that in the case in which edges appear along the boundary, an additional 
boundary condition should be imposed. Indeed, as Koiter (1964) pointed out, a force (per unit length) \textit{tangential} to the edge should be specified according to the relation: 
$Q=\frac{1}{2}\llbracket m_{(nn)}\rrbracket$, where $\llbracket\rrbracket$ denotes the jump of the enclosed quantity through the edge. This tangential line load along the edge is the 
counterpart of the concentrated normal force which may be specified at the corner of the edge of a Kirchhoff plate or shell. Moreover, it is worth noticing that at first sight, it might seem 
plausible that the surface tractions (i.e. the force-traction and the couple-traction) can be prescribed arbitrarily on the external surface of the body through relations \eqref{tr-T}, which 
stem from the equilibrium of the material tetrahedron. However, in 
this case, the resulting number of six traction boundary conditions (three force-tractions and three couple-tractions) would be in contrast with the five geometric boundary conditions that 
can be imposed (Koiter, 1964). Indeed, since the rotation vector $\omega_{q}$ in couple-stress elasticity is not independent of the displacement vector, see Eq. \eqref{eq16}$_1$; the normal 
component of the rotation is fully specified by the distribution of tangential displacements over the boundary. Therefore, only the three displacement and the two tangential rotation 
components can be prescribed independently. As a consequence, only five surface tractions (i.e. the work conjugates of the above five independent kinematical quantities) can be specified at a 
point of the bounding surface of the body, i.e. Eqs. \eqref{tr-P}. On the other hand, in the Cosserat (micropolar) theory, the traction boundary conditions are six since the rotation is fully 
independent of the displacement vector (see e.g. Nowacki, 
1986). In the latter case, the tractions can directly be derived from the equilibrium of the material tetrahedron, so \eqref{tr-T} are the pertinent traction boundary conditions.

For a linear and isotropic material behavior, the strain-energy density has the following form
\begin{equation}
\label{s.energy}
W\equiv W(\varepsilon_{pq},\kappa_{pq})=\frac{1}{2}\lambda\varepsilon_{pp}\varepsilon_{qq}+\mu\varepsilon_{pq}\varepsilon_{pq}+2\eta\kappa_{pq}\kappa_{pq}+2\eta'\kappa_{pq}\kappa_{qp},
\end{equation}
where $(\lambda,\mu,\eta,\eta')$ are material constants. Then, Eq. \eqref{s.energy} leads, through the standard variational manner, to the following constitutive equations
\begin{equation}
\label{eq22}
\tau_{pq}\equiv \sigma_{(pq)}=\frac{\partial{W}}{\partial\varepsilon_{pq}}=\lambda\delta_{pq}\varepsilon_{kk}+2\mu\varepsilon_{pq}, \quad 
m_{pq}=\frac{\partial{W}}{\partial\kappa_{pq}}=4\eta\kappa_{pq}+4\eta'\kappa_{qp}.
\end{equation}
In view of \eqref{eq22}, the moduli $(\lambda,\mu)$ have the same meaning as the Lam$\acute{e}$ constants of classical elasticity theory and are expressed in dimensions of $[force][length]^{-2}$, whereas the moduli $(\eta,\eta')$ account for couple-stress effects and are expressed in dimensions of $[force]$. Further, following Mindlin and Tiersten (1962), we assume $W$ to be a positive definite function of its arguments, so that
\begin{equation}
    \label{eq24}
		     3\lambda+2\mu>0, \quad \mu>0, \quad \eta>0, \quad -1<\frac{\eta'}{\eta}<1.
 \end{equation}
Incorporating now the constitutive relations \eqref{eq22} into the equation of motion \eqref{eq13} and using the geometric relations \eqref{eq14} and \eqref{eq16}, one may obtain the equations of motion in terms of the displacement vector
\begin{equation}
    \label{eq25}
		    \mu\nabla^2\bm{u}+(\lambda+\mu)\nabla(\nabla\cdot\bm{u})-\mu\ell^2\nabla^2[\nabla^2\bm{u}-\nabla(\nabla\cdot\bm{u})]=\rho\ddot{\bm{u}}+\frac{I}{4}\nabla\times\nabla\times\ddot{\bm{u}},
 \end{equation}
where $\ell\equiv (\eta/\mu)$ is a characteristic material length, and the absence of body forces and couples is assumed. In the limit $\ell\rightarrow 0$, the Navier-Cauchy equations of classical linear isotropic elasticity are recovered from \eqref{eq25}. Indeed, the fact that Eqs. \eqref{eq25} have an increased order w.r.t. their limit case (recall that the Navier-Cauchy equations are PDEs of the second order) and the coefficient $\ell$ multiplies the higher-order term reveals the singular-perturbation character of the couple-stress theory and the emergence of associated boundary-layer effects. 

Next, by taking the divergence and the curl of \eqref{eq25} we obtain the equations governing the propagation of dilatation and rotation, respectively
\begin{equation}
    \label{eq26}
		    c^2_p\nabla^2(\nabla\cdot\bm{u})=\nabla\cdot\ddot{\bm{u}},
\end{equation}
\begin{equation}
    \label{eq27}
		    c^2_s (1-\ell^2\nabla^2)\nabla^2(\nabla\times\bm{u})=(1-h^2\nabla^2)\nabla\times\ddot{\bm{u}},
\end{equation}
where $c_{p}=[(\lambda+2\mu)/\rho]^{1/2}$ and $c_{s}=(\mu/\rho)^{1/2}$  are the velocities of the pressure (P) and shear (S) waves, respectively, in the classical elasticity theory, and $h=(I/4\rho)^{1/2}$ is a characteristic intrinsic material length associated with the micro-inertia of the material (Mishuris et al., 2012). An interrelation of the two characteristic microstructural lengths $\ell$ and $h$ was given by Georgiadis and Velgaki (2003) by comparing the forms of dispersion curves of Rayleigh waves in couple-stress theory with micro-rotational inertia with the ones obtained by the discrete particle theory (atomic-lattice approach). It is interesting to note that Eq. \eqref{eq26} governing the propagation of dilatational waves is the same as in the classical theory. On the other hand, unlike the corresponding case of classical elastodynamics, Eq. \eqref{eq27} is of the fourth order. This implies that for shear waves, wave signals emitted from a disturbance point propagate at different velocities. The last 
statement can be supported by considering a time-harmonic plane wave solution and determining dispersion relations. Specifically, we consider a plane wave solution of Eq. \eqref{eq27} in the following form
\begin{equation}
    \label{eq28}
		    \bm{u}=A\bm{d}\exp\Bigl[i\bigl(\xi\left(\bm{n}\cdot\bm{x}\right)-wt\bigr)\Bigr],
\end{equation}
where $A$ denotes the amplitude, $(\bm{d},\bm{n})$ are unit vectors defining the directions of motion and propagation, respectively, $\bm{x}$ is the position vector, $\xi$ is the wavenumber, $w$ is the circular frequency of the plane wave, and $i^2=-1$. Then, on substituting \eqref{eq28} into Eq. \eqref{eq27}, we obtain the following dispersion relation for the shear waves
\begin{equation}
    \label{eq29}
		    w^2=c_s^2 \xi^2(1+\ell^2 \xi^2)(1+h^2 \xi^2)^{-1}.
\end{equation}
Accordingly, the phase velocity $V_s$ of the shear waves in couple stress elasticity takes the following form
\begin{equation}
    \label{eq30}
		    V_s \equiv \frac{w}{\xi}=c_s(1+\ell^2 \xi^2)^{1/2}(1+h^2 \xi^2)^{-1/2}.
\end{equation}
Equation \eqref{eq30} shows that the propagation velocity of these waves depends on the respective wavenumber. Hence, shear waves are dispersive in couple-stress elasticity, while the longitudinal waves remain non-dispersive as in the classical theory (Toupin, 1962; Graff and Pao, 1967). 

To investigate further upon the nature of the dispersion relation in couple-stress elasticity, we consider the group velocity $\upsilon=dw/d\xi$   at which the energy propagates in a dispersive medium (Achenbach, 1973). In particular, according to \eqref{eq29} and \eqref{eq30}, we obtain
\begin{equation}
    \label{eq31}
		    \upsilon_{s}=V_s+(\ell^2-h^2)c_{s}\xi^2(1+\ell^2 \xi^2)^{-1/2}(1+h^2 \xi^2)^{-3/2}.
\end{equation}
The following three cases are then distinguished: $(i)$ For $\ell<h$, Eq. \eqref{eq31} implies that $\upsilon_{s}<V_s$, thus the dispersion for shear waves is normal. $(ii)$ For $\ell>h$, we have $\upsilon_{s}>V_s$ indicating that the dispersion is anomalous. $(iii)$ For $\ell=h$ or $(\ell,h)\rightarrow 0$ (i.e. no material microstructure), the shear wave velocity degenerates into the non-dispersive velocity of classical elastodynamics.

\section{Basic equations in plane-strain}
\label{sec3}		
\noindent
For a body that occupies a domain in the $(x,y)$-plane under conditions of plane strain, the displacement field takes the general form
\begin{equation}
    \label{3.1}
		   u_x\equiv u_x(x,y)\neq0, \quad  u_y\equiv u_y(x,y)\neq0, \quad u_z\equiv 0.
\end{equation}
Further, Muki and Sternberg (1965) showed that, if the general 3D equations of Section 2 are combined with the restrictions in \eqref{3.1} and the normalization $\mu_{kk}=0$ of the couple-stress field is adopted, all field quantities are independent of the coordinate $z$. It is noteworthy that, contrary to the respective plane-strain case in the conventional theory, this independence is not obvious within the couple-stress theory (Muki and Sternberg, 1965). Notice further, that except for $\omega_{z} \equiv \omega$ and $(\kappa_{xz},\kappa_{yz})$, all other components of the rotation vector and the curvature tensor vanish identically. The aforementioned restrictions describe the so called \textit{first planar} problem, which is a generalization of the classical in plane elasticity problem (Ostoja-Starzewski and Jasiuk, 1995). In view of the above, the following kinematic relations are obtained
\begin{equation}
    \label{3.2}
		  \varepsilon_{xx}=\partial{_x}u_{x},\quad\varepsilon_{yy}=\partial{_y}u_{y}, \quad  \varepsilon_{xy}=\varepsilon_{yx}=\frac{1}{2}(\partial{_x}u_{y}+\partial{_y}u_{x}),
\end{equation}
\begin{equation}
    \label{3.3}
		  \omega=\frac{1}{2}(\partial{_x}u_{y}-\partial{_y}u_{x}), \quad \kappa_{xz}=\partial{_x}\omega, \quad \kappa_{yz}=\partial{_y}\omega.
\end{equation}
Accordingly, the constitutive equations in \eqref{eq22} furnish the non-vanishing components of the symmetric part of stress and the couple-stress, respectively. Vanishing body forces and body couples are assumed in what follows. Then, the antisymmetric part of the stresses is found from \eqref{eq9}. In view of the above, the following expressions are written
\begin{equation}
    \label{3.4}
		 \tau_{xx}=(\lambda+2\mu)\varepsilon_{xx}+\lambda\varepsilon_{yy}, \quad \tau_{yy}=(\lambda+2\mu)\varepsilon_{yy}+\lambda\varepsilon_{xx}, \quad \tau_{yx}=\tau_{xy}=2\mu\varepsilon_{xy}
\end{equation}
\begin{equation}
    \label{3.5}
		 m_{xz}=4\mu\ell^2\kappa_{xz},\quad m_{yz}=4\mu\ell^2\kappa_{yz},
\end{equation}
\begin{equation}
    \label{3.6}
		 \alpha_{xx}=\alpha_{yy}=0,\quad\alpha_{yx}=-\alpha_{xy}=2\mu\ell^2\nabla^2\omega-\frac{I}{2}\ddot{\omega},
\end{equation}
and consequently the (asymmetric) stresses are given as
\begin{equation}
    \label{3.7}
		  \sigma_{xx}=\tau_{xx},\quad\sigma_{yy}=\tau_{yy},\quad\sigma_{yx}=\tau_{yx}+\alpha_{yx},\quad\sigma_{xy}=\tau_{xy}+\alpha_{xy}.
\end{equation}
The equation of motion in \eqref{eq25} leads to the following system of coupled PDEs of the fourth order for the displacement components $(u_{x},u_{y})$
\begin{equation}
    \label{3.8}
		 \mu\nabla^2u_x+(\lambda+\mu)\partial_xe-\mu\ell^2\nabla^2[\nabla^2u_x-\partial_xe]=\rho\ddot{u}_x+\frac{I}{2}\partial_y{\ddot{\omega}},
\end{equation}
\begin{equation}
    \label{3.9}
		 \mu\nabla^2u_y+(\lambda+\mu)\partial_ye-\mu\ell^2\nabla^2[\nabla^2u_y-\partial_ye]=\rho\ddot{u}_y-\frac{I}{2}\partial_x{\ddot{\omega}},
\end{equation}
where $\nabla^2\equiv\partial_x^2()+\partial_y^2()$ and $e=\partial_x{u_x}+\partial_y{u_y}$ is the dilatation. Although the above system is much more complicated than that in the respective case of classical elastodynamics, uncoupling by the use of Lam$\acute{e}$-type potentials still proves to be successful. The potentials $\phi(x,y,t)$ and $\psi(x,y,t)$ are defined in terms of the displacement components as
\begin{equation}
\label{3.10}
u_x=\frac{\partial{\phi}}{\partial x}+\frac{\partial{\psi}}{\partial y}, \quad u_y=\frac{\partial{\phi}}{\partial y}-\frac{\partial{\psi}}{\partial x}.
\end{equation}
Consequently, the dilation and the rotation vector become
\begin{equation}
    \label{3.12}
		 e=\nabla^2\phi,\qquad\omega=-\frac{1}{2}\nabla^2\psi.
\end{equation}
Incorporating the above relations into \eqref{3.8} and \eqref{3.9}, we derive, after some tedious but straightforward algebra, the following uncoupled PDEs, which constitute our field equations. The potential $\phi$ is governed by a second order PDE, whereas the potential $\psi$ satisfies a PDE of the fourth order, i.e.
\begin{equation}
    \label{fielda}
		 c_p^2\nabla^2\phi=\ddot{\phi},
\end{equation}
\begin{equation}
    \label{fieldb}
		 c_s^2\left(1-\ell^2\nabla^2\right)\nabla^2\psi=\left(1-h^2\nabla^2\right)\ddot{\psi}.
\end{equation}
The fact that \eqref{3.10}-\eqref{fieldb} comprise the complete solution of the equations of motion \eqref{3.8} and \eqref{3.9} may be proved by following exactly the steps in Sternberg's (1960) form of the proof for the case $\ell=0$ and $h=0$. In fact, it is only necessary to replace in Sternberg's proof the standard wave operator governing the shear wave propagation: $\square_2^2=\nabla^2-c_s^{-2}\partial{_t^2}$ by the modified operator in couple-stress elasticity with micro-inertia: $\lozenge_2^2=\left(1-\ell^2\nabla^2\right)\nabla^2-c_s^{-2}\left(1-h^2\nabla^2\right)\partial{_t^2}$ (see also Mindlin and Tiersten, 1962). 

Accordingly, the stresses and couple-stresses take the following form in terms of the Lam$\acute{e}$ potentials
\begin{equation}
    \label{3.15}
		 \sigma_{xx}=\lambda\nabla^2\phi+2\mu\left(\frac{\partial^2{\phi}}{\partial{x^2}}+\frac{\partial^2{\psi}}{\partial{x}\partial{y}}\right),
\end{equation}
\begin{equation}
    \label{3.16}
		 \sigma_{yy}=\lambda\nabla^2\phi+2\mu\left(\frac{\partial^2{\phi}}{\partial{y^2}}-\frac{\partial^2{\psi}}{\partial{x}\partial{y}}\right),
\end{equation}
\begin{equation}
    \label{3.17}
		 \sigma_{yx}=\mu\left[2\frac{\partial^2{\phi}}{\partial{x}\partial{y}}+\frac{\partial^2{\psi}}{\partial{y^2}}-\frac{\partial^2{\psi}}{\partial{x^2}}\right]-\mu\ell^2\nabla^4\psi+\frac{I}{4}\nabla^2\ddot{\psi},
\end{equation}

\begin{equation}
    \label{3.18}
		 \sigma_{xy}=\mu\left[2\frac{\partial^2{\phi}}{\partial{x}\partial{y}}+\frac{\partial^2{\psi}}{\partial{y^2}}-\frac{\partial^2{\psi}}{\partial{x^2}}\right]+\mu\ell^2\nabla^4\psi-\frac{I}{4}\nabla^2\ddot{\psi},
\end{equation}
\begin{equation}
    \label{3.19}
		 m_{xz}=-2\mu\ell^2\partial_x\nabla^2\psi,
\end{equation}
\begin{equation}
    \label{3.20}
		 m_{yz}=-2\mu\ell^2\partial_y\nabla^2\psi.
\end{equation}
Regarding the traction boundary conditions in the plane strain case, we note that these are defined through Eqs. \eqref{tr-P} by taking also into account that the normal component of the couple-stress $m_{(nn)}$ is zero. Indeed, since the components $(m_{xx},m_{yy},m_{yx},m_{xy})$ of the couple-stress tensor vanish identically in the plane-strain case (recall that in this case, all field quantities are independent upon the $z$ coordinate), we conclude that $m_{(nn)}\equiv n_{p}n_{q}m_{pq}=0$. This, in turn, implies that no edge forces $Q=(1/2) \llbracket m_{(nn)} \rrbracket$ appear at the corners of a boundary or section in plane strain. However, we remark that these edge forces should be considered in antiplane strain problems, where, in general, $m_{(nn)}\neq 0$. In addition, pertinent edge conditions should be taken into account in both plane and antiplane strain problems, in the more general theory of dipolar gradient elasticity (see e.g. Gourgiotis et al., 2010; Sciarra and Vidoli, 
2012a).
\subsection{Propagation of Rayleigh waves in couple-stress elasticity}
\noindent We close this Section with a brief study of the propagation of Rayleigh waves in couple-stress elasticity with micro-rotational inertia. A more detailed study can be found in Georgiadis and Velgaki (2003). To this purpose, we consider the following two-dimensional time-harmonic response in the half-space $y\geq 0$
\begin{equation}
    \label{3.21}
		 \phi\left(x,y,t\right)=\Phi(y)\cdot \exp{\bigl[i(\xi x-wt)\bigr]},
\end{equation}
\begin{equation}
    \label{3.22}
		 \psi\left(x,y,t\right)=\Psi(y)\cdot \exp{\bigl[i(\xi x-wt)\bigr]}.
\end{equation}
Then, substitution of the above equations into \eqref{fielda} and \eqref{fieldb} yields
\begin{equation}
    \label{3.23}
		 \frac{d^2{\Phi}}{dy^2}-\beta_p^2\Phi=0,
\end{equation}
\begin{equation}
    \label{3.24}
		 \frac{d^4\Psi}{dy^4}-(\beta_s^2+\gamma_s^2)\frac{d^2\Psi}{dy^2}+\beta_s^2\gamma_s^2\Psi=0,
\end{equation}
where the following definitions have been employed
\begin{equation}
    \label{3.25}
		 \beta^2_p=\xi^2-\frac{w^2}{c_p^2},
\end{equation}
\begin{equation}
    \label{3.26}
		 \beta^2_s=\xi^2-\sigma^2_s \qquad \textup{with} \qquad \sigma_s=\frac{1}{\sqrt{2}\ell}\left[\Delta_s-\left(1-\frac{w^2h^2}{c^2_s}\right)\right]^{1/2},
\end{equation}
\begin{equation}
    \label{3.27}
		 \gamma^2_s=\xi^2+\tau^2_s \qquad \textup{with} \qquad \tau_s=\frac{1}{\sqrt{2}\ell}\left[\Delta_s+\left(1-\frac{w^2h^2}{c^2_s}\right)\right]^{1/2},
\end{equation}
and
\begin{equation}
    \label{3.28}
\Delta_s=\left[\left(1-\frac{w^2h^2}{c^2_s}\right)^{2}+\frac{4\ell^2w^2}{c_s^2}\right]^{1/2}.
\end{equation}
The criterion for surface waves is that the displacement potentials decay exponentially with the distance $y$ from the free surface. It is noted that in the context of classical elasticity 
these waves are non-dispersive at any frequency. From the solution of the ODEs \eqref{3.23} and \eqref{3.24}, we obtain the displacement potentials $(\phi,\psi)$ as
\begin{equation}
    \label{3.29}
\phi=A\exp{\bigl[-\beta_p y+i(\xi x-wt)\bigr]},
\end{equation}
\begin{equation}
    \label{3.30}
\psi=B\exp{\bigl[-\beta_s y+i(\xi x-wt)\bigr]}+C\exp{\bigl[-\gamma_s y+i(\xi x-wt)\bigr]}.
\end{equation}
Since $\Delta_s>0$ always, the quantities $ (\sigma_s,\tau_s)$ defined in \eqref{3.26} and \eqref{3.27} are real and greater than zero. Consequently, it can be readily shown that $\gamma_s^2$ 
is always greater than zero, whereas $ \beta_p^2$ and $\beta_s^2$ have real values greater than zero if only the phase velocity of the Rayleigh waves is  $c_{ph}=w/\xi<\min\{V_s,c_p\}$, where 
$V_s$ is the shear wave velocity for a couple-stress material defined in Eq. \eqref{eq30}.

Now, the pertinent boundary conditions for a traction-free boundary defined by the plane $(x,y=0)$ with $\bm{n}=(0,1)$ take the following form
\begin{equation}
    \label{3.31}
\sigma_{yy}(x,0)=0,\qquad\sigma_{yx}(x,0)=0,\qquad m_{yz}(x,0)=0.
\end{equation}
Substitution of expressions \eqref{3.29} and \eqref{3.30} into the boundary conditions \eqref{3.31} results in an eigenvalue problem for the wave amplitudes $(A,B,C)$. Accordingly, the 
vanishing of the pertinent determinant provides the dispersion equation for the propagation of Rayleigh waves in a material characterized by couple-stress elasticity. The analytical 
expression of the determinant is very lengthy and is omitted here. A numerical solution was derived using the symbolic computer program MATHEMATICA$^{\textup{TM}}$.

\begin{figure}[!htcb]
\centering
\subfloat[]{\includegraphics[scale=0.36]{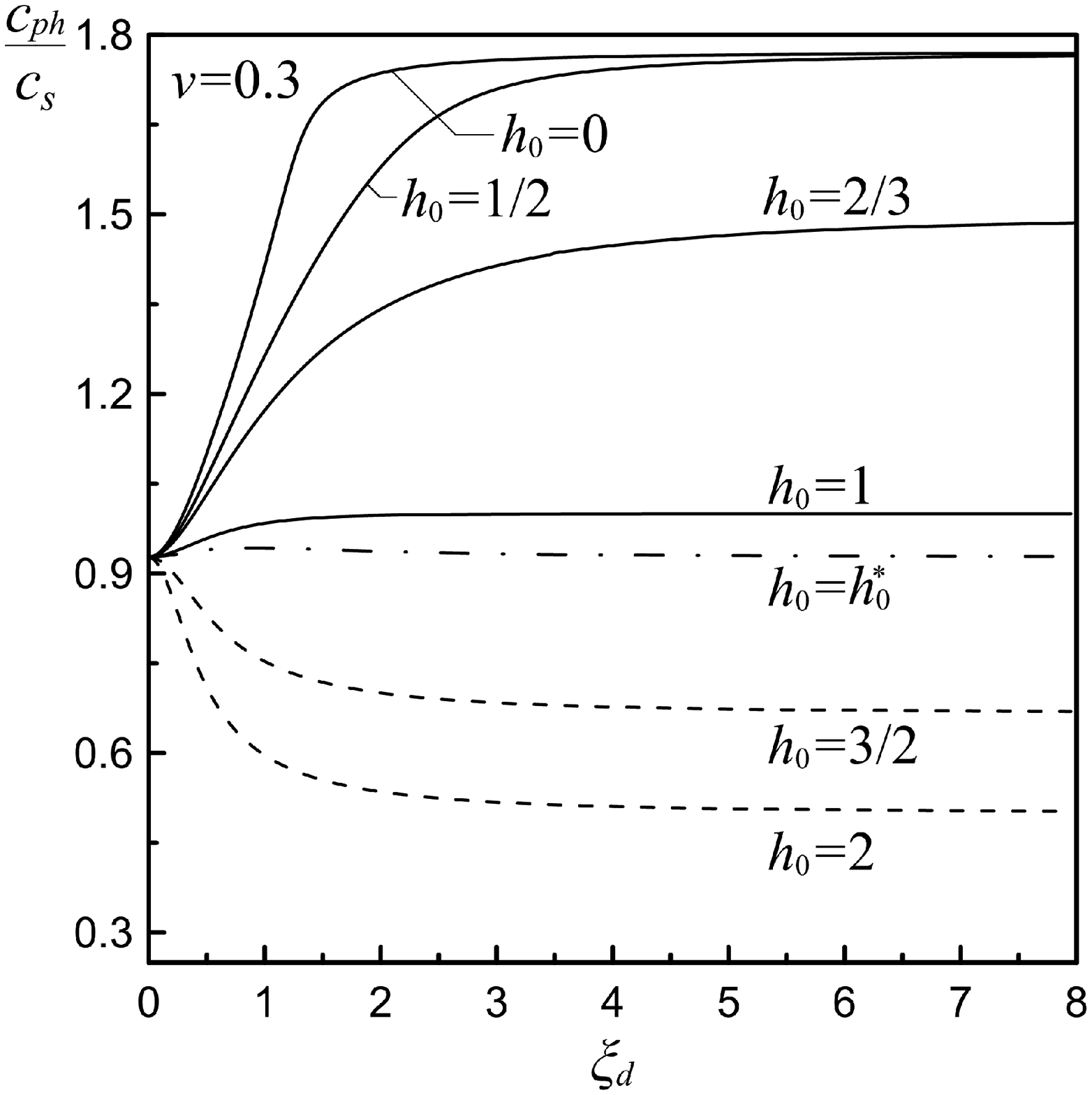}}
\quad
\subfloat[]{\includegraphics[scale=0.36]{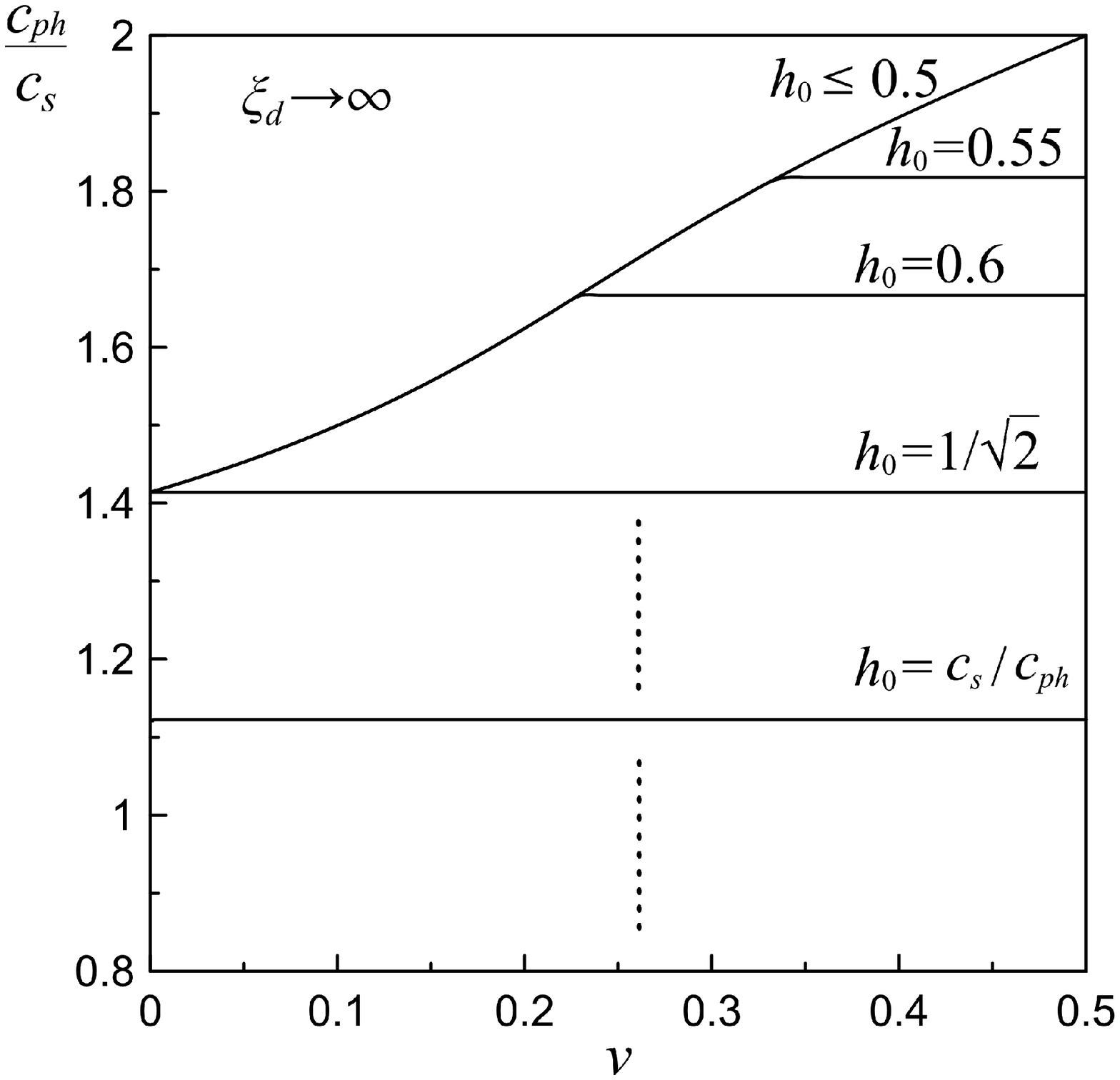}}
\caption{(a) Dispersive character of Rayleigh waves in couple-stress elasticity. (b) Limit of the Rayleigh velocity in couple-stress elasticity with respect to the Poisson's ratio as $\xi_d\to\infty$.}
\label{fig01}
\end{figure}

Figure \ref{fig01}a shows the variation of the normalized phase velocity $c_{ph}/c_s$ of the Rayleigh waves in couple-stress elasticity versus the normalized wavenumber $\xi_d=\xi\ell$ for a 
material with Poisson's ratio $\nu=0.3$. It is observed that for small wavenumbers (low frequencies) the phase velocity of the Rayleigh wave tends to the respective classical value: 
$c_R=0.927c_s$. This is to be expected intuitively since for relatively long wavelengths the wave should not `see' the material microstructure. On the other hand, for large wave numbers (high 
frequencies) the dispersive Rayleigh wave speed depends strongly upon the ratio of the characteristic material lengths $h_0=h/\ell$. In particular, it is observed that as $\xi_d$ increases, the 
phase velocity reaches a plateau attaining a constant value which depends upon $h_0$ and the Poisson's ratio $\nu$. Indeed, as it is illustrated in Figure \ref{fig01}b, the following three 
cases are distinguished as $\xi_d\rightarrow\infty$: (i) $h_0\leq{0.5}$; in this case the limit value 
of the normalized phase velocity depends only upon the Poisson ratio. (ii) $0.5<h_0<1/\sqrt{2}$; in this case $c_{ph}/c_s$ depends upon both $h_0$ and $\nu$. In fact, for $\nu>\nu^*(h_0)$ the 
ratio becomes constant and tends to the value $1/h_0$. (iii) $h_0\geq \sqrt{1/2}$; in this case we obtain that $c_{ph}/c_s=1/h_0$ always, independently of the Poisson's ratio.
%
%
Moreover, it is worth noting that the curve $h_0^*=c_s/c_R$ in Figure \ref{fig01}a, differs slightly from the classical elasticity solution,which, in turn, implies that in this case the Rayleigh waves 
are \textit{almost} non-dispersive. The constant $h_0^*$ depends, according to its definition, only upon the Poisson's ratio; in fact for
a material with $0 \leq \nu \leq 0.5$, we have $0.874 \leq c_R/c_s \leq 0.955$, and, consequently, $h_0^*$ ranges from $1.046 \leq h_0^* \leq 1.144$ (see e.g. Achenbach, 1973).
Finally, it should be remarked that a propagation velocity $V$ is characterized as sub-Rayleigh, which is our case of interest here, when $V<c_{ph}$. 
Consequently, when $0 \leq h_0 \leq h_0^*$ it suffices that the propagation velocity is less than the classical Rayleigh velocity i.e. $V<c_R$, whereas for $h_0>h_0^*$ the propagation is sub-Rayleigh provided that $V<c_s/h_0$.


\section{Formulation of the crack problem}
\label{sec4}

\begin{figure}[!htcb]
\centering
\includegraphics[scale=0.36]{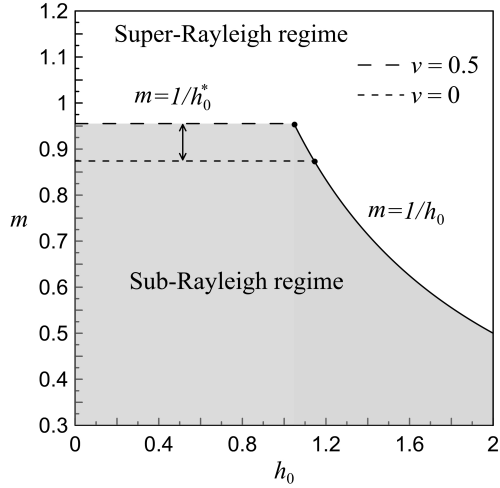}
\caption{Sub-Rayleigh and Super-Rayleigh regimes in the $m - h_0$ plane.}
\label{fig02}
\end{figure}

\noindent
Consider now a semi-infinite crack in a body of infinite extent under plane strain conditions. The body is governed by the equations of couple-stress elasticity. The crack propagates with constant sub-Rayleigh velocity straight along the $x$-axis and is subjected to a distribution of shear stresses along the crack faces moving with the same velocity. 
We now introduce the standard steady-state assumption for moving sources according to which a steady stress and displacement field is created in the medium w.r.t. an observer situated in a frame of reference attached to the tip of the crack, if the crack has been moving steadily (with a velocity $V$, say) for a sufficiently long time. In this way, any transients can reasonably be avoided (therefore gaining considerable simplification in the analysis). Specifically, upon introducing the Galilean transformation
\begin{equation}
    \label{galilean}
X=x-Vt,\qquad Y=y,
\end{equation}
the time derivative in the fixed Cartesian system becomes: $\partial_t=-V\partial_X$, which, in turn, implies that the field and the boundary conditions become independent of the time $t$, and the variables $(x,t)$ enter the problem only in the combination $x-Vt$. Furthermore, in the new moving Cartesian coordinate system, partial derivatives w.r.t. $t$ are neglected and the field equations \eqref{fielda} and \eqref{fieldb} for the Lam$\acute{e}$ potentials $\phi\left(X,Y\right)$ and $\psi\left(X,Y\right)$ can now be written as
\begin{equation}
    \label{steady-fielda}
(1-m^2c^2)\frac{\partial^2\phi}{\partial{X^2}}+\frac{\partial^2\phi}{\partial{Y^2}}=0,
\end{equation}
\begin{equation}
    \label{steady-fieldb}
(1-m^2)\frac{\partial^2\psi}{\partial{X^2}}+\frac{\partial^2\psi}{\partial{Y^2}}-\ell^2\left[\left(1-m^2h_0^2\right)\frac{\partial^2{(\nabla^2\psi)}}{\partial{X^2}}+\frac{\partial^2{(\nabla^2\psi)}}{\partial{Y^2}}\right]=0,
\end{equation}
where $c=c_s/c_p=\left[\left(1-2\nu\right)/2\left(1-\nu\right)\right]^{1/2}<1$. Also, $m=V/c_s$ and $mc=V/c_p$ are the two Mach numbers. It is worth noting that $m<m_R$ with $m_R=\min\{1/h_0,1/h^*_0\}$ and $h^*_0=c_s/c_R$, in order for the crack to propagate with a sub-Rayleigh speed (see also Section 2). Figure \ref{fig02} depicts the sub-Rayleigh and the super-Rayleigh regimes in couple-stress elasticity. 

\noindent
Due to the anti-symmetry with respect to the $Y=0$ plane, the problem can be viewed as a half-plane problem in the region $Y\geq0$ under the following boundary conditions
\begin{alignat}{2}
&\sigma_{yx}(X,Y\!=\!0)=-T(X)\qquad    & \textup{for} \qquad               -\infty<&X<0,    \label{bc1} \\ 
&\sigma_{yy}(X,Y\!=\!0)=0\qquad        & \textup{for} \qquad               -\infty<&X<\infty, \label{bc2} \\ 
&m_{yz}(X,Y\!=\!0)=0\qquad             & \textup{for} \qquad               -\infty<&X<\infty, \label{bc3}\\ 
&u_{x}(X,Y\!=\!0)=0\qquad              & \textup{for} \qquad    \phantom{-\infty}0<&X<\infty \label{bc4}, 
\end{alignat}
where $T(X)$ is the distribution of shear tractions along the crack faces.


\section{Full field solution}
\label{sec5}

\noindent
An exact solution of the boundary value problem described above will be obtained here through the Fourier transform and the Wiener-Hopf technique (Noble, 1958). The direct and inverse Fourier transforms are defined as
\begin{equation}
\label{dir}
f^*(s,Y)=\int_{-\infty}^{+\infty} f(X,Y) e^{isX}\,dX, \quad f(X,Y)=\frac{1}{2\pi}\int_{\mathcal{L}} f^*(s,Y) e^{-isX}ds,
\end{equation}
where $\mathcal{L}$ denotes the inversion path within the region of analyticity of the function in the complex $s$-plane. Transforming the field equations \eqref{steady-fielda} and \eqref{steady-fieldb} with \eqref{dir}$_1$, we obtain the following ODEs
\begin{equation}
    \label{tr-fielda}
\frac{d^2\phi^*}{dY^2}-(1-m^2c^2)s^2\phi^*=0,
\end{equation}
\begin{equation}
    \label{tr-fieldb}
\ell^2\frac{d^4\psi^*}{dY^4}-\Bigl[1+\left(2-m^2h_0^2\right)\ell^2s^2\Bigr]\frac{d^2\psi^*}{dY^2}+\Bigl[\left(1-m^2\right)+\left(1-m^2h_0^2\right)\ell^2s^2\Bigr]s^2\psi^*=0.
\end{equation}
The above equations have the following general solutions that will be required to be \textit{bounded} as $Y\to+\infty$
\begin{equation}
    \label{sol-trfielda}
\phi^*(s,Y)=A(s)e^{-\alpha\,Y},
\end{equation}
\begin{equation}
    \label{sol-trfieldb}
\psi^*(s,Y)=B(s)e^{-\beta\,Y}+C(s)e^{-\gamma\,Y},
\end{equation}
where
\begin{equation}
    \label{defa}
\alpha\equiv\alpha(z)=\frac{\left[\left(1-m^2c^2\right)z^2\right]^{1/2}}{\ell},
\end{equation}
\begin{equation}
    \label{defbeta}
\beta\equiv\beta(z)=\frac{\left[1+\left(2-m^2h_0^2\right)z^2+\chi(z)\right]^{1/2}}{\sqrt{2}\ell},
\end{equation}
\begin{equation}
    \label{defgamma}
\gamma\equiv\gamma(z)=\frac{\left[1+\left(2-m^2h_0^2\right)z^2-\chi(z)\right]^{1/2}}{\sqrt{2}\ell},
\end{equation}
\begin{equation}
    \label{defchi}
\chi\equiv\chi(z)=\left[1+2\left(2-h_0^2\right)m^2z^2+m^4h_0^4z^4\right]^{1/2},
\end{equation}
with $z=s\ell$ being a dimensionless complex variable. 

The complex function $\chi(z)$ has four branch points at $\pm{ib_1}$ and $\pm{ib_2}$, where
\begin{equation}
    \label{b1b2}
b_{1,2}=\frac{\left(2-h_0^2\mp2\left(1-h_0^2\right)^{1/2}\right)^{1/2}}{mh_0^2}.
\end{equation}
In particular, when $h_0<1$, $b_{1,2}$ are always real and the branch points are located along the imaginary axis. In this case, the branch cuts are chosen to run from $\pm{ib_1}$ to $\pm{ib_2}$ (see Fig A.1a in the Appendix A). On the other hand, when $h_0>1$, $b_{1,2}$ are complex and the branch points are symmetrically located with respect to the real axis at the four quadrants of the complex $z$-plane (Fig. A.1b). Accordingly, $z=0$ and $ z=\pm{ib_0}$ with
\begin{equation}
    \label{b0}
b_{0}=\frac{\left(1-m^2\right)^{1/2}}{\left(1-m^2h_0^2\right)^{1/2}},
\end{equation}
are additional branch points of the functions $\beta(z)$ or $\gamma(z)$ depending on the values of the parameters $m$ and $h_0$ (the branch cuts of these functions are given in Appendix A). Note, that $b_0$ is always real when $m<m_R$. In any case, the specific introduction of the branch cuts secures that the functions $(\chi,\beta,\gamma)$, are single-valued and positive along the real axis.

Moreover, the transformed expressions for the stresses and displacements that enter the boundary conditions take the following form
\begin{equation}
   \begin{split}
    \label{tr-bc1}
\sigma^*_{yx}(s,Y)=-\mu\biggl[\ell^2\frac{d^4\psi^*}{dY^4}&-\bigl(1+(2-m^2h_0^2)\ell^2s^2\bigr)\frac{d^2\psi^*}{dY^2}\\
&-\bigl(1-(1-m^2h_0^2)\ell^2s^2\bigr)s^2\psi^*+2is\frac{d\phi^*}{dY}\biggr],
   \end{split}
\end{equation}
\begin{equation}
    \label{tr-bc2}
\sigma^*_{yy}(s,Y)=(\lambda+2\mu)\frac{d^2\phi^*}{dY^2}-\lambda s^2\phi^*+2i\mu s \frac{d\psi^*}{dY},
\end{equation}
\begin{equation}
    \label{tr-bc3}
m^*_{yz}(s,Y)=-2\mu\ell^2\left(\frac{d^3\psi^*}{dY^3}-s^2\frac{d\psi^*}{dY}\right),
\end{equation}
\begin{equation}
    \label{tr-bc4}
u^*_{x}(s,Y)=\frac{d\psi^*}{dY}-is\phi^*.
\end{equation}
Next, in preparation for formulating a Wiener-Hopf equation, the unilateral Fourier transforms of the unknown stress $\sigma_{yx}(X\!>\!0,Y\!=\!0)$ ahead of the crack tip, and the unknown crack-face displacement $u{_x}(X\!<\!0,Y\!=\!0)$ are defined as follows
\begin{equation}
\label{tr-stress+}
\Sigma^{+}(s)=\int_{0}^{\infty}\sigma_{yx}(X,Y=0)e^{isX}\,dX, \quad \sigma_{yx}(X,Y\!=\!0)=\frac{1}{2\pi}\int_{\mathcal{L}}\Sigma^{+}(s)e^{-isX}\,ds,
\end{equation}
and
\begin{equation}
\label{tr-u-}
U^{-}(s)=\int_{-\infty}^{0}u_{x}(X,Y=0)e^{isX}\,dX, \quad u_{x}(X,Y\!=\!0)=\frac{1}{2\pi}\int_{\mathcal{L}}U^{-}(s)e^{-isX}\,ds,
\end{equation}
where the inversion path is considered to lie inside the region of analyticity of each transformed function. In particular, we assume the following \textit{finiteness conditions} to hold: $|\sigma_{yx}(X,Y\!=\!0)|<M$ for $X\!\to\!+\infty$ and $|u_{x}(X,Y\!=\!0)|<N$ for $X\!\to\!-\infty$, where $(M,N)$ are positive constants. Consequently, the transformed function $\Sigma^{+}(s)$ is analytic and defined in the upper half-plane $\textup{Im}(s)>0$, while $U^{-}(s)$ is analytic and defined in the lower half-plane $\textup{Im}(s)<0$.

Enforcing the boundary conditions \eqref{bc2}-\eqref{bc4}, taking into account also \eqref{tr-bc2}-\eqref{tr-bc4} and \eqref{tr-u-}$_1$, results in the following equations for the unknown functions $A(s)$, $B(s)$ and $C(s)$
\begin{equation}
    \label{A}
A(s)=\frac{2i}{m^2s}U^{-}(s),
\end{equation}
\begin{equation}
    \label{B}
B(s)=-\frac{(2-m^2)(\gamma^2-s^2)}{m^2\beta(\beta^2-\gamma^2)}U^{-}(s),
\end{equation}
\begin{equation}
    \label{C}
C(s)=\frac{(2-m^2)(\beta^2-s^2)}{m^2\gamma(\beta^2-\gamma^2)}U^{-}(s).
\end{equation}
Consequently, Eqs. \eqref{A}-\eqref{C}, together with \eqref{bc1} and \eqref{tr-bc1} provide the final Wiener-Hopf equation of the problem, connecting the two unknown functions $\Sigma^+(s)$ and $U^-(s)$
\begin{equation}
    \label{WH1}
\Sigma^{+}(s)-T^{-}(s)=\frac{\mu(2-m^2)^2\,s^2}{m^2 \left[ (s\ell)^2 \right]^{1/2} } K(s\ell) U^{-}(s),
\end{equation}
where $T^{-}(s)$ is the Fourier transform of the loading along the crack-faces. The kernel function $K(s\ell)$ is given as
\begin{equation}
    \label{defK}
K(z)=\frac{\theta^2+\left[z^2\right]^{1/2}\theta+m^2}{(\beta+\gamma)\, \theta} - \ell d,
\end{equation}
where
\begin{equation}
    \label{deftheta}
\theta\equiv \theta(z)=\ell^2\,\frac{\beta(z)\gamma(z)} {\left[z^2\right]^{1/2}}=\left(1-m^2+(1-m^2h_0^2)z^2\right)^{1/2},
\end{equation}
and $d \equiv d(m,\nu)=4\left(1-m^2c^2\right)^{1/2}/\left(2-m^2\right)^2$. The problem has now been reduced to the determination of the unknown functions $\Sigma^{+}(s)$ and $U^-(s)$ from the single functional equation \eqref{WH1}.
 

\subsection{Wiener-Hopf factorization}

\noindent
To proceed further a product-factorization of the kernel $K(z)$ is required. First it is checked that $K(z)$ has no zeros and no poles in the complex plane in the sub-Rayleigh regime ($m \leq m_R$). This was verified by using the symbolic computer program MATHEMATICA$^{\textup{TM}}$. 

Moreover, the kernel function exhibits the following asymptotic behavior
\begin{gather}
\lim_{|z|\to\infty}K(z)=\ell(1-d)+O(z^{-2}), \label{limKinf} \\
\lim_{|z|\to0}K(z)=\ell\left[\left(1-m^2\right)^{-1/2}-d\right]+O(z^2). \label{limK0}
\end{gather}
This leads us to introduce a modified kernel given as $N(z)=K(z)/\left[\ell(1-d)\right]$, which possesses the desired asymptotic property $\lim_{|z|\to\infty}N(z)=1$. Indeed, this new form of the kernel facilitates its product splitting by the use of Cauchy's integral theorem (Noble, 1958; Roos, 1969). The functional equation \eqref{WH1} takes now the form
\begin{equation}
    \label{WH2}
\Sigma^{+}(s)-T^{-}(s)=\frac{\mu\ell(1-d)\left(2-m^2\right)^2\,s^2}{m^2 \left[s\ell\right]_+^{1/2}\cdot\left[s\ell\right]_-^{1/2} } N(s\ell) U^{-}(s),
\end{equation}
where the function $\left[\left(s\ell\right)^2\right]^{1/2}\!\equiv\left[z^2\right]^{1/2}$ is written as a product of two analytic functions in the upper and lower half-plane, respectively (see e.g. Mishuris et al., 2012)
\begin{equation}
    \label{decompz}
\left[z^2\right]^{1/2}=\left[z\right]_+^{1/2}\cdot\left[z\right]_-^{1/2}.
\end{equation}

\begin{figure}[!htcb]
\centering
\includegraphics[scale=0.4]{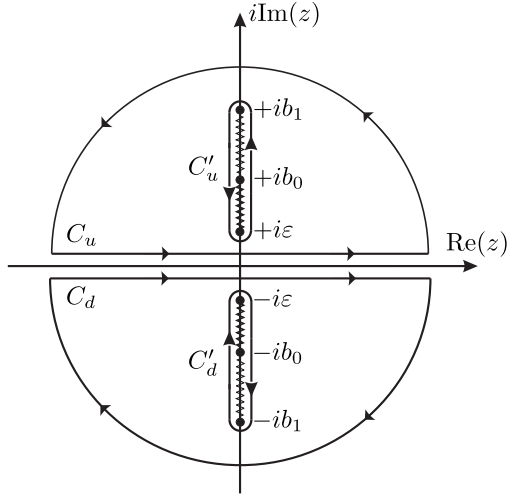}
\caption{Contour integration for the factorization of the kernel function $N(z)$.}
\label{fig03}
\end{figure}

In addition, the modified kernel splits up as
\begin{equation}
    \label{decompN}
N(z)=N^{+}(z)\cdot N^-(z),
\end{equation}
where 
\begin{equation}
    \label{Nplus}
N^+(z)=\exp{\left\{\frac{1}{2\pi i}\bigintssss_{C_d}\frac{\log\left[N(\zeta)\right]}{\zeta-z}\,d\zeta\right\}},
\end{equation}
\begin{equation}
\label{Nminus}
N^-(z)=\exp{\left\{-\frac{1}{2\pi i}\bigintssss_{C_u}\frac{\log\left[N(\zeta)\right]}{\zeta-z}\,d\zeta\right\}}.
\end{equation}
The use of the Cauchy integral theorem is depicted in Figure \ref{fig03}. The functions $N^+(z)$ and $N^-(z)$ are analytic and nonzero in the half-planes $\textup{Im}(z)>-\varepsilon$ and 
$\textup{Im}(z)<\varepsilon$, respectively, with $\varepsilon$ being a real number such that $\varepsilon\to0$. $\!$ In fact, introducing $\varepsilon$ facilitates the introduction of the 
branch cuts for $N(z)$ (see e.g. Georgiadis, 2003). The original integration paths $C_d$ and $C_u$ extend parallel to the real axis in the complex $z$-plane. According to the Cauchy integral 
theorem, Jordan's lemma, and by taking into account that $N(z)$ has no poles or zeros in the finite complex plane, we are allowed to deform the original integration paths and shrink them to $(C'_d,C'_u)$ 
contours around the branch cuts of $N(z)$ extending along: $\pm \varepsilon \leq \textup{Im}(z) \leq \pm b_0$, and  $\pm b_0 \leq \textup{Im}(z) \leq \pm b_1$ (Fig. \ref{fig03}). Note that the second branch of $N(z)$ exists only in 
certain cases, depending on the values of $(m,h_0)$. A more detailed discussion about the branch points and the pertinent branch cuts of the kernel function $N(z)$ is given in Appendix B.

This eventually leads to the following forms of the sectionally analytic functions $N^{\pm}(z)$ (the analytic derivation is provided in Appendix B) 
\begin{equation}
\label{N+-}
N^{\pm}(z)=\exp{\left\{-\frac{1}{\pi}\left[\bigintssss_{0}^{ib_0}\!\!\tan^{-1}\!\left[\frac{\textup{Im}\left(N(\zeta)\right)}{\textup{Re}\left(N(\zeta)\right)}\right] \frac{d\zeta}{\zeta \pm z}+a\bigintssss_{ib_0}^{ib_1}\!\!\tan^{-1}\!\left[\frac{\textup{Im}\left(N(\zeta)\right)}{\textup{Re}\left(N(\zeta)\right)}\right] \frac{d\zeta}{\zeta \pm z}\right]\right\}},
\end{equation}
with the properties $N^+(-z)=N^-(z)$ and $N^{\pm}(z)=1$ for $|z|\to\infty$. The constant $a$ takes the values 0 or 1 depending on the branches of the kernel function $N(z)$, and is defined in Appendix B. Moreover, it is noted that when $z$ approaches the branch cuts in the  $z$-plane from one side or the other, the integrals defining the functions $N^+(z)$ or $N^-(z)$ become Cauchy singular integrals. However, these integrals are not singular simultaneously. In fact, considering that the function $N(z)$ does not have any poles or zeros in the complex $z$-plane, we can extend by analytic continuation the functions $N^+(z)$ and $N^-(z)$ below and above the real axis, respectively, excluding their pertinent branch cuts. In this way, we may calculate these exceptional cases without resorting to principal values by employing $N^{\pm}(z)=N(z)/N^{\mp}(z)$, where we choose in the denominator the $N^{\pm}(z)$ function that is not defined by a singular integral.

In view of the above, Eq. \eqref{WH2} can now be rewritten as
\begin{equation}
    \label{WH3}
\frac{\Sigma^{+}(s)\,\left[s\ell\right]_+^{1/2}}{N^{+}(s\ell)}=\frac{\mu\ell(1-d)\left(2-m^2\right)^2\,s^2}{m^2\left[s\ell\right]_-^{1/2} } N^{-}(s\ell) U^{-}(s)+\frac{T^{-}(s)\,\left[s\ell\right]_+^{1/2}}{N^{+}(s\ell)}.
\end{equation}

\subsection{Solution of the Wiener-Hopf equation}
\noindent
We now assume the following form of the loading applied on the crack faces
\begin{equation}
\label{load}
T(X)=\frac{T_0}{L}\,e^{X/L} \quad \textup{with} \quad -\infty<X<0,
\end{equation}
where $T_0$ and $L$ are positive constants having pertinent dimensions. Transforming \eqref{load} with \eqref{dir}$_1$ we obtain
\begin{equation}
\label{tr-load}
T^-(s)=\frac{T_0}{1+isL},
\end{equation}
so that \eqref{WH3} becomes
\begin{equation}
    \label{WH4}
\frac{\Sigma^{+}(s)\,\left[s\ell\right]_+^{1/2}}{N^{+}(s\ell)}=\frac{\mu\ell(1-d)\left(2-m^2\right)^2\,s^2}{m^2\left[s\ell\right]_-^{1/2} } N^{-}(s\ell) U^{-}(s)+\frac{T_0\,\left[s\ell\right]_+^{1/2}}{N^{+}(s\ell)(1+isL)}.
\end{equation}
The sum-splitting of the second term in the RHS of \eqref{WH4} is required to complete the decoupling process. This is be obtained by inspection as
\begin{equation}
    \label{defM}
M(z)\equiv\frac{T_0\,\left[z\right]_+^{1/2}}{N^{+}(z)(1+iz\ell^{-1}L)}=M^{+}(z)+M^{-}(z),
\end{equation}
where
\begin{equation}
    \label{Mplus}
M^{+}(z)=\frac{T_0}{(1+iz\ell^{-1}L)}\left[\frac{\left[z\right]_+^{1/2}}{N^{+}(z)}-\frac{[i\ell/L]_+^{1/2}}{N^{+}(i\ell/L)}\right],
\end{equation}
\begin{equation}
    \label{Mminus}
M^{-}(z)=\frac{T_0}{(1+iz\ell^{-1}L)}\frac{[i\ell/L]_+^{1/2}}{N^{+}(i\ell/L)},
\end{equation}
with $M^{+}(z)$ being an analytic function in the same upper half-plane where $N^{+}(z)$ is defined, while $M^{-}(z)$ is an analytic function in the half-plane where $\textup{Im}(z)<i\ell/L$. Equations \eqref{Mplus} and \eqref{Mminus}, when combined, allow the final re-arrangement of the Wiener-Hopf equation
\begin{equation}
    \label{WH5}
\frac{\Sigma^{+}(s)\,\left[s\ell\right]_+^{1/2}}{N^{+}(s\ell)}-M^+(s\ell)=\frac{\mu\ell(1-d)\left(2-m^2\right)^2\,s^2}{m^2\left[s\ell\right]_-^{1/2} } N^{-}(s\ell) U^{-}(s)+M^-(s\ell)\equiv E(s\ell).
\end{equation}
The above functional equation defines the function $E(z)$ only on the real line. In order to evaluate this function, it is first necessary to examine the asymptotic behavior of the functions $\Sigma^+(s)$ and $U^-(s)$. In particular, guided by the results concerning the modification of stress singularities in the presence of couple stresses in plane-strain crack problems subjected to a quasi-static shear loading (Huang et al., 1997; Gourgiotis et al., 2011), we assume that the stress and crack-face displacement along the crack faces exhibit the following asymptotic behavior
\begin{equation}
    \label{asymptSyx}
\sigma_{yx}(X,Y=0)=O(X^{-1/2}) \quad \textup{as} \quad X\to+0,
\end{equation}
\begin{equation}
    \label{asymptux}
u_{x}(X,Y=0)=O((-X)^{1/2}) \quad \textup{as} \quad X\to-0.
\end{equation}
Further, we consider at this point the transformation formula $X^\kappa\overset{FT}{\leftrightarrow} i^{k+1} \Gamma(\kappa+1)s^{-\kappa-1}$, where $\Gamma()$ is the Gamma function with $\kappa>-1$ ($\kappa\neq 0,+1,+2,...,$). The symbol $\overset{FT}{\leftrightarrow}$ means that the quantities on either side of the arrow are connected through the unilateral Fourier-transform. Then, employing theorems of the Abel-Tauber type (see e.g. Roos, 1969), we obtain the following asymptotic behavior in the transform domain
\begin{equation}
    \label{asympt-tr-Syx}
\Sigma^+(s)=O(s^{-1/2}) \quad \textup{as} \quad |s|\to+\infty \quad \textup{with}\quad \textup{Im}(s)>0,
\end{equation}
\begin{equation}
    \label{asympt-tr-ux}
U^-(s)=O(s^{-3/2}) \quad \textup{as} \quad |s|\to+\infty \quad \textup{with}\quad \textup{Im}(s)<0.
\end{equation}
In light of the above, and bearing in mind that: $N^{\pm}(s\ell) \to 1$ and $M^{\pm}(s\ell) \to 0$ as $|s|\to+\infty$, we conclude that the first member of \eqref{WH5} is a \textit{bounded} nonzero analytic function at infinity for $\textup{Im}(s)>0$, whereas the second member is a \textit{bounded} nonzero analytic function at infinity for $\textup{Im}(s)<0$. Then, based on the theorems of analytic continuation, the two members define one and the same analytic function $E(z)$ over the entire complex  $z$-plane. Moreover, Liouville's theorem leads to the conclusion that $E(z)=E_0$, where $E_0$ is a constant. 

The transformed shear stress is now given by \eqref{WH5} as
\begin{equation}
    \label{full-tr-syx}
\Sigma^+(s)=[E_0+M^+(s\ell)]\,N^+(s\ell)\left[s\ell\right]_+^{-1/2}.
\end{equation}
The constant $E_0$ can be determined from simple equilibrium considerations. In particular, from the requirement that the upper part of the body $Y\geq 0$ is in equilibrium in the 
\textup{moving} framework, we have: $\int_{-\infty}^{\infty}\sigma_{yx}(X,Y=0)\,dX=0$, which, by taking into account Eqs. \eqref{tr-stress+}$_1$ and \eqref{tr-load}, can be written as 
$-T_0+\Sigma^+(0)=0$. Further, in view of \eqref{Mplus} and \eqref{full-tr-syx}, we obtain the limit
\begin{equation}
    \label{limit S+}
\Sigma^+(0) \equiv \lim_{|s|\to 0}\Sigma^+(s)=\ell^{-1/2}\left[\frac{\left(1-m^2\right)^{-1/2}-d}{1-d}\right]^{1/2}\left[E_0-\frac{T_0\left[i\ell/L\right]_+^{1/2}}{N^+(i\ell/L)}\right]\,s^{-1/2}+T_0+O(s^{1/2}).
\end{equation}
Thus, the only possibility left from \eqref{limit S+} is the vanishing of the term inside the second bracket, i.e.
\begin{equation}
    \label{E0}
E_0=\frac{T_0\left[i\ell/L\right]_+^{1/2}}{N^+(i\ell/L)}.
\end{equation}
Alternatively, the constant $E_0$ can be evaluated from the conditions at remote regions in the physical plane where the couple-stress effects are diminished. First, we observe that the exponentially decaying tractions in \eqref{load} can be replaced by an equivalent concentrated load of intensity: $\int_{-\infty}^{0}\sigma_{yx}(X,Y=0)\,dX=-T_0$ at distance $L$ from the crack-tip. Thus, according to Saint-Venant's principle and in view of the classical solution of the steady-state problem of concentrated shear tractions along the crack faces (see e.g. Freund, 1990), we anticipate that for $X\to+\infty$ the shear stress ahead of the crack-tip will behave as in the classical theory i.e. as $\sim X^{-3/2}$. This, in turn, implies the following asymptotic behavior in the transformed domain: $\Sigma^+(s)=O(s^{1/2})$ as $|s|\to 0$. Using this result in conjunction with Eq. \eqref{limit S+} provides the same value for the constant $E_0$.

The final transformed expressions (valid for all $s$ in the pertinent half-plane of convergence) for the stress ahead of the crack-tip and the crack-face displacement then become
\begin{equation}
    \label{ftr-S}
\Sigma^+(s)=\left[\frac{T_0\left[i\ell/L\right]_+^{1/2}}{N^+(i\ell/L)}+M^+(s\ell)\right]\,N^+(s\ell)\left[s\ell\right]_+^{-1/2}, \quad \textup{Im}(s)>0,
\end{equation}
\begin{equation}
    \label{ftr-U}
U^{-}(s)=\frac{m^2}{\mu\ell (1-d)\left(2-m^2\right)^2} \left[\frac{T_0\left[i\ell/L\right]_+^{1/2}}{N^+(i\ell/L)}-M^-(s\ell)\right] \frac{\left[s\ell\right]_-^{1/2}}{s^2 N^{-}(s\ell)}, \quad \textup{Im}(s)<0.
\end{equation}
The limits of the latter expressions for $|s|\to\infty$ are found to be
\begin{equation}
    \label{ftr-asympt-S}
\Sigma^+(s)=\frac{T_0\left[i\right]_+^{1/2}}{\sqrt{L}\cdot N^+(i\ell/L)} s^{-1/2}+O(s^{-3/2}),
\end{equation}
\begin{equation}
    \label{ftr-asympt-U}
U^{-}(s)=\frac{T_0 m^2}{\mu \sqrt{L} (1-d)\left(2-m^2\right)^2} \frac{\left[i\right]_+^{1/2}}{N^+(i\ell/L)} s^{-3/2}+O(s^{-5/2}),
\end{equation}
which, accordingly, provide the following near-tip field as 
\begin{equation}
    \label{fasympt-syx}
\sigma_{yx}(X\to+0,Y=0)=\frac{T_0}{\sqrt{\pi L}\,N^+(i\ell/L)} X^{-1/2},
\end{equation}
\begin{equation}
    \label{fasympt-ux}
u_{x}(X\to-0,Y=0)=\frac{2T_0\,m^2}{\mu \sqrt{\pi L} (d-1)\left(2-m^2\right)^2 \, N^+(i\ell/L)} \left(-X\right)^{1/2}.
\end{equation}
%

\section{Results}

\subsection{Analytical representation of displacements, stresses and couple-stresses}
\noindent The shear stress ahead of the crack tip and the crack opening displacement can be obtained from \eqref{ftr-S} and \eqref{ftr-U}, respectively, by employing the inverse Fourier 
transform according to \eqref{dir}$_2$. For a crack propagating with a sub-Rayleigh speed, the functions in \eqref{ftr-S} and \eqref{ftr-U} do not have branch points or poles along the real 
line, consequently, the path of integration $\mathcal{L}$ coincides with the real line $s$. In this case we obtain
\begin{equation}
    \label{syx}
\sigma_{yx}(X)=\frac{1}{2\pi}\int_{-\infty}^{\infty} \Sigma^+(s) e^{-iXs} ds, \qquad X>0.
\end{equation}
\begin{equation}
    \label{ux}
u_x(X)=\frac{1}{2\pi}\int_{-\infty}^{\infty} U^-(s) e^{-iXs} ds, \qquad X<0,
\end{equation}
Alternatively, taking into account that $\Sigma^+(s)$ and $U^-(s)$ are analytic in the upper and lower half-planes respectively, we can deform the original integration paths by making use of Jordan's lemma, around the pertinent branch cuts of these functions. In particular, the branch cuts for the functions $U^-(s)$ and $\Sigma^+(s)$ can be chosen to be along the negative ($\textup{Im}(s)<0)$ and the positive ($\textup{Im}(s)>0$) imaginary axis, respectively.

\begin{figure}[!htcb]
\centering
\includegraphics[scale=0.272]{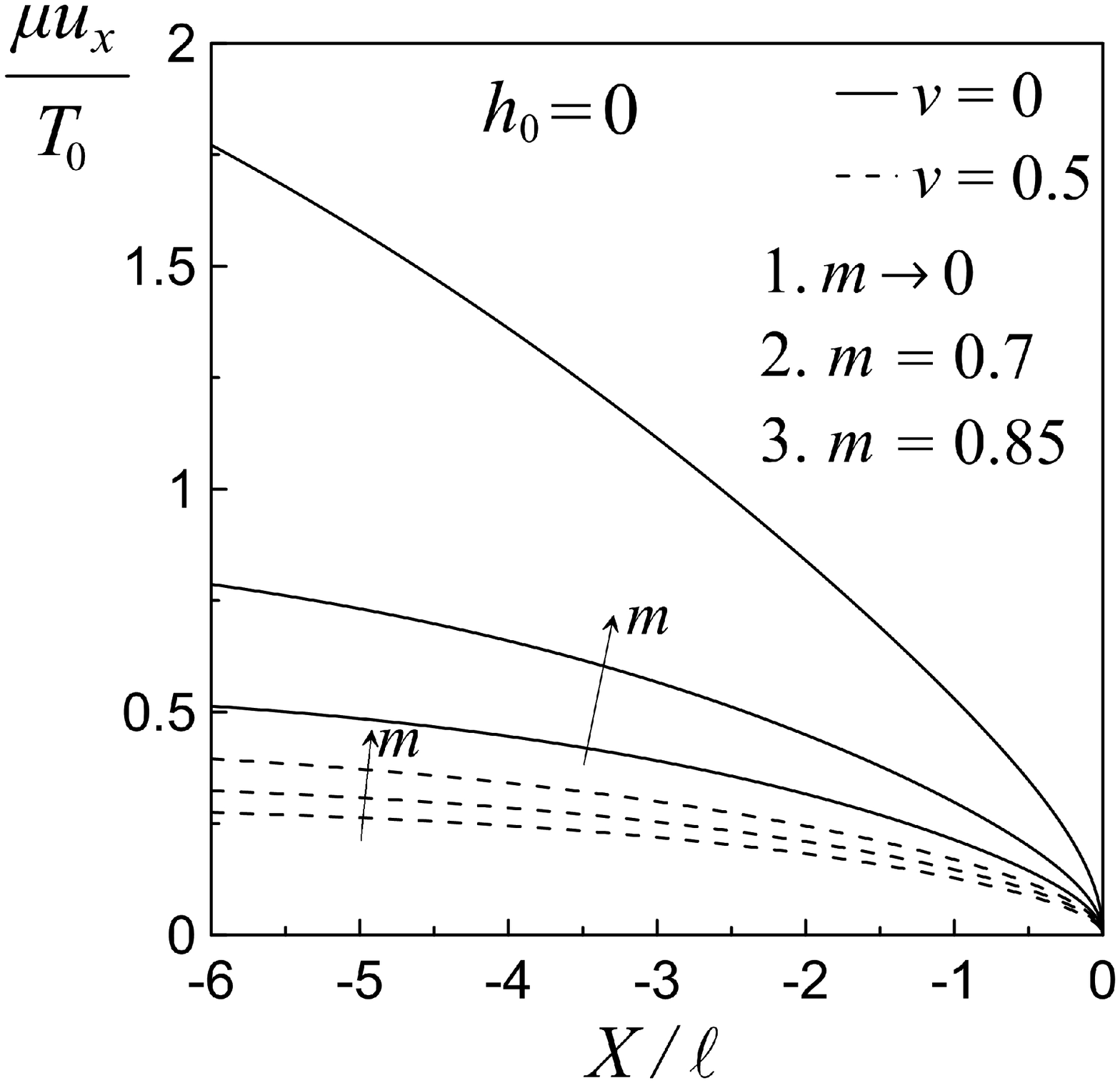}
\quad
\includegraphics[scale=0.272]{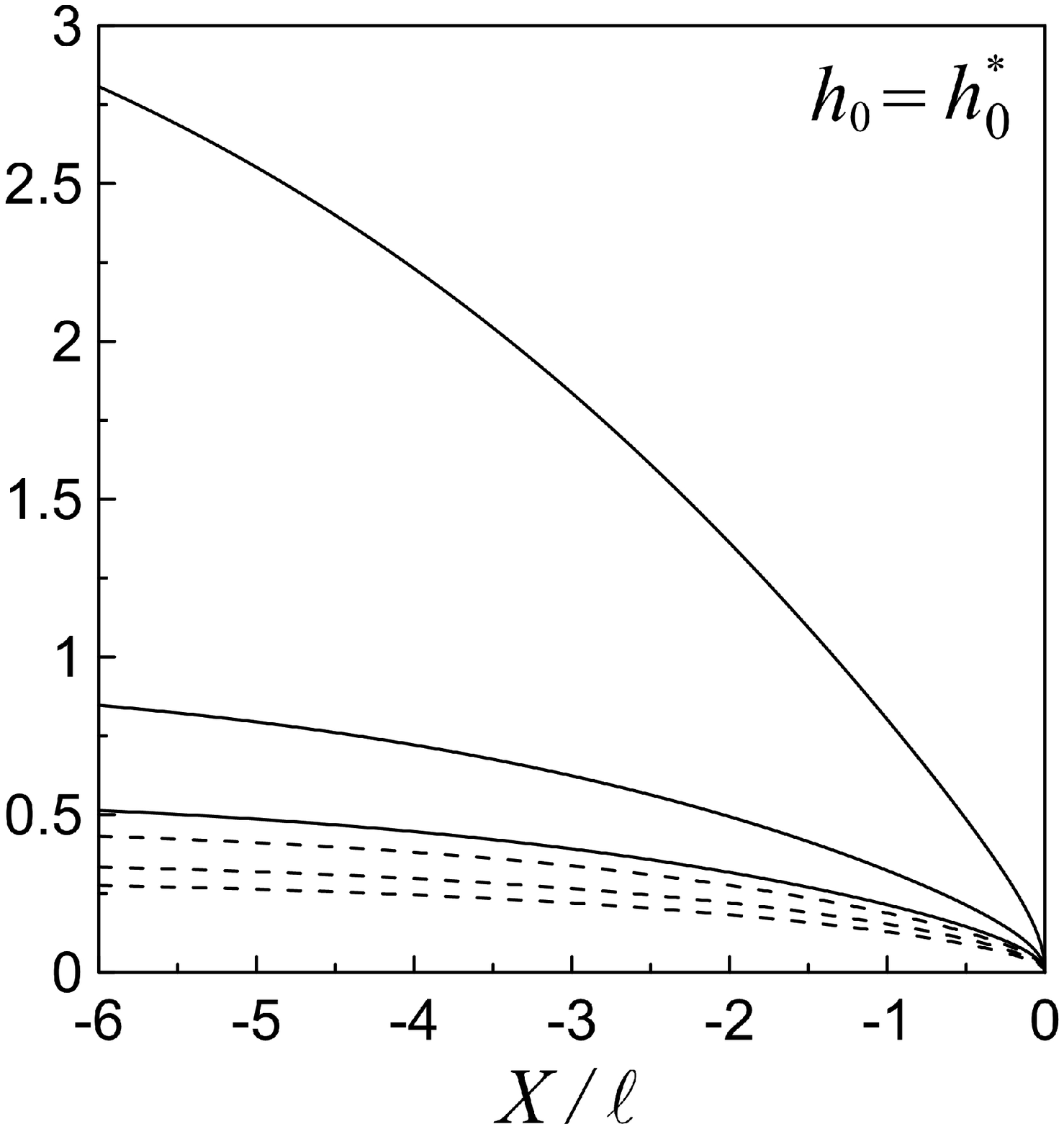}
\quad
\includegraphics[scale=0.272]{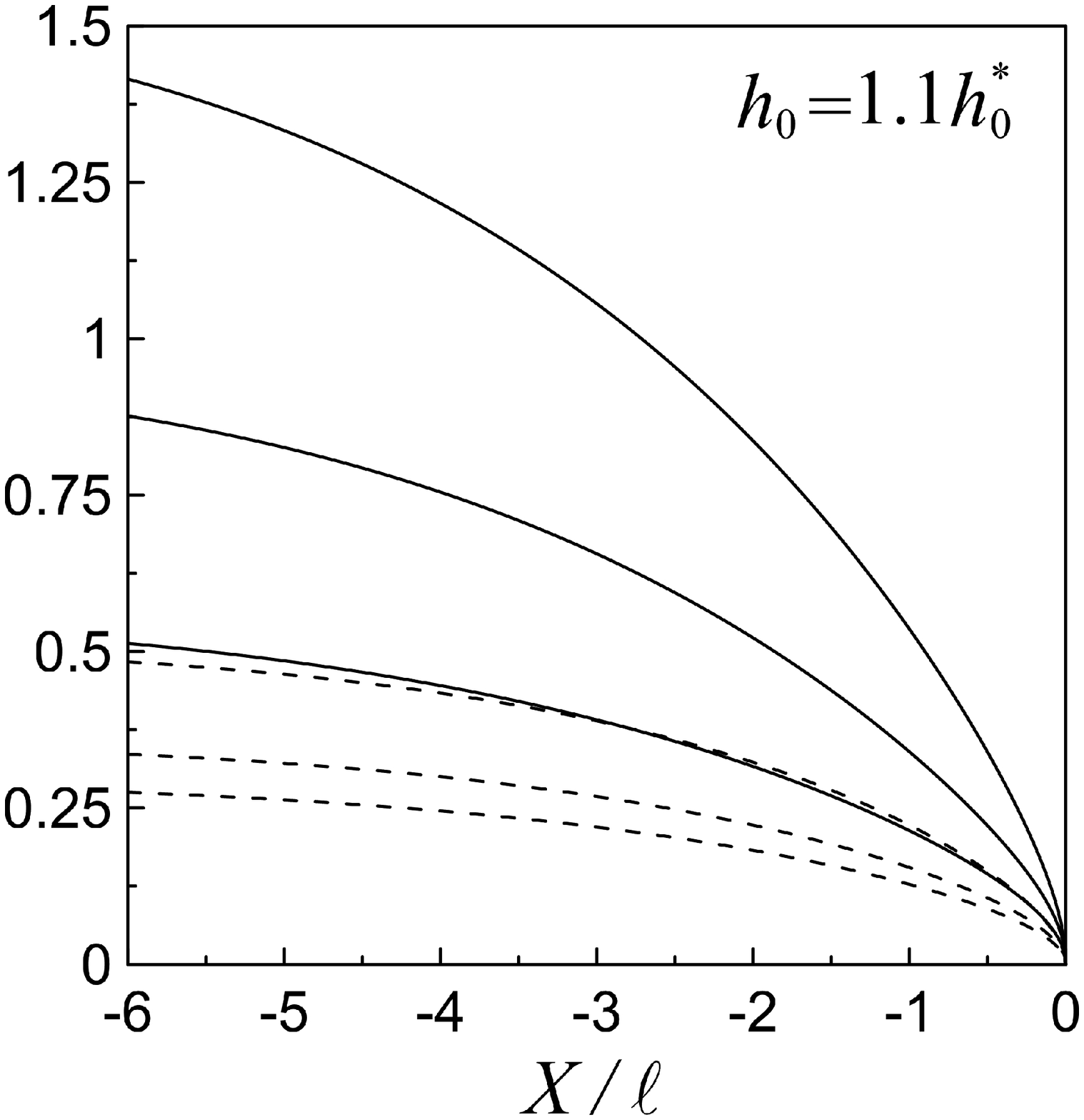}
\caption{Variation of the crack sliding displacement along the upper crack face for three different values: $h_0=\{0,h_0^*,1.1h_0^*\}$ and various crack speeds.}
\label{fig04}
\end{figure}

\begin{figure}[!htcb]
\centering
\includegraphics[scale=0.36]{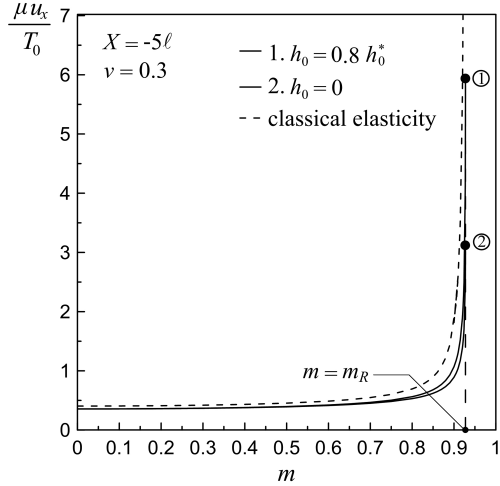}
\caption{Variation of the crack sliding displacement at the point $X=-5\ell$ in couple-stress elasticity and classical elasticity with respect to the normalized crack speed.}
\label{fig05}
\end{figure}

The variations of the normalized tangential displacement on the upper crack face $Y=+0$, $X\leq 0$, versus normalized distance $X/\ell$ to the crack tip are shown in Fig. \ref{fig04}, for a 
material 
with $L/\ell=10$ and a selected range of values of $m$, $h_0$ and $\nu$. It is noted that the specific value $L/\ell=10$ was chosen for most of the graphs in this paper, since for small values of this ratio the couple-stress effects are more pronounced. In fact, the effect of the ratio $L/\ell$ on the dynamic solution is the same as in the stationary mode II crack studied previously by Gourgiotis et al. (2012). Now, as is in classical elasticity, the crack tip profile is blunted in the couple-stress solution. Moreover, it observed, that the magnitude of the sliding displacement between the crack-faces increases as the propagation speed approaches the pertinent Rayleigh velocity, for every set of material parameters. However, the displacement in couple-stress elasticity remains always \textit{bounded} in the sub-Rayleigh regime. This result is in marked contrast with the classical elastodynamic solution, where the crack face displacements become infinite in magnitude as we approach the Rayleigh speed (Freund, 1990). This is illustrated more clearly in Fig. \ref{fig05}, where the variation of the crack face displacement at the point $X=-5\ell$ is shown with respect to the normalized propagation speed for $h_0=\{0, 0.8h_0^*\}$. Since, $h_0<h_0^*$, in both cases, the Rayleigh velocity takes the value $m_R=1/h_0^*=0.927$ for $\nu=0.3$. As noted earlier, when $m\to m_R$, the displacement in classical elasticity becomes unbounded, whereas the respective displacement in the couple-stress theory increases significantly but remains \textit{finite}. In addition, we note that for every propagating velocity $m$, the crack-face displacements are smaller than the respective ones in the classical theory. 
This finding shows that the crack becomes stiffer due to the effects of the microstructure, here brought into play by rotational gradients. This trend was also observed for the stationary mode II crack in couple-stress elasticity (Huang et al., 1999; Gourgiotis and Georgiadis, 2007), and proves that the microstructure may shield the crack tip from fracture. Finally, we remark that the influence of the 
Poisson's ratio $\nu$ on the solution becomes significant as the speed of the 
crack increases, and is more pronounced as $\nu \to 0$.

\begin{figure}[!htcb]
\centering
\includegraphics[scale=0.274]{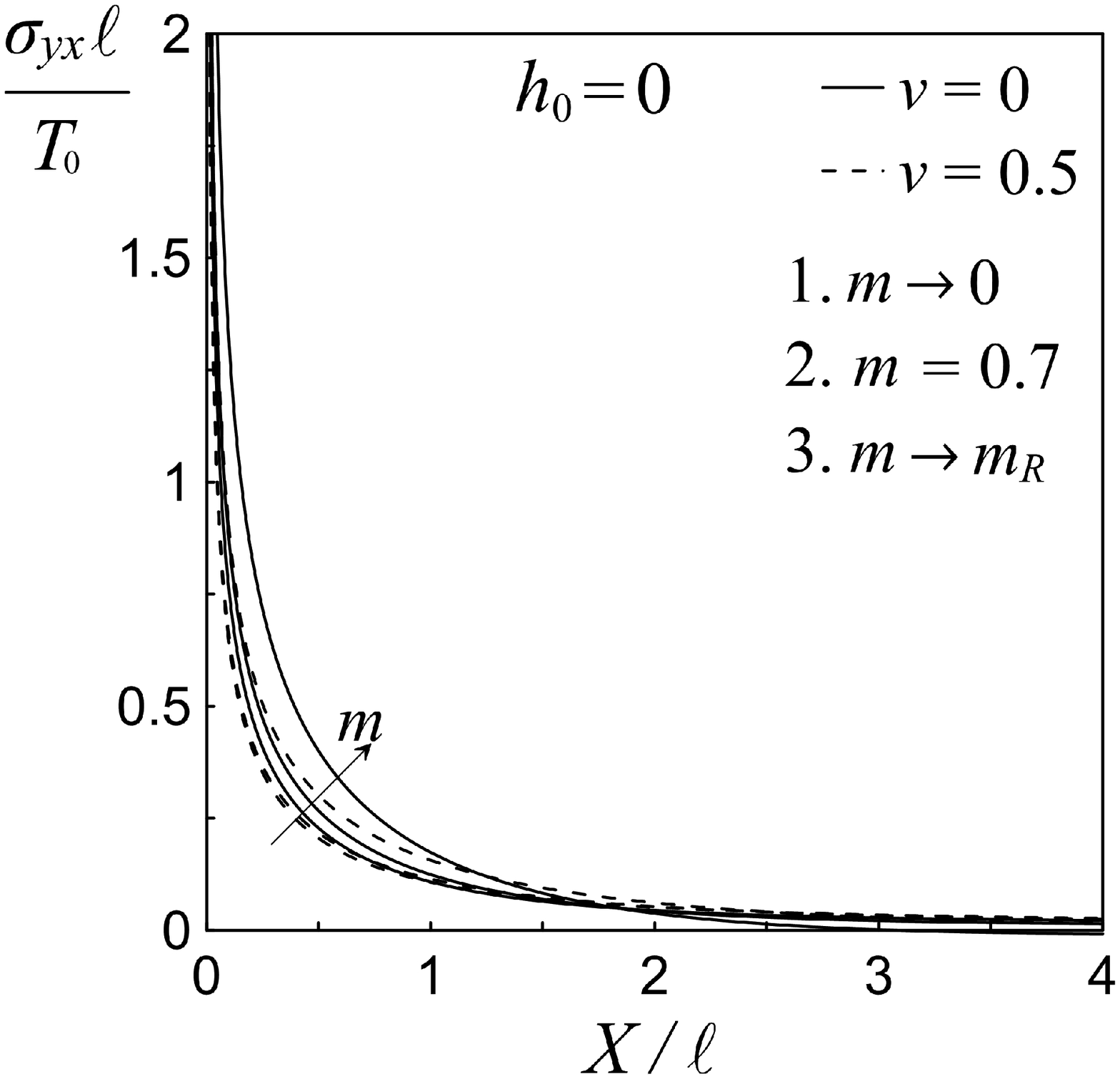}
\quad
\includegraphics[scale=0.274]{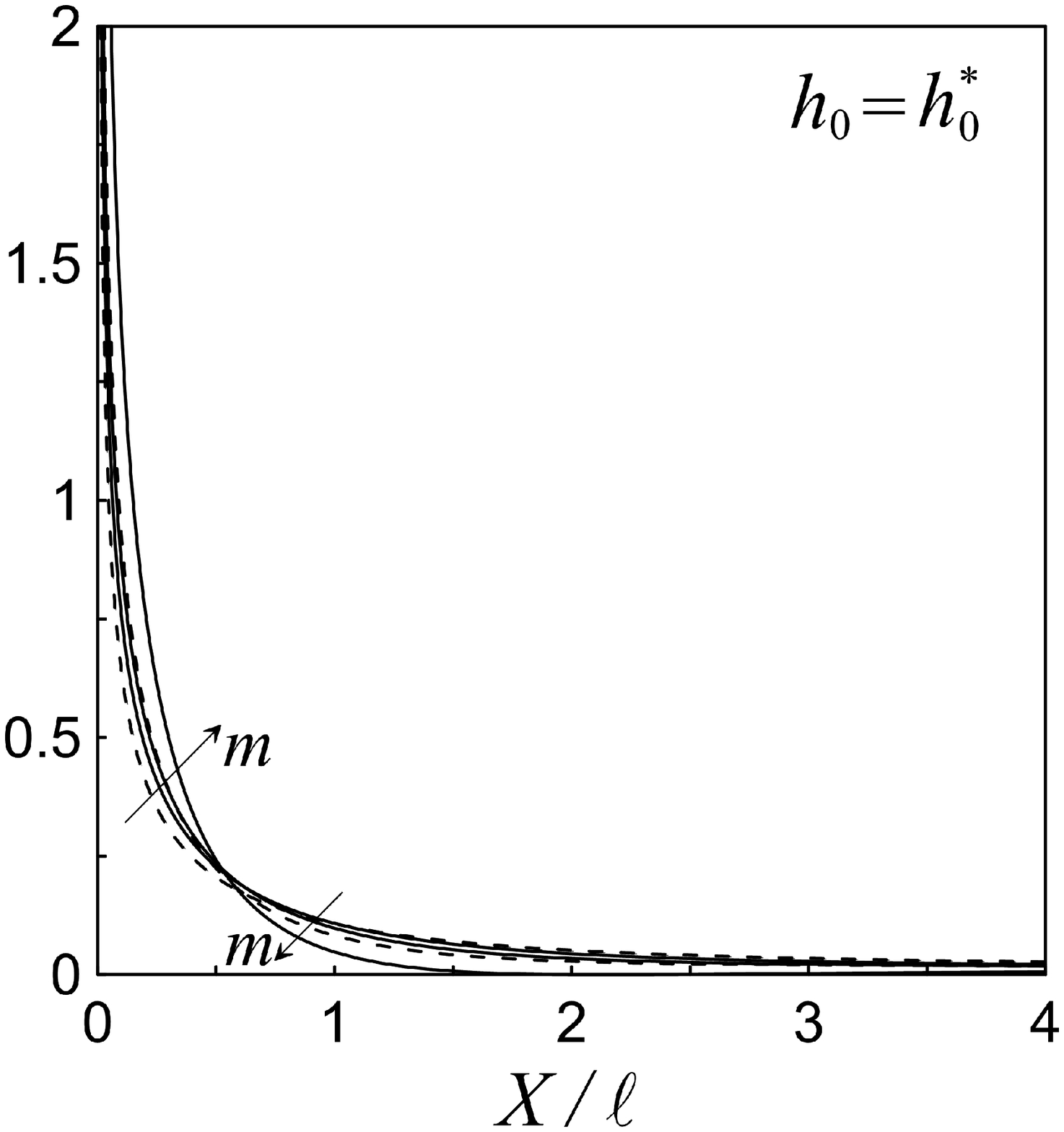}
\quad
\includegraphics[scale=0.274]{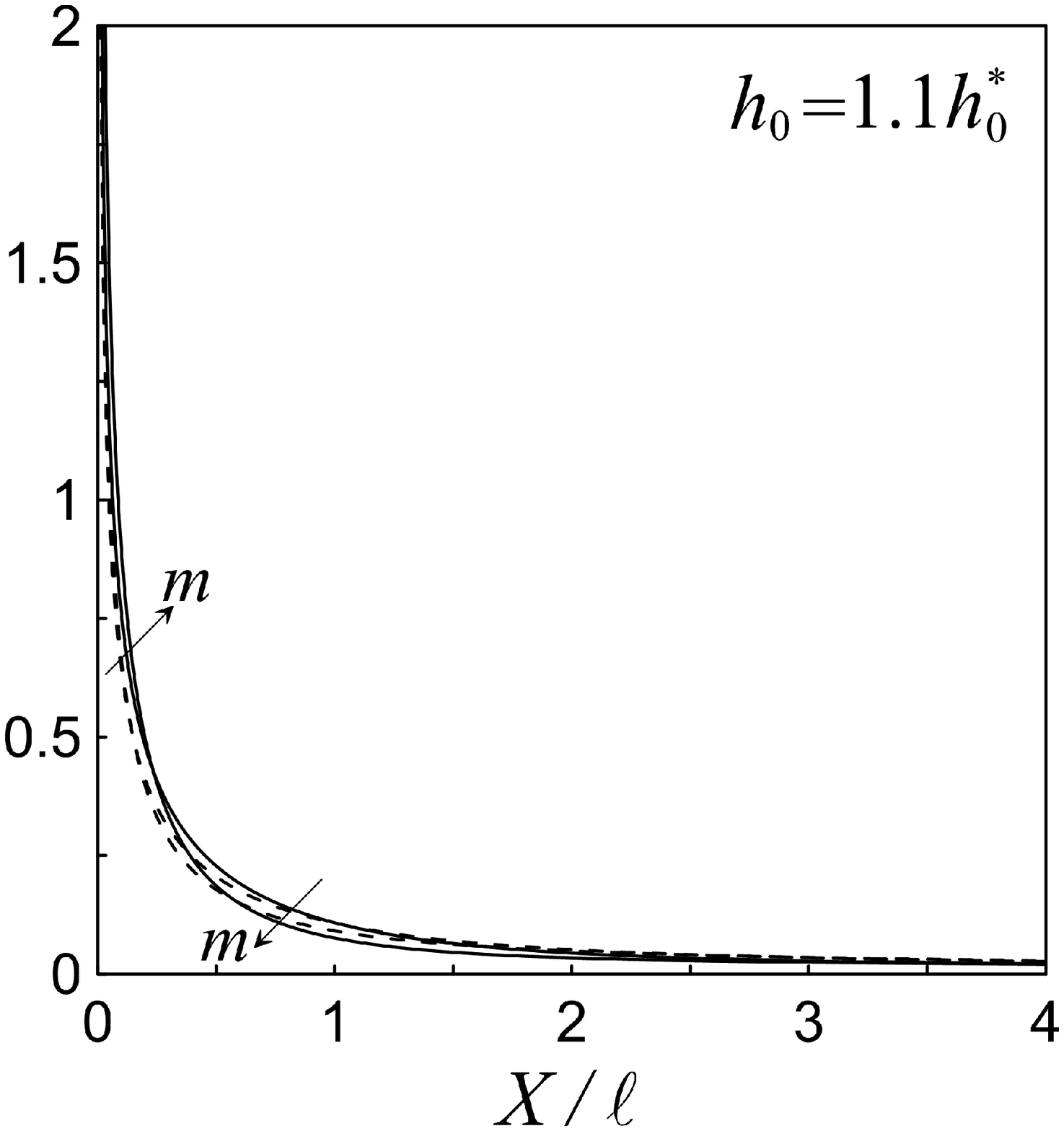}
\caption{Variation of the shear stress ahead of the crack-tip in couple-stress elasticity for three different values: $h_0=\{0,h_0^*,1.1h_0^*\}$ and various crack speeds.}
\label{fig06}
\end{figure}

\begin{figure}[!htcb]
\centering
\subfloat[]{\includegraphics[scale=0.36]{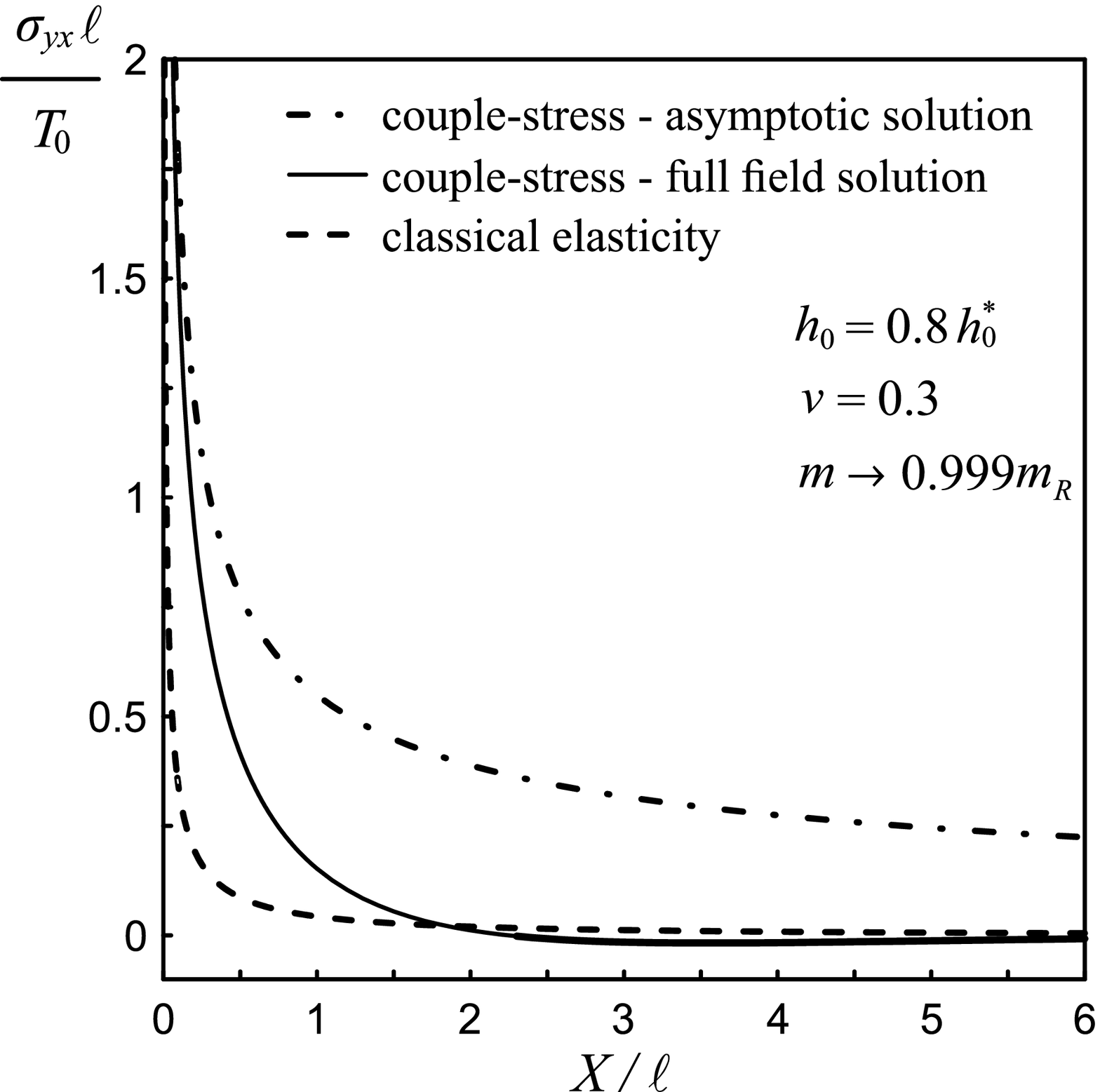}}
\quad
\subfloat[]{\includegraphics[scale=0.36]{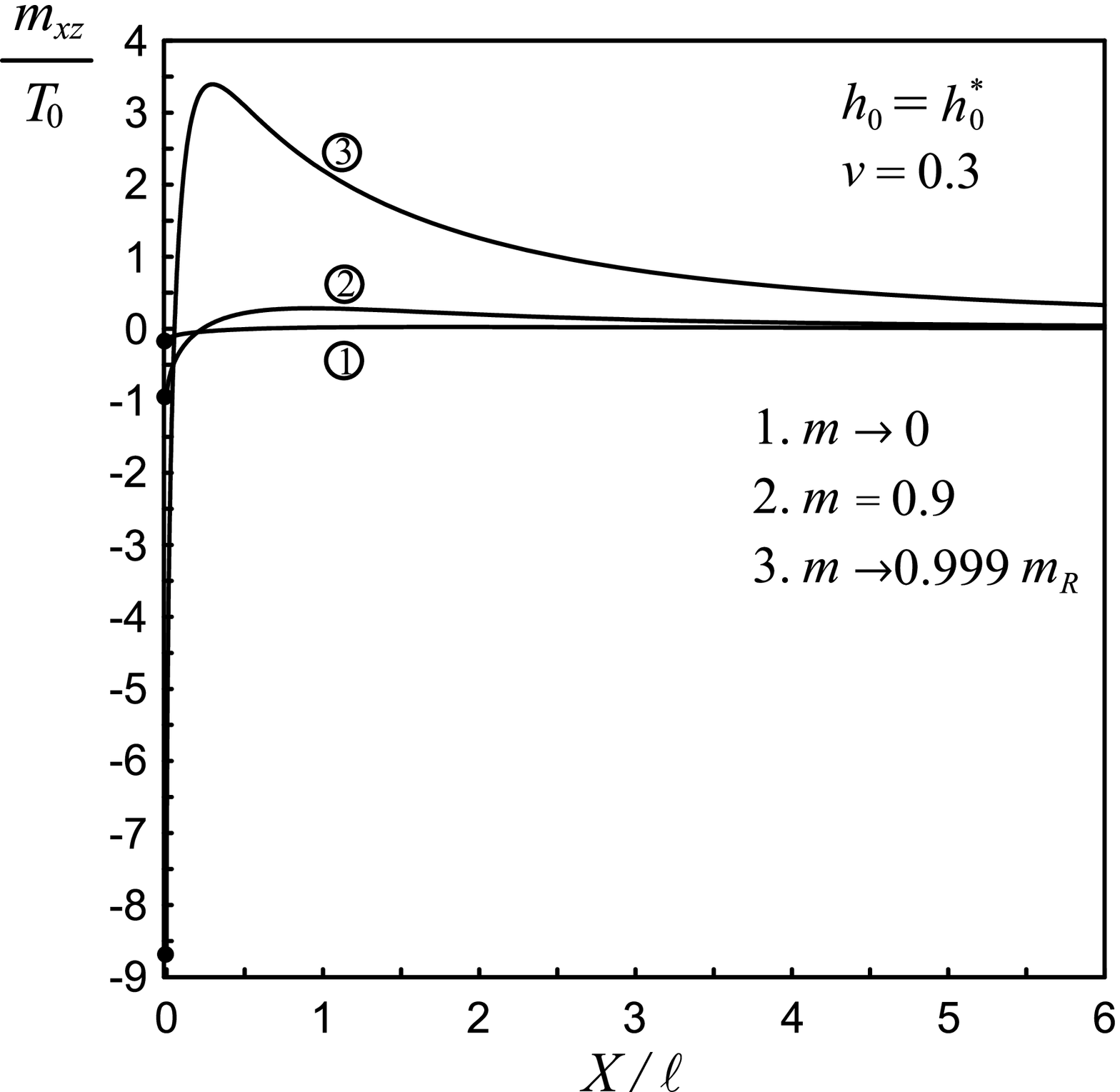}}
\caption{(a) Variation of the shear stress ahead of the crack-tip in couple-stress elasticity as the crack speed approaches the Rayleigh velocity. The near-tip asymptotic field and the 
classical elasticity solution are also shown. (b) Variation of the couple-stress $m_{xz}$ ahead of the crack-tip for a material with $\nu=0.3$ and $h_0=h_0^*$ and various crack speeds.}
\label{fig07}
\end{figure}

Next, the variation of the normalized shear stress ahead of the crack-tip with respect to the normalized distance $X/\ell$ is depicted in Fig. \ref{fig06}, for a material with $L/\ell=10$, 
$h_0=\{0, h_0^*,1.1h_0^*\}$ and $\nu=\{0, 0.5\}$. It is observed that the effect of the Poisson's ratio becomes more significant when $h_0 \to 0$ (i.e. in the case of no micro-inertia). 
Further, it is remarked that as $m \to 0$ the steady state solution approaches the static solution, irrespectively of the value of the ratio $h_0$, confirming thus previous results obtained 
by Gourgiotis et al. (2012) for the stationary crack. On the other hand, as $m \to m_R$, the zone within which the couple-stress effects become important increases. Interestingly, when the 
propagation speed approaches the Rayleigh velocity, the shear stress takes on negative values very close to the crack-tip within the range $2 \ell \leq X \leq 6\ell$, as is shown in Figure 
\ref{fig07}a for a material with $h_0=0.8h_0^*$ ($m_R=0.927$) and $\nu=0.3$. However, these negative 
values are very small in magnitude and appear \textit{only} in the case $m \to m_R$ and $h_0 \leq h_0^*$. In Figure \ref{fig07}a the near-tip couple-stress asymptotic solution in \eqref{fasympt-syx} and 
the classical elasticity solution are also depicted. It is noticed that the near-tip asymptotic field dominates within a zone of $0.5\ell$ to the crack tip, but the couple-stress effects are 
significant within a zone of $2\ell$. Moreover, as we move away from the crack-tip ($X \geq 8\ell$) the full field solution approaches gradually the classical elastodynamic solution for every 
crack speed. 

Finally, Figure \ref{fig07}b shows distribution the couple-stress $m_{xz}$ normalized by $T_0$, versus the distance $X/\ell$ to the crack tip for a material with $\nu=0.3$ and $h_0=h_0^*$. 
It is observed that the couple stress decays rapidly as the distance from the crack-tip increases. Also, it is interesting to note that $m_{xz}$ is \textit{bounded} at the crack-tip, as in the stationary crack case 
(Huang et al., 1997; Gourgiotis and Georgiadis, 2007).  In particular, the couple-stress $m_{xz}$ takes a finite negative value at the crack tip then increases to a bounded positive maximum, and 
diminishes monotonically to zero for $X >> \ell$, thus recovering the classical solution of linear elasticity. In general, the couple-stress $m_{xz}$ becomes more pronounced as the crack propagation speed approaches the pertinent Rayleigh velocity $m_R$ (curve 3).


\subsection{The dynamic stress intensity factor}
\noindent According to Eq.\eqref{fasympt-syx} and the definition $K_{II}=\lim_{X\to+0} (2\pi X)^{1/2} \sigma_{yx}(X,Y=0)$, the dynamic $(d)$ stress intensity factor (SIF) in couple-stress 
elasticity becomes
\begin{equation}
    \label{KIId}
K^{d}_{II}=\frac{\sqrt{2}T_0}{\sqrt{L}\cdot N^+(i\ell/L)}.
\end{equation}
In the limit case $m\to 0$, the dynamic SIF \eqref{KIId} degenerates to its static counterpart $K^{s}_{II}$ in couple-stress elasticity derived previously by Gourgiotis et al. (2012). The 
details of the pertinent limit procedure and an explicit expression for $K^{s}_{II}$ are given in Appendix C.

Furthermore, the ratio of the dynamic SIF in couple-stress elasticity to the respective one in the classical theory $(cl.)$ is
\begin{equation}
    \label{ratioK}
\frac{K^{d}_{II}}{K^{d_{cl.}}_{II}}=\frac{1}{N^+(i\ell/L)},
\end{equation}
where $K^{d_{cl.}}_{II}=\sqrt{2}T_0/\sqrt{L}$. 

\begin{figure}[!htcb]
\centering
\includegraphics[scale=0.36]{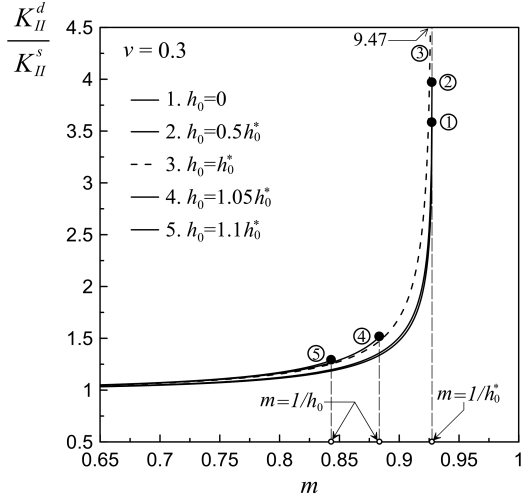}
\caption{Variation of the ratio of the dynamic SIF to the static SIF in couple-stress elasticity with respect to normalized crack-speed.}
\label{fig08}
\end{figure}

It is noted that in the classical elastodynamic theory the distribution of the shear stress ahead of the crack-tip does \textit{not} depend upon the crack speed $V$. Consequently, the dynamic 
stress intensity factor is identical to the corresponding static result for any velocity $V$: $K_{II}^{d_{cl.}}=K_{II}^{s_{cl.}}$ (Freund, 1990). This result is in marked contrast with the one 
obtained here in the context couple-stress elasticity, where the dynamic SIF in \eqref{KIId} depends upon the speed of the propagation through the function $N^+(i\ell/L)$. An analogous result 
was found also by Itou (1981) for a propagating Yoffe crack in couple-stress elasticity with no micro-inertia. This dependence is clearly depicted in Fig. \ref{fig08} for a material with $L/\ell=10$ 
and $\nu=0.3$ ($h_0^*=1.078$). Indeed, for curves 1-3, we have $h_0\leq h_0^*$, which, in turn, implies that 
the normalized crack velocity should be $m \leq 1/h_0^*$ in order for the crack to propagate with a sub-Rayleigh speed (see also Fig. \ref{fig02}). On the other hand, when $h_0>h_0^*$ (curves 4 and 5), 
the requirement for the crack-speed to be sub-Rayleigh is $m<1/h_0$. It is noteworthy that as the crack speed approaches the pertinent Rayleigh limit: $m \to m_R$, a significant but 
\textit{finite} increase of the ratio $K_{II}^{d}/K_{II}^{s}$ is observed. This increase is more pronounced as $h_0\to h_0^*$, where $K_{II}^{d} = 9.47 K_{II}^{s}$ (curve 3).

\begin{figure}[!htcb]
\centering
\subfloat[]{\includegraphics[scale=0.36]{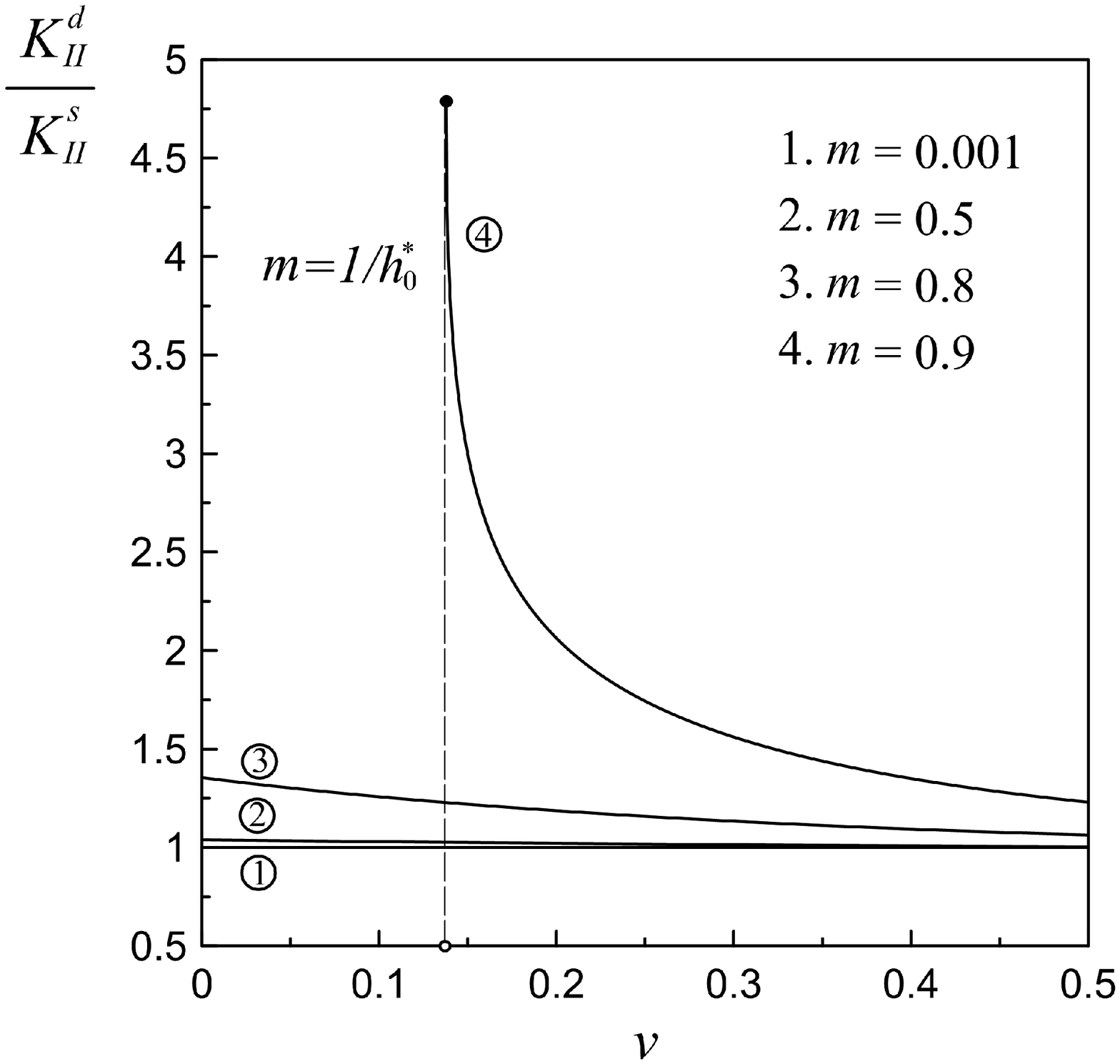}}
\quad
\subfloat[]{\includegraphics[scale=0.36]{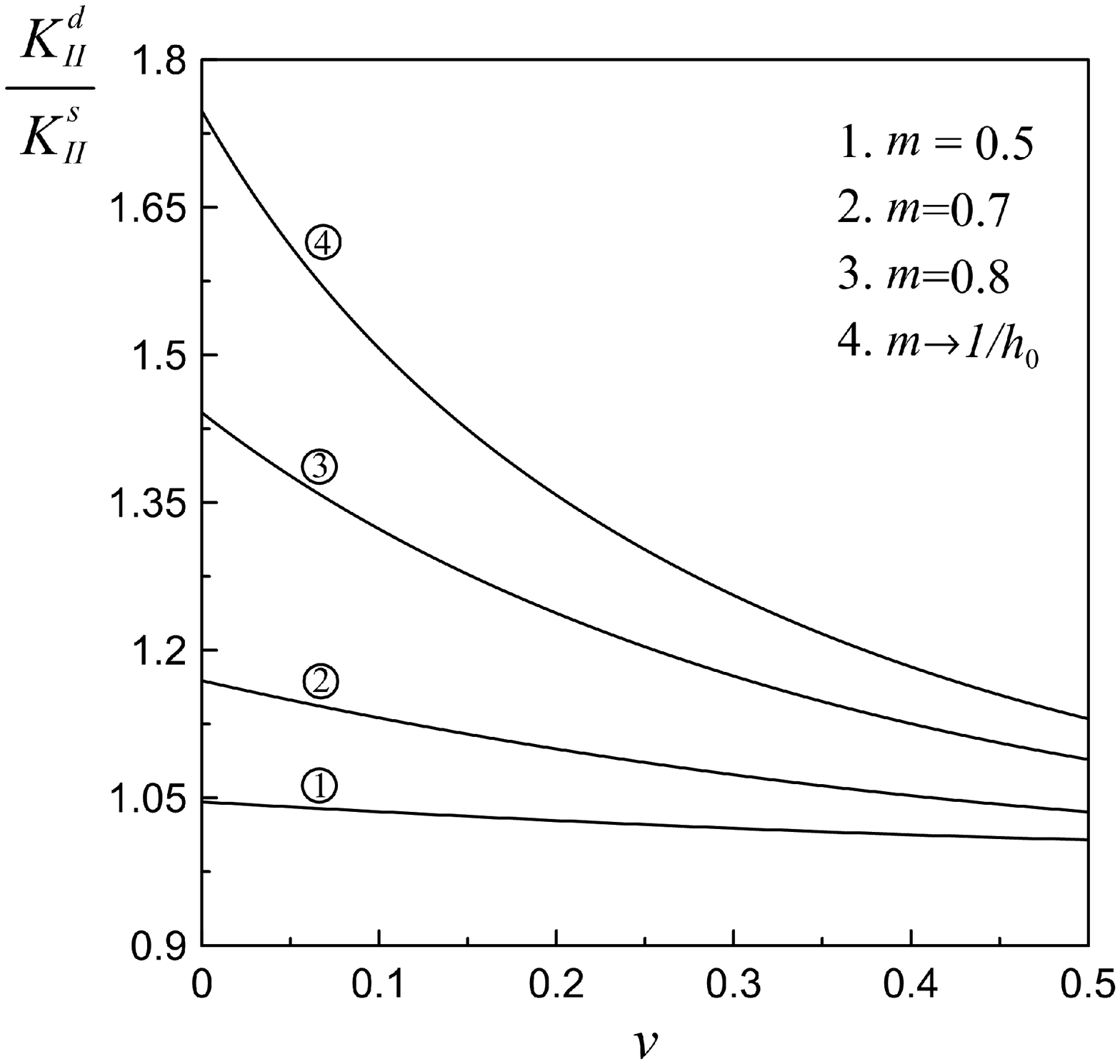}}
\caption{Variation of the ratio of the dynamic SIF to the static SIF in couple-stress elasticity with respect to the Poisson's ratio $\nu$ for: (a) $h_0=0.8$ and (b) $h_0=1.2$.}
\label{fig09}
\end{figure}

Figure \ref{fig09} illustrates the effect of Poisson's ratio on the variation of the ratio of the SIFs $K_{II}^{d}/K_{II}^{s}$, for two values of $h_0$. In particular, in Fig. \ref{fig09}a 
we have $h_0=0.8<h_0^*$ for all $\nu$ (recall that $1.046<h_0^*<1.144$) and, consequently, $m<1/h_0^*$ for the crack to propagate with sub-Rayleigh speed. This requirement is always satisfied 
for curves 1-3, however, for $m=0.9$ (curve 4), this inequality is satisfied only in the range $0.137<\nu<0.5$. In fact, as $\nu \to 0.137$ the propagation speed approaches the sub-Rayleigh limit speed, 
resulting to a steep increase of the dynamic SIF. On the other hand, for propagation velocities $m \leq 0.5$ the ratio varies slowly. Further, in Fig. \ref{fig09}b we have $h_0=1.2>h_0^*$ for 
all $\nu$, and thus, $m<1/h_0$. As $m \to 1/h_0$ an increase in the ratio $K_{II}^{d}/K_{II}^{s}$  is noted again, however, the increase is more moderate compared to the case $h_0 \leq h_0^*$.

\begin{figure}[!htcb]
\centering
\includegraphics[scale=0.36]{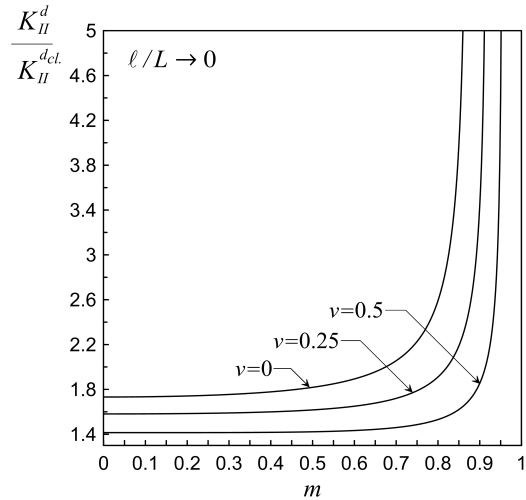}
\caption{Variation of the ratio of the dynamic SIFs in couple-stress and classical elasticity with respect to normalized crack-speed in the limit case $\ell/L \to 0$.}
\label{fig.new}
\end{figure}

Finally, we remark that in the limit case $\ell/L \to 0$ the ratio of the SIFs in couple-stress and in classical theory is not unity but exhibits an increase which becomes more pronounced as the crack-tip velocity approaches the pertinent Rayleigh limit (Fig.~\ref{fig.new}). In fact, by taking into account \eqref{limK0}, \eqref{decompN} and bearing in mind that $N^+(0)=N^-(0)$, we derive that
\begin{equation}
    \label{limit.ratioK}
\lim_{\ell/L \to 0} \frac{K^{d}_{II}}{K^{d_{cl.}}_{II}} =\frac{1}{N^+(0)}=\frac{\left(1-d\right)^{1/2}}{\left[\left(1-m^2\right)^{-1/2}-d\right]^{1/2}}.
\end{equation}
In particular, when $m \to 0$ the ratio of the SIFs becomes: $(3-2\nu)^{1/2}$, recovering the result derived previously by Gourgiotis et al. (2012) in the \emph{quasi-static} case (see also Huang et al., 1999). It is worth noting that a similar increase of the SIF was previously observed by Sternberg and Muki (1967) for the quasi-static mode I case in couple-stress elasticity. On the other hand, in the micropolar (Cosserat) theory, Antipov (2012) showed that the ratio of the SIFs for the respective quasi-static mode II problem tends to unity when the pertinent characteristic micropolar lengths tend to zero. As Sternberg and Muki (1967) pointed out the aggravation of the SIF can be attributed to the severe boundary layer effects of couple-stress elasticity in singular stress-concentration problems. Indeed, in contrast to the micropolar theory, the field equations in the standard couple-stress theory are of the fourth-order (due to the dependence of the rotation upon the displacement vector - see \eqref{eq25}) having a singular perturbation character associated with the emergence of strong boundary-layer effects.

   \subsection{The dynamic energy release rate}
\begin{figure}[!htcb]
\centering
\includegraphics[scale=0.7]{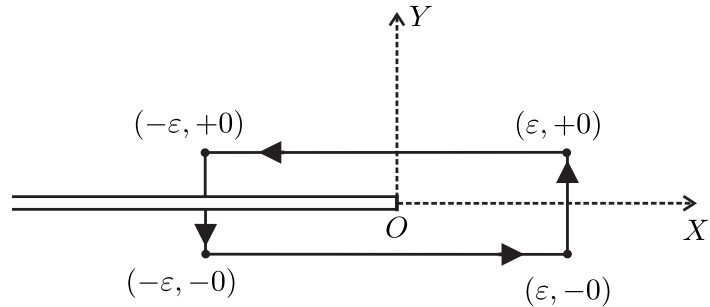}
\caption{Rectangular-shaped contour surrounding the crack-tip.}
\label{fig10}
\end{figure}

\noindent In this Section, we evaluate the dynamic $J$-integral (energy release rate) in the context of couple-stress elasticity for a steady-state propagating mode II crack. The  $J$-integral in the quasi-static case was first established by Atkinson and Leppington (1974) (see also Atkinson and Leppington, 1977; Lubarda and Markenscoff, 2000). By following relative concepts from Rice (1968a,b) and Freund (1990), the extended definition of the dynamic $J$-integral in couple-stress elasticity takes the following form
\begin{equation}
\label{def-J}
J^{d} = \lim_{\Gamma\to0} \frac{1}{V} \bigintssss_{\Gamma} \left\{[W+T]Vn_{x}+P_q^{(n)} \frac{\partial u_q}{\partial t} + 
R_q^{(n)} \frac{\partial \omega_q}{\partial t}\right\} d\Gamma,
\end{equation}
where $\Gamma$ is a piecewise smooth simple 2D contour surrounding the crack-tip, $V$ is the crack-tip velocity, $W$ is the strain-energy density \eqref{s.energy}, $T$ is the kinetic energy 
density including micro-rotational inertia terms defined in \eqref{k.energy}, and $(P_q^{(n)},R_q^{(n)})$ are the reduced force-traction and tangential couple-traction defined in 
\eqref{tr-P}. It is noted that in the general transient case the $J$-integral is not path-independent (Freund, 1990). However, for the case of steady crack growth considered here, it can 
be readily proved following an analogous procedure as the one proposed by Grentzelou and Georgiadis (2008) for the stationary crack problem, that the $J$-integral remains path-independent. 
Moreover, we note that in the expression for the dynamic $J$-integral in \eqref{def-J} the edge forces are not included since the latter are zero under plane-strain conditions (see also Section 3). 
Utilizing the Galilean transformation in \eqref{galilean}, the $J$-integral in 
the steady-state case becomes 
\begin{equation}
\begin{split}
\label{J}
J^{d} 
&= \bigintssss_{\Gamma} \left\{ [W+T]n_{x}-P_q^{(n)}\frac{\partial u_q}{\partial X}-R_q^{(n)}\frac{\partial \omega_q}{\partial X} \right\} d\Gamma \\[3mm]
&=\bigintssss_{\Gamma} \left\{ [W+T]dY-\left[P_q^{(n)}\frac{\partial u_q}{\partial X}+R_q^{(n)}\frac{\partial \omega_q}{\partial X}\right] d\Gamma \right\}.
\end{split}
\end{equation}
For the evaluation of the  $J$-integral, we consider the rectangular-shaped contour $\Gamma$ (surrounding the crack-tip) with vanishing 'height' along the $Y$-direction and with $\varepsilon\to 0$ (see Fig. \ref{fig10}). Such a contour was first introduced by Freund (1972) in examining the energy flux into the tip of a rapidly extending crack and it is particularly convenient in computing energy quantities in the vicinity of crack tips. In fact, this type of contour permits using solely the asymptotic near-tip stress and displacement fields. It is noted that upon this choice of contour, the integral $\int_{\Gamma} [W+T]dY$ in \eqref{J} becomes zero if we allow the 'height' of the rectangle to vanish. In this way, the expression for the $J$-integral becomes
\begin{equation}
\label{J2}
J^{d} = -2\lim_{\varepsilon\to 0} \bigintssss_{-\varepsilon}^{\varepsilon} \left[P_q^{(n)}\frac{\partial u_q}{\partial X}+R_q^{(n)}\frac{\partial \omega_q}{\partial X}\right]dX.
\end{equation}

\noindent
Further, we note that due to the anti-symmetry conditions that prevail in the mode II case, the normal stress $\sigma_{yy}(X,Y=0)$ and the couple-stress $m_{yz}(X,Y=0)$ are \textit{zero} along the whole crack line. In view of the above, the dynamic $J$-integral for the mode II plane-strain case assumes the following form 
\begin{align}
\label{J3}
J^{d} &= -2\lim_{\varepsilon\to+0} \bigintssss_{-\varepsilon}^{\varepsilon} \Biggl\{ \sigma_{yy}(X,+0)\cdot\frac{\partial u_y(X,+0)}{\partial X} + 
\sigma_{yx}(X,+0)\cdot\frac{\partial u_x(X,+0)}{\partial X} \nonumber \\
& \hspace{28mm} + m_{yz}(X,+0)\cdot\frac{\partial\omega(X,+0)}{\partial X} \Biggr\} dX,  \\
&=-2\lim_{\varepsilon\to+0} \bigintssss_{-\varepsilon}^{\varepsilon} \sigma_{yx}(X,+0)\cdot\frac{\partial u_x(X,+0)}{\partial X}dX. \nonumber
\end{align}
Now, by using the asymptotic solution \eqref{fasympt-syx} and \eqref{fasympt-ux}, we finally obtain
\begin{equation}
\label{J4}
J^{d} = \frac{2T_0^2 m^2}{\mu\pi L(d-1)\left(2-m^2\right)^2 \left[N^+(i\ell/L)\right]^2} \lim_{\varepsilon\to+0} 
\int_{-\varepsilon}^{\varepsilon} (X_+)^{-1/2}(X_-)^{-1/2} dX,
\end{equation}
where for any real $\lambda$ with the exception of $\lambda=-1,-2,-3,...$, the following definitions of singular distributions (of the bisection type) have been employed
\begin{equation}
    \label{distributions}
X_+^{\lambda}=\begin{cases}
X^{\lambda}, &\text{for $X>0$}\\
0, &\text{for $X<0$}
\end{cases}
\qquad
\text{and}
\qquad
X_-^{\lambda}=\begin{cases}
0, &\text{for $X>0$}\\
|X|^{\lambda}, &\text{for $X<0$}
\end{cases} .
\end{equation}
Moreover, the product of distributions inside the integral in \eqref{J4} is obtained through the use of Fisher's theorem (Fisher, 1971), i.e. the operational relation: 
\begin{equation}
    \label{fisher}
(X_-)^{\lambda} (X_+)^{-1-\lambda}=-\pi\delta(X)\left[2\sin(\pi\lambda)\right]^{-1} \quad \text{with} \quad \lambda\neq -1,-2,-3,... 
\end{equation}
with $\delta(X)$ being the Dirac delta distribution. Then, taking into account the properties of the Dirac delta distribution, we derive the final expression for the $J$-integral
\begin{equation}
\label{J5}
J^{d} = \frac{T_0^2 m^2 \left(1-m^2\right)^{1/2}}{\mu L \, Q(m)\left[N^+(i\ell/L)\right]^2},
\end{equation}
where $Q(m)=\left(d-1\right)\left(1-m^2\right)^{1/2}\left(2-m^2\right)^2=\left(1-m^2\right)^{1/2}\left[4\left(1-m^2c^2\right)^{1/2}-\left(2-m^2\right)^{2}\right]$. 

On the other hand, the respective $J$-integral in the classical theory of elasticity is
\begin{equation}
    \label{Jclas}
J_{cl.}^{d}=\frac{T_0^2\,m^2\left(1-m^2\right)^{1/2}}{\mu L\,R(m)},
\end{equation}
where $R(m)=4\left(1-m^2\right)^{1/2}\left(1-m^2c^2\right)^{1/2}-\left(2-m^2\right)^{2}$ is the classical Rayleigh function (see e.g. Ravi-Chandar, 2004). 

Combining now the expressions derived previously, we obtain the following relationship between the corresponding energy release rate (ERR) and SIF in the couple-stress and classical elasticity theories
\begin{equation}
    \label{ERRs}
J^{d}=\frac{m^2\left(1-m^2\right)^{1/2}}{2\mu\,Q(m)}\,[K_{II}^{d}]^2, \quad J_{cl.}^{d}=\frac{m^2\left(1-m^2\right)^{1/2}}{2\mu\,R(m)}\,[K_{II}^{d_{cl.}}]^2.
\end{equation}
Accordingly, the ratio of the ERRs in the two theories becomes
\begin{equation}
    \label{ratios}
\frac{J^{d}}{J_{cl.}^{d}}=\frac{R(m)}{Q(m)}\,\frac{[K_{II}^{d}]^2}{[K_{II}^{d_{cl.}}]^2}=\frac{R(m)}{Q(m)}\,\frac{1}{\left[N^+(i\ell/L)\right]^2}.
\end{equation}
It can be readily shown that as $\ell/L \to 0$, the ratio of the ERRs tends to unity. Indeed, by taking into account \eqref{limK0}, \eqref{decompN}, we derive that
\begin{equation}
    \label{limit.Nplus}
\lim_{\ell/L \to 0} {\left[N^+(i\ell/L)\right]^2}=\left[N^+(0)\right]^2=N(0)=\frac{\left(1-m^2\right)^{-1/2}-d}{1-d}=\frac{R(m)}{Q(m)},
\end{equation}
which, in turn, implies that 
\begin{equation}
    \label{limit.ratio}
\lim_{\ell/L \to 0} {\frac{J^{d}}{J_{cl.}^{d}}}=1.
\end{equation}
It is worth noting that when $m \to 0$, we have $Q(m)/R(m) \to (3-2\nu)$, and thus we recover the quasi-static limit (Gourgiotis et al., 2012).

As the speed of the crack reaches the \emph{classical} Rayleigh wave velocity (i.e. $m_R=1/h_0^* = c_R/c_s$) the Rayleigh function becomes zero and, consequently, the ERR in \eqref{Jclas} becomes infinite. This is a common feature of the steady-state problems in classical elasticity where the SIF does not depend upon the crack-tip velocity. On the 
other hand, for $\ell\neq0$ (i.e. when couple-stress effects are taken into account) the denominator in \eqref{J5} is never zero in the range $m\in\ [0,m_R]$, which, in turn, implies that the 
ERR in couple-stress elasticity, remains always \textit{finite} even when the crack propagates with the Rayleigh speed. 

\begin{figure}[!htcb]
\centering
\includegraphics[scale=0.36]{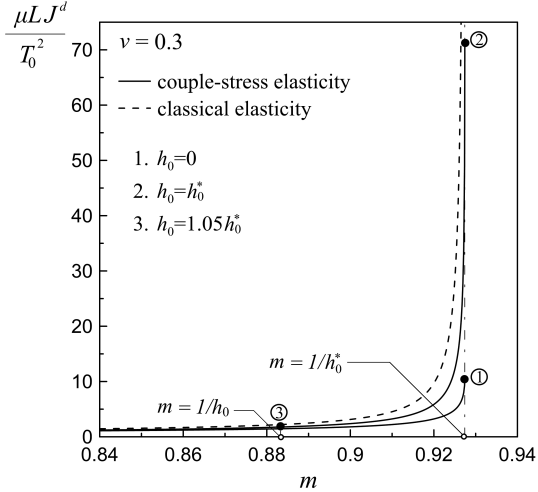}
\caption{Distribution of the dynamic ERRs in couple-stress elasticity and classical elasticity with respect to the normalized crack-tip velocity for three different values: $h_0=\{0,h_0^*,1.05h_0^*\}$.}
\label{fig11}
\end{figure}

Figure \ref{fig11} depicts the variation of the normalized ERR in couple-stress elasticity and in classical elasticity with respect to the normalized crack propagation velocity $m$ for a 
material with $L/\ell=10$ and $\nu=0.3$. For curves 1 and 2 the sub-Rayleigh velocity limit is $m_R=1/h_0^*=0.927$ with $h_0^*=1.078$, whereas for curve 3 ($h_0=1.05h_0^*$) we have: $m_R=1/h_0$. As the 
propagation speed approaches the pertinent sub-Rayleigh limit, an increase is observed in the dynamic ERR in couple-stress elasticity, however, contrary to the classical elasticity case (dashed 
curve), the ERR remains always bounded. The increase becomes more significant when $h_0=h_0^*$ (curve 2). This is attributed to the fact that in this case the Rayleigh waves travel 
\textit{almost} non-dispersively in an infinite medium governed by couple-stress elasticity, thus, resembling the classical elasticity situation (see Fig. \ref{fig01} and the relevant 
discussion in Section 3). Finally, applying the classical Griffith criterion for the stability of the crack 
propagation: $J^d=J^d_{cr}$, where $J^d_{cr}$ is the critical value depending on the material properties, we conclude that, for $J^d_{cr}$ constant, the steady state motion studied here is 
likely to be unstable and that the crack will accelerate rapidly up to the pertinent limiting Rayleigh velocity as in the classical theory of elasticity (see e.g. Willis, 1971).  

\begin{figure}[!htcb]
\centering
\subfloat[]{\includegraphics[scale=0.36]{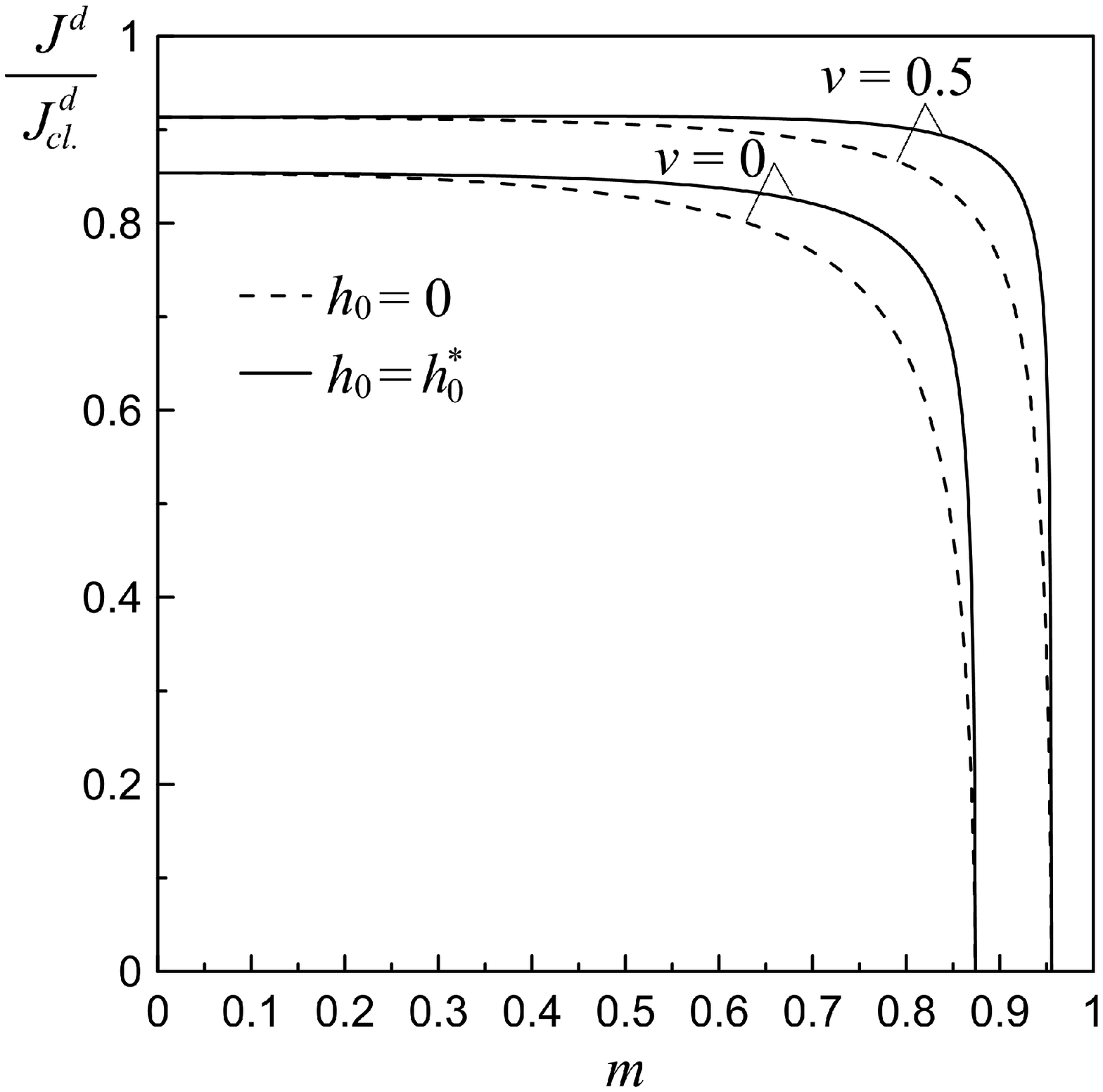}}
\quad
\subfloat[]{\includegraphics[scale=0.36]{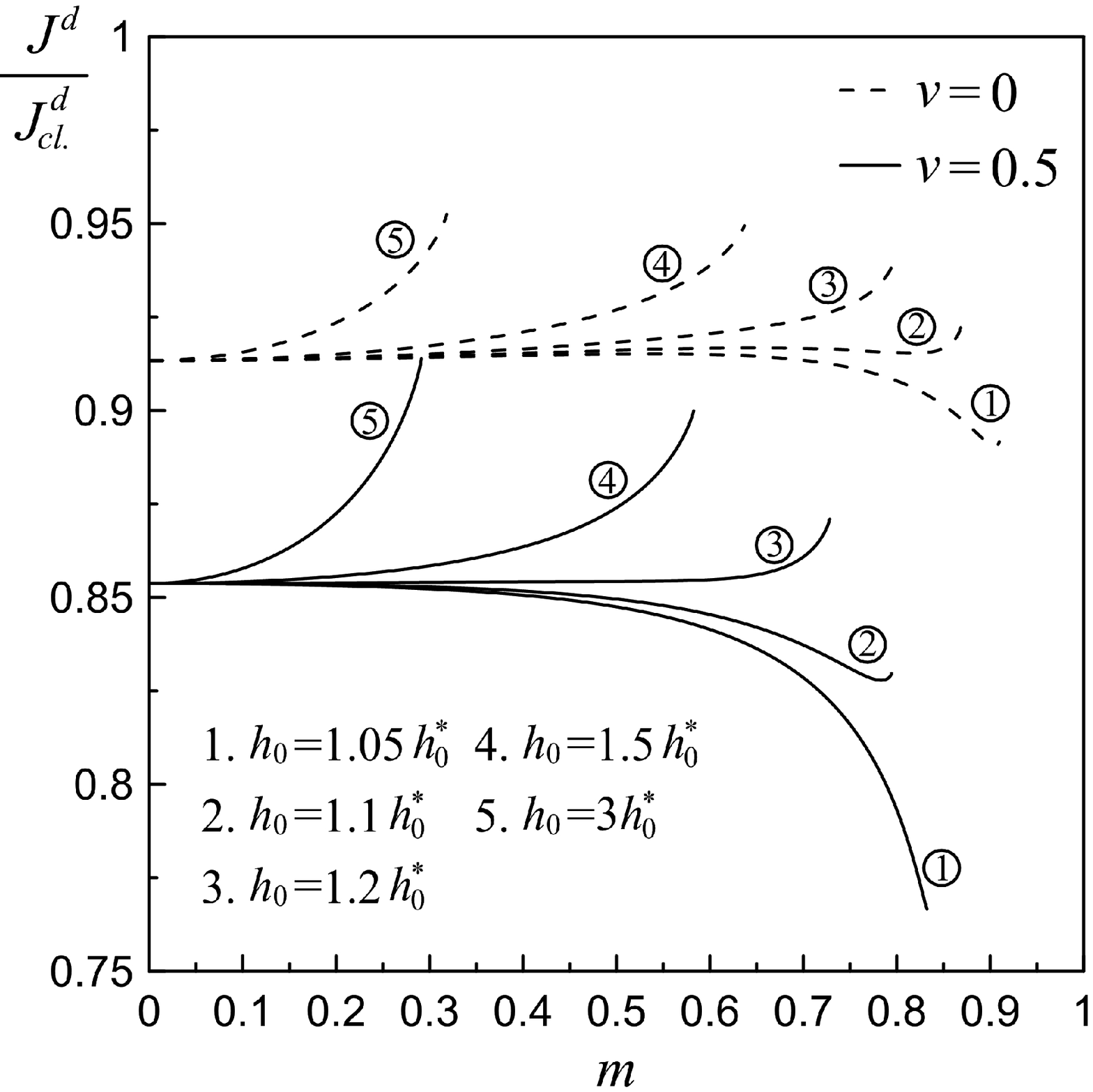}}
\caption{Variation of the ratio of the dynamic ERR in couple-stress elasticity to the ERR in classical elasticity with respect to the normalized crack-tip velocity: (a) for $h_0 \leq h_0^*$ and (b) 
for $h_0 > h_0^*$.}
\label{fig12}
\end{figure}

The variation of the ratio of the ERR in couple-stress elasticity to the respective one in the classical theory versus the normalized velocity is shown in Figures \ref{fig12}a and 
\ref{fig12}b for the cases: 
$h_0 \leq h_0^*$ and $h_0>h_0^*$, respectively. In particular, in the case $h_0 \leq h_0^*$, the ratio of the ERRs is a monotonically decreasing function of the crack speed (Fig. 
\ref{fig12}a). In fact, as the crack speed approaches the sub-Rayleigh limit velocity ($m \to 1/h_0^*$), the ratio becomes zero since at that speed the classical ERR is infinite. It is worth noting 
that this trend for the ratio of the ERRs was also observed in the case of a mode II crack propagating with a sub-Rayleigh speed in a plane triangular-cell lattice (Kulakhmetova et al., 1984; 
Nieves et al., 2013). The plane triangular-cell lattice employed in these works consists of point particles (connected by massless elastic bonds) which interact only with forces and 
not internal moments, thus, corresponding to the case of zero micro-rotational inertia ($h_
0=0$) in the present study. Moreover, in the latter studies, the global ERR for the crack propagating through the homogenized medium is naturally identified as the classical (far-field) ERR 
(Eq. \eqref{Jclas} in our study), whereas the local ERR for the crack propagating through the lattice (i.e. the energy spent on fracture itself) corresponds to the ERR in couple-stress 
elasticity (Eq. \eqref{J5} in our study) evaluated at the crack-tip (see Fig. \ref{fig10}).

In the case $h_0 > h_0^*$, a different trend is observed for the variation of the ratio $J^d/J_{cl.}^d$  (Fig. \ref{fig12}b). Specifically, as $h_0$  increases, the ratio becomes eventually 
a monotonically increasing function of the normalized speed. However, in all cases, the ratio remains below unity. Moreover, since $h_0 > h_0^*$, the pertinent Rayleigh limiting speed is defined as 
$m_R=1/h_0$, which is evidently less than the classical Rayleigh velocity ($1/h_0^*$). Therefore, the curves plotted in Figure \ref{fig12}b terminate before they reach the zero value.

\begin{figure}[!htcb]
\centering
\includegraphics[scale=0.36]{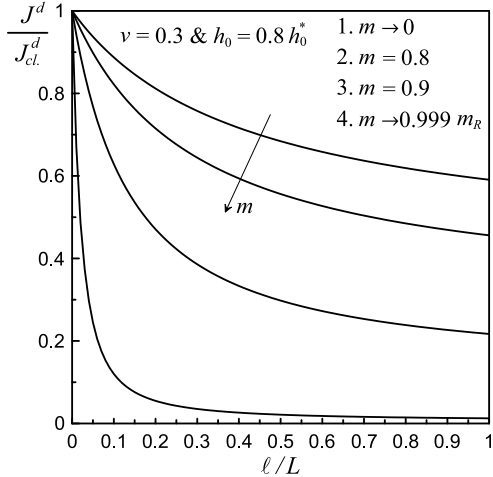}
\caption{Variation of the ratio of the dynamic ERRs in couple-stress and classical elasticity with respect to the ratio of the characteristic lengths $\ell/L$.}
\label{fig13}
\end{figure}

Finally, Figure \ref{fig13} displays the dependence of the ratio $ J^{d}/ J_{cl.}^{d}$ upon the ratio of lengths $\ell/L$. It is observed that the ratio $ J^{d}/ J_{cl.}^{d}$  decreases 
monotonically with increasing values of $\ell/L$. This decrease is more significant as the propagation speed approaches the Rayleigh limit velocity $m_R$. This finding shows that the couple-stress 
theory predicts a strengthening effect since a reduction of the crack driving force takes place as the material microstructure becomes more pronounced. In fact, as the propagation velocity increases, the strengthening effect is more evident. In the case when $m \to m_R$ (curve 4) the ratio of the ERRs tends eventually to zero for each $\ell>0$. Also, as $\ell/L \to 0$, the $J$-integral in couple-stress elasticity tends continuously to its counterpart in the classical theory: $J^{d}/ J_{cl.}^{d} \to 1$. As a final remark, we note that in the case $m \to 0$, the steady state solution approaches the static solution, confirming previous results obtained by Gourgiotis et al. (2012) for the stationary crack.

\section{Conclusions}
\noindent
In the present work, which is closely allied in scope to Mishuris et al. (2012), we examined the plane-strain problem of a semi-infinite mode II crack propagating steadily with constant sub-Rayleigh speed in a body with microstructure governed by couple-stress elasticity. The case of shear loading was chosen since, in principle, couple-stress effects are predominant in this type of deformation (Huang et al., 1999; Gourgiotis and Georgiadis, 2007; Gourgiotis et al., 2012). An exact full field solution was obtained by using the Fourier transform and the Wiener-Hopf technique. It was shown that the mode II problem is reduced to a scalar Wiener-Hopf equation. It is worth noting that this is also the case in the (unconstrained) Cosserat elasticity, where Antipov (2012) showed that the quasi-static mode II problem can be reduced to a scalar Riemann-Hilbert problem (RHP), while the opening mode is governed by an order-2 vector RHP. Moreover, the effect of rotational micro-inertia was also considered in our study since previous experience with couple-stress analyses of surface waves and anti-plane crack problems showed that this term is important, especially at high frequencies. It is remarked that this is the first study in the literature that includes micro-inertial effects in a dynamic plane-strain crack problem in the context of a generalized continuum theory. Our goal was to determine possible deviations from the predictions of classical linear elasticity when a more refined theory is employed to attack plane-strain crack propagation problems. By including pertinent characteristic material lengths, the theory of couple-stress elasticity utilized here accounts for effects of microstructure encompassing, thus, the analytical possibility of size effects, which are absent in the classical theory. 

Our results differ in several important respects from the predictions of standard linear fracture mechanics. It was shown that the microstructural parameters, such as the characteristic material length $\ell$ and the micro-rotational inertia $h$, and on the other hand the maximum crack speed, strongly influence the fracture process near a rapidly moving crack. Indeed, it was observed that depending on the values of the microstructural ratio $h_0=h/\ell$, the actual limiting crack speed under plane strain loading conditions, can be much lower than the classical Rayleigh wave speed predicted by the standard elastodynamic theory. Experimental findings (Livne, et al., 2010) shed light to the influence of the fracture process on the limiting value. Specifically, for the steady-state crack propagation problem examined here, we have found that both the SIF and the ERR ($J$-integral) depend upon the crack speed. In particular, our results showed that as the crack speed approaches the pertinent Rayleigh velocity both 
quantities increase significantly but remain always finite. These findings are in marked contrast with the classical elasticity results where it is known that: (i) the respective SIF for the steady-state crack propagation problem does not depend upon the crack-tip speed, and (ii) the ERR becomes infinite as the propagating crack reaches the Rayleigh velocity. Further, as in the case of the stationary crack, it was observed that the ERR decreases monotonically with increasing values of the ratio of the characteristic material length $\ell$ over the pertinent geometrical length $L$. This means that a decrease of the value of $J$-integral is noticed in comparison with the classical theory. Therefore, the couple-stress predicts a strengthening effect since a reduction of the crack driving force takes place as the material microstructure becomes more pronounced. 

In light of the above, we conclude that couple-stress elasticity can provide a clearer picture of the failure process near a rapidly moving crack tip than classical elasticity. Motivated by experimental observations (Rosakis et al., 1999), we intend, in a future work, to extend the present study to the case of intersonic mode II crack propagation with the aim of gaining further insight into the physical mechanisms governing the fracture process of microstructured materials. 

\begin{appendices}
\numberwithin{figure}{section}
\numberwithin{equation}{section}
\section[Branch cuts for the functions]{Branch cuts for the functions $\chi(z)$, $\beta(z)$ and $\gamma(z)$}

\noindent
The complex function $\chi(z)$ has four branch points (BPs) at $\pm ib_1$ and $\pm ib_2$, where $(b_1,b_2)$ are given in Eq. \eqref{b1b2}. In the case $h_0\leq1$, $(b_1,b_2)$ are always real, consequently, the BPs are located along the imaginary axis (Fig. A.1a). On the other hand, when  $h_0>1$, $(b_1,b_2)$ are complex and the BPS are symmetrically located with respect to the real axis at the four quadrants of the complex $z$-plane (Fig. A.1b).
\begin{figure}[!htcb]
\centering
\includegraphics[scale=0.42]{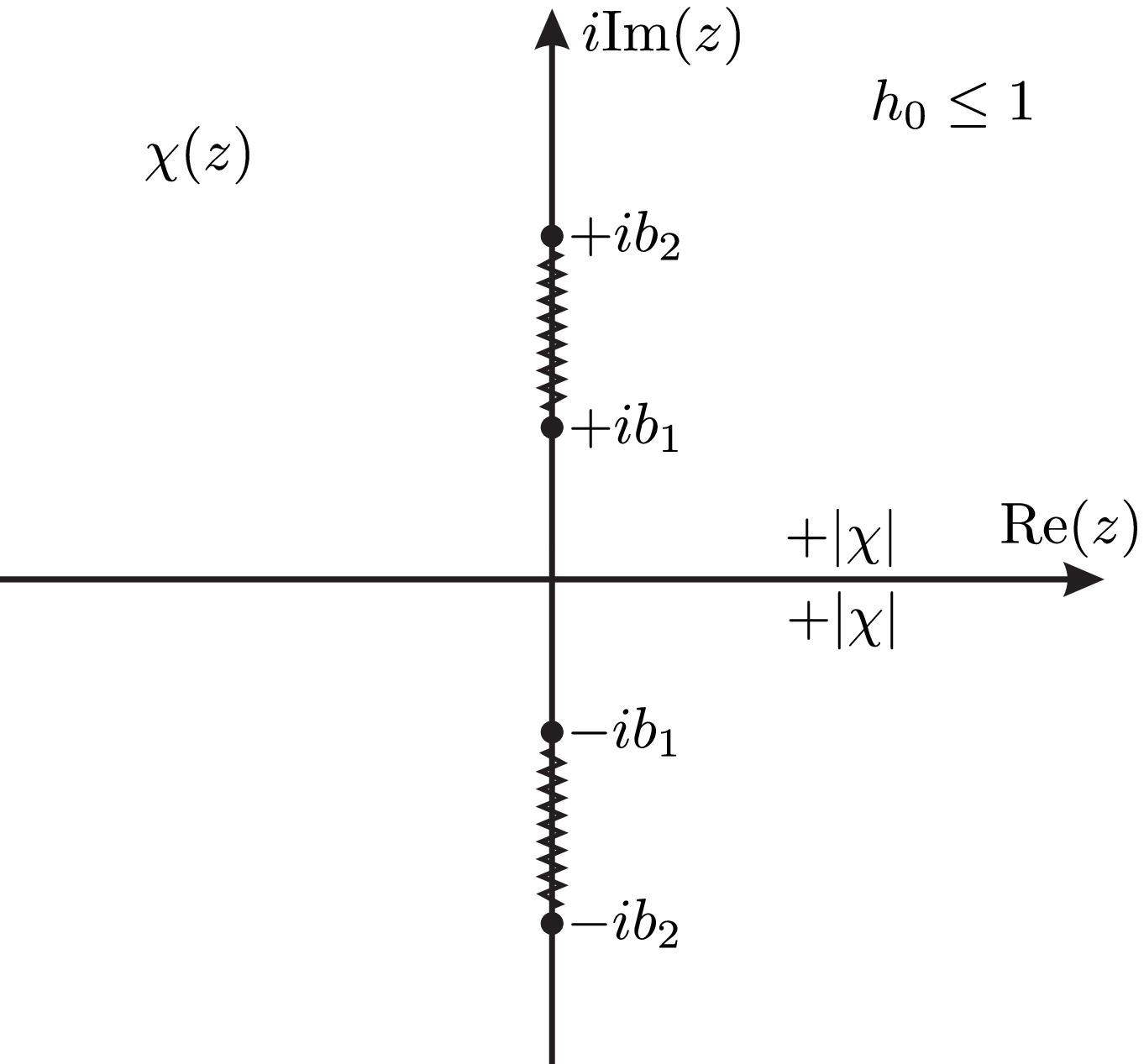}
\quad
\includegraphics[scale=0.42]{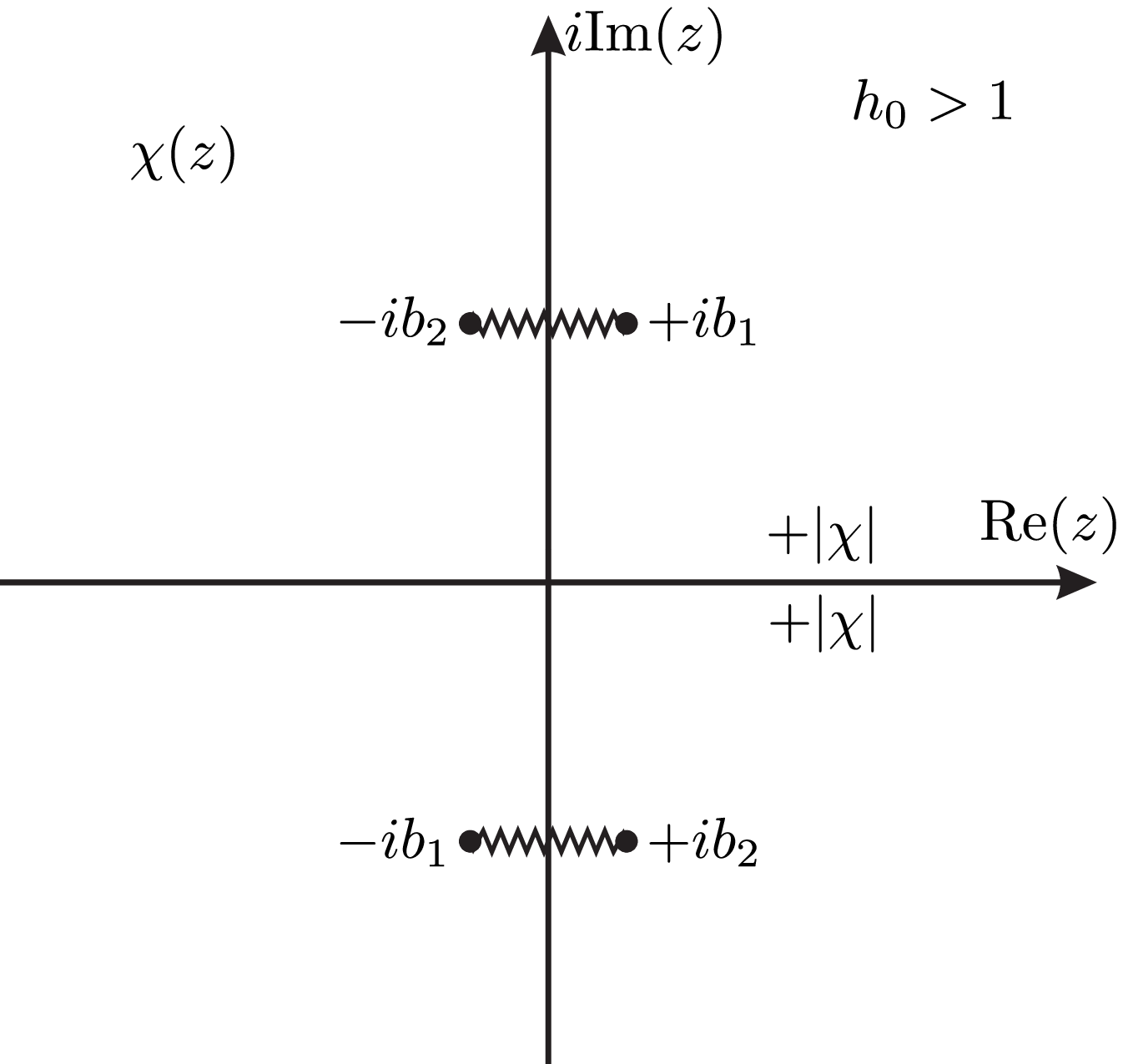}
\caption{Branch cuts for the function $\chi(z)$}
\end{figure}

The complex functions $\beta(z)$ and $\gamma(z)$ are four-valued functions (having four Riemann sheets). Therefore, the BPs $\pm ib_1$ and $\pm ib_2$ are double valued BPs for $\beta(z)$ and $\gamma(z)$. Also, $z=0$ and $z= \pm ib_0$ are additional BPs. However, $z=\infty$ is not a BP as it can be readily shown by utilizing the transformation $z=1/t$ with $t \to 0$. Now, depending on the parameters $h_0$ and $m$, the branch cuts for the functions $\beta(z)$ and $\gamma(z)$ are illustrated in Figures A.2 and A.3, respectively, with $\varepsilon$ being a real number such that $\varepsilon\to+0$. In fact, introducing $\varepsilon$ facilitates the introduction of the branch cuts corresponding to the branch point (BP) $z=0$ of the function $\gamma(z)$. It is also noted that in all cases the BPs of the function $\chi(z)$ are also BPs of $\beta(z)$ and $\gamma(z)$. We remark that the specific introduction of the branch cuts secures that the functions $(\chi,\beta,\gamma)$ are always single-valued and positive along the real axis. Finally, we note that $m<m_R$ with $m_R=\min\{1/h_0,1/h_0^*\}$ in order for the crack to propagate with a sub-Rayleigh speed.
\begin{figure}[!htcb]
\centering
\includegraphics[scale=0.32]{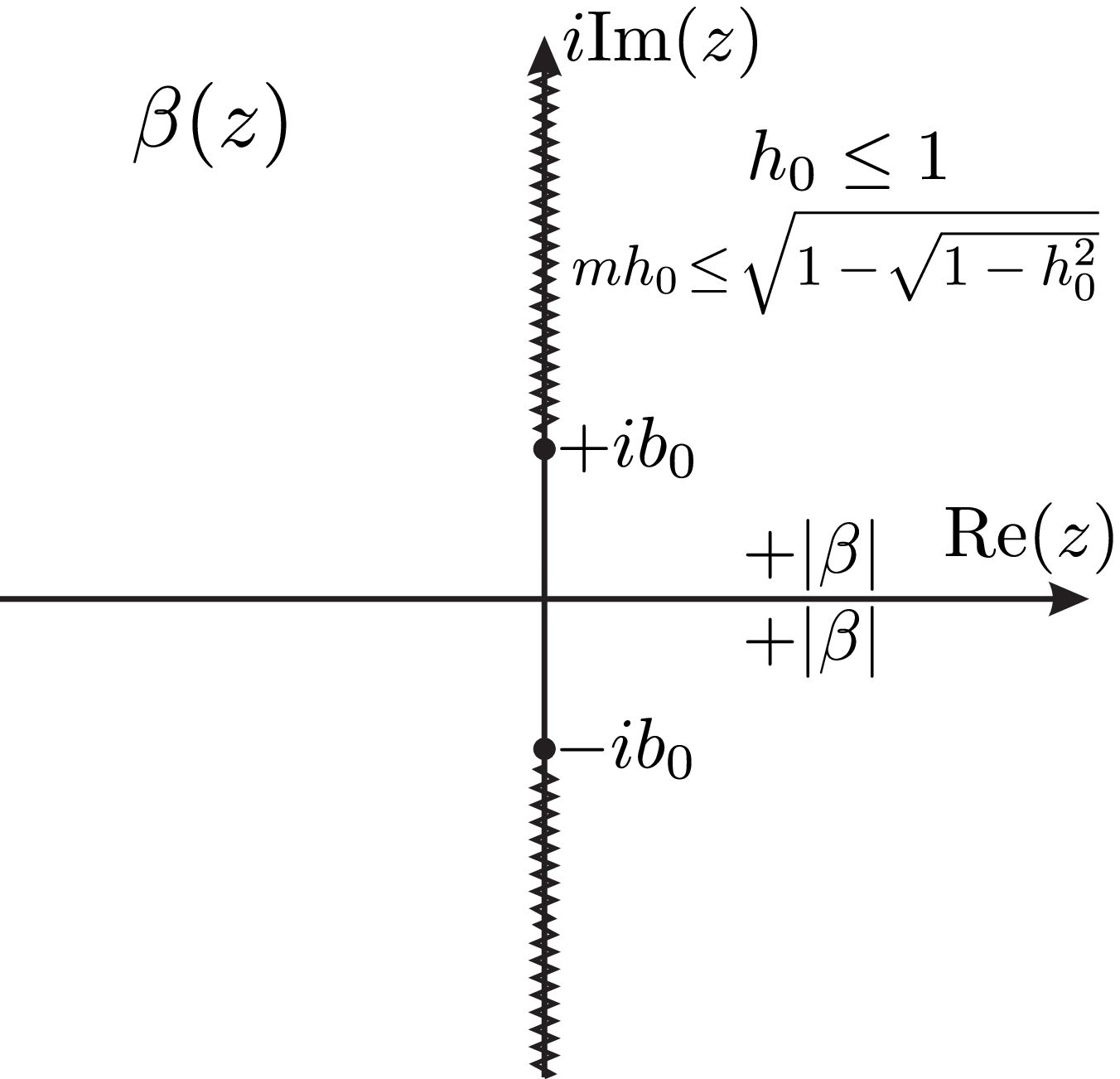}
\quad
\includegraphics[scale=0.32]{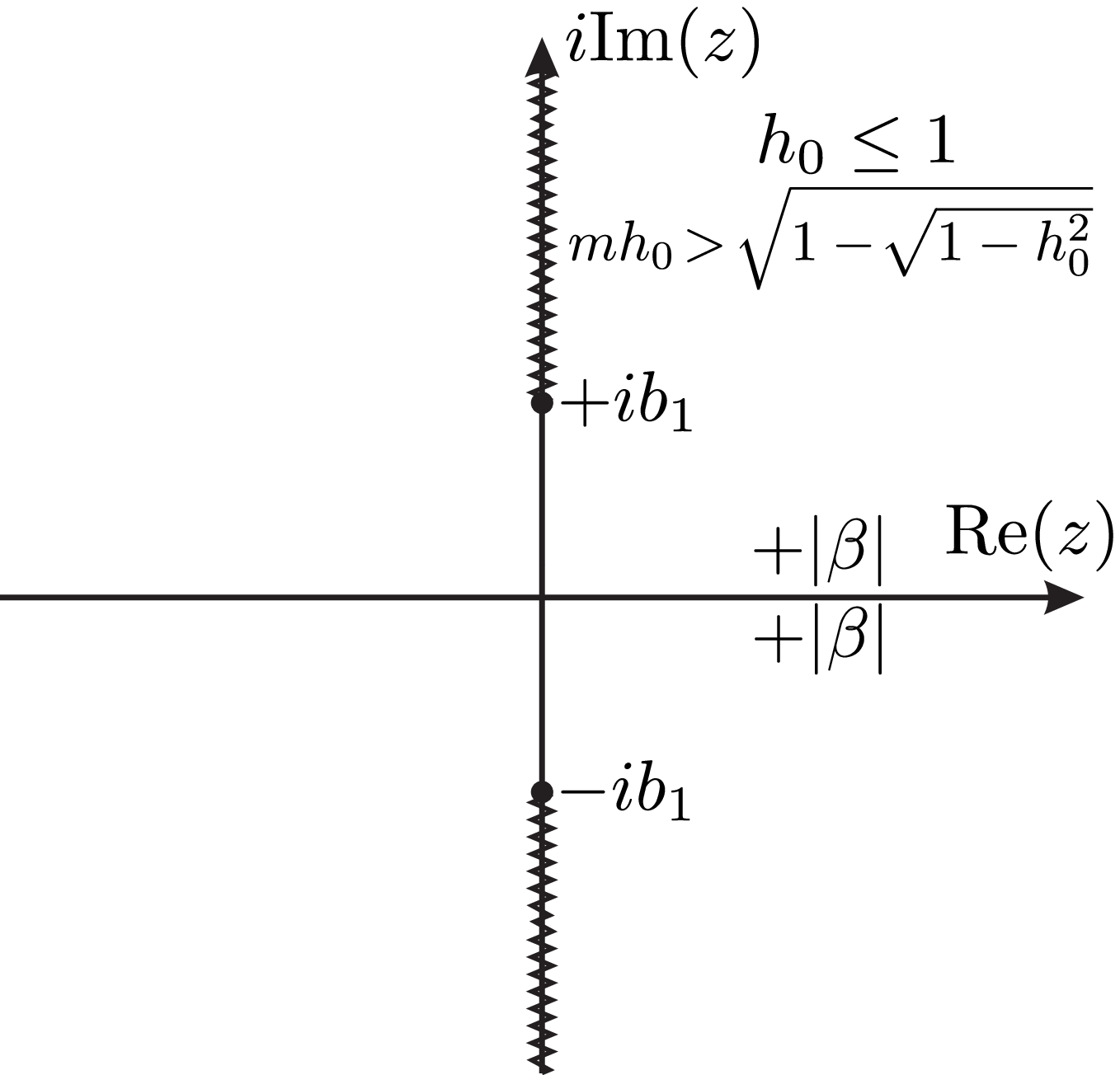}
\quad
\includegraphics[scale=0.32]{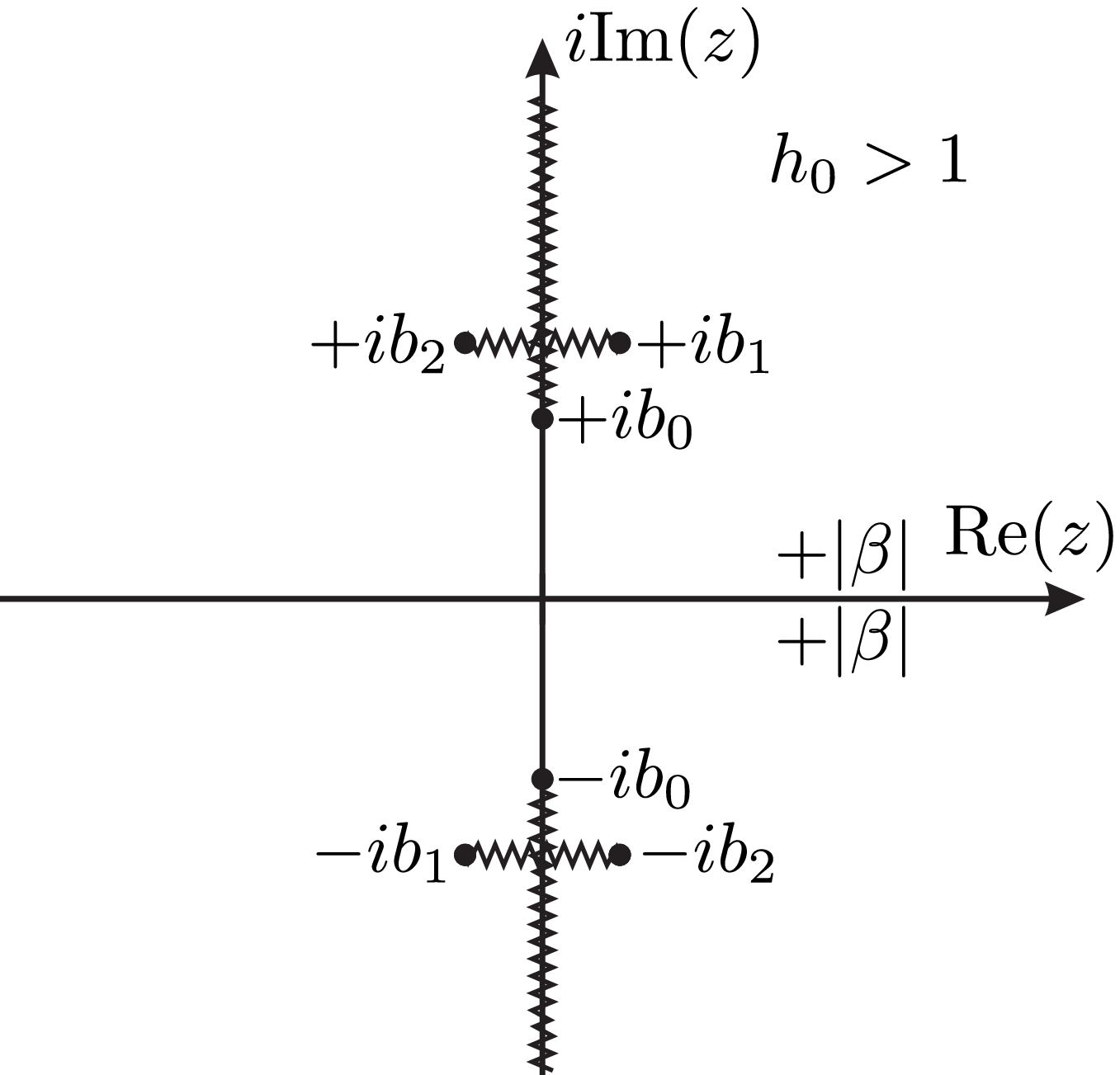}
\caption{Branch cuts for the function $\beta(z)$}
\end{figure}
\begin{figure}[!htcb]
\centering
\includegraphics[scale=0.32]{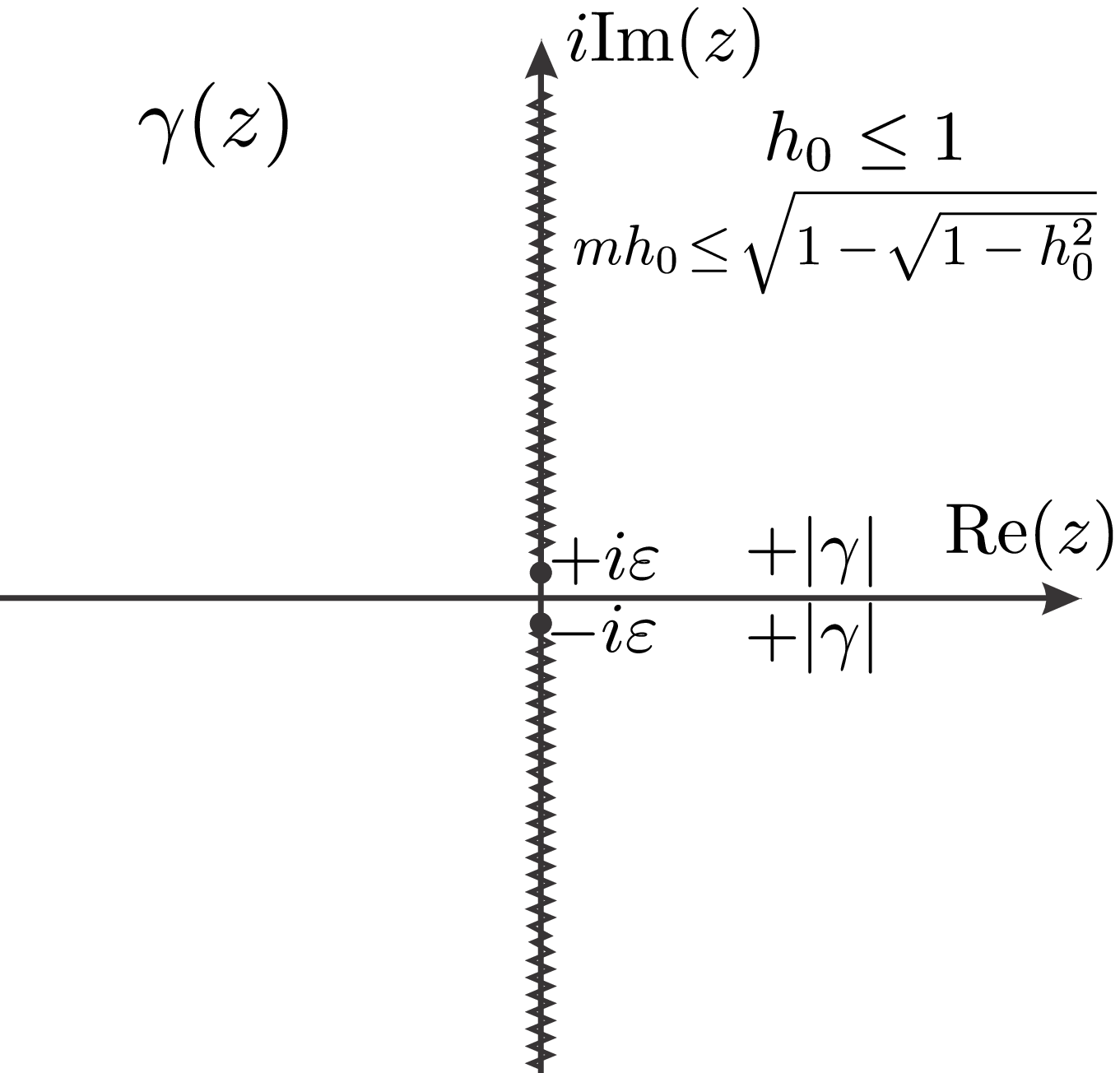}
\quad
\includegraphics[scale=0.32]{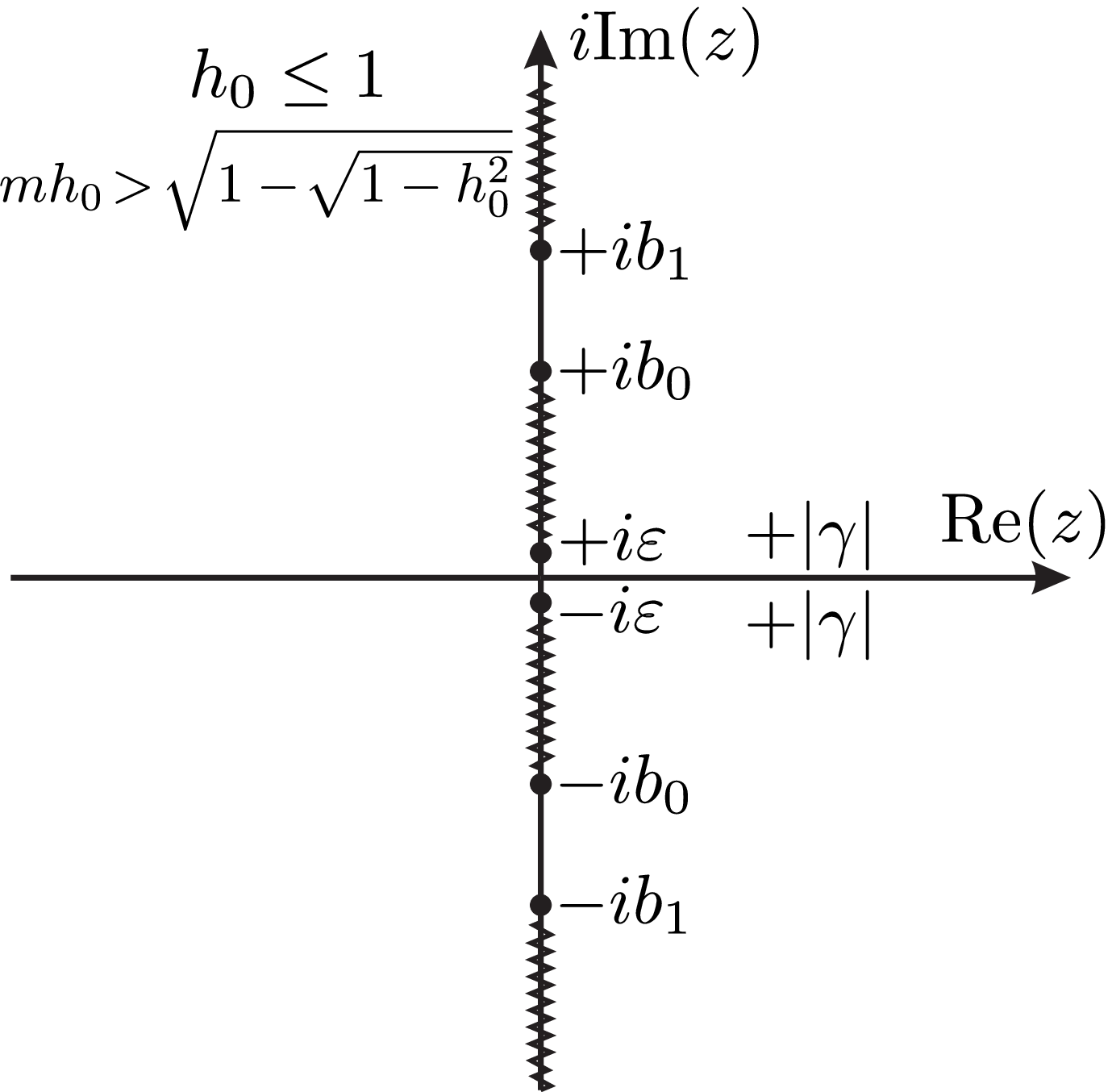}
\quad
\includegraphics[scale=0.32]{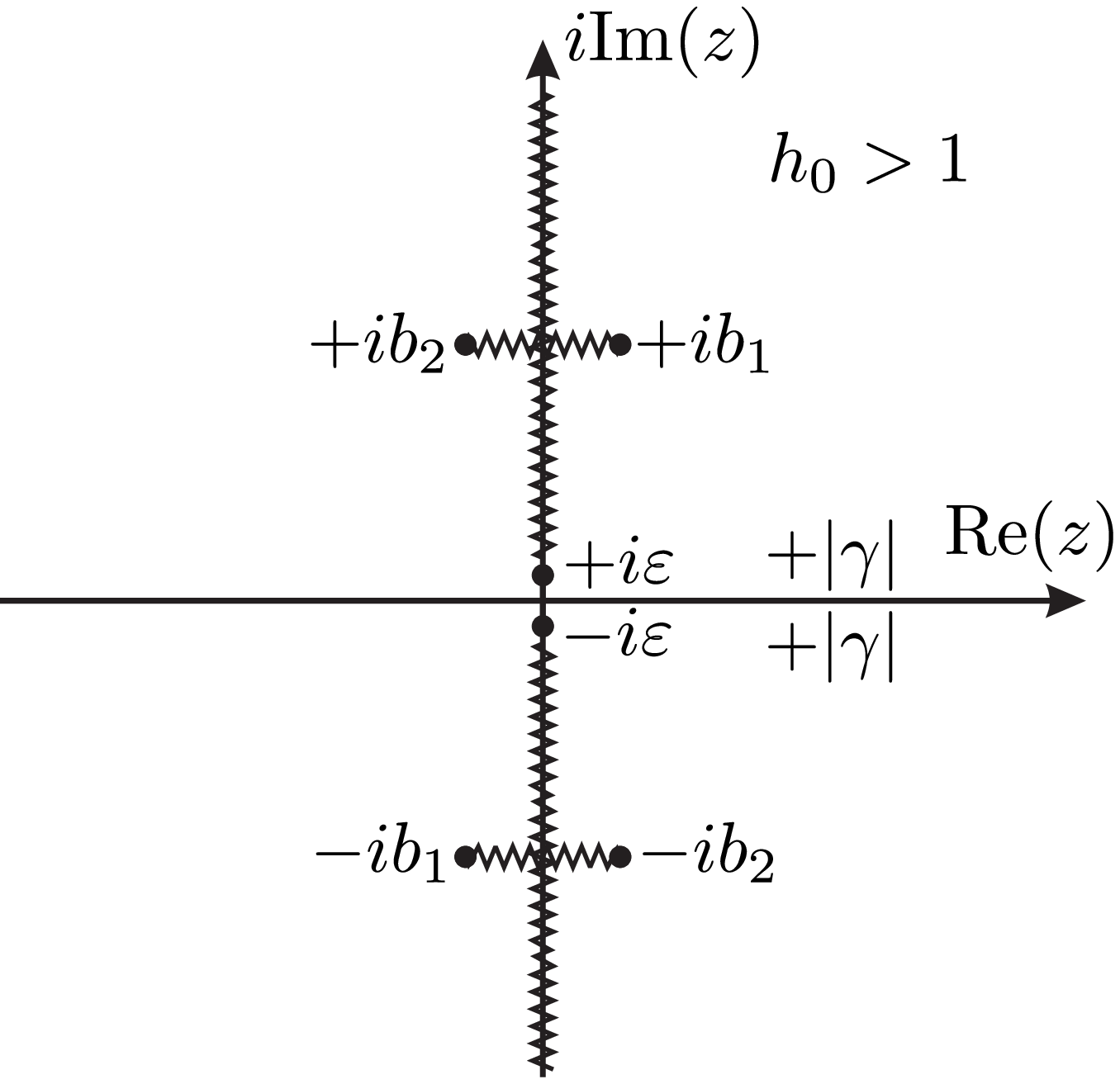}
\caption{Branch cuts for the function $\gamma(z)$}
\end{figure}
%

\section[Factorization of the kernel function]{Factorization of the kernel function $N(z)$}

\noindent
The kernel $N(z)$ is given by the expression
\begin{equation}
    \label{N}
N(z)=\frac{1}{1-d}\left[\frac{\theta^2+\left[z^2\right]^{1/2}\theta+m^2}{\ell\theta(\beta+\gamma)}-d\right],
\end{equation}
where the complex function $\theta\equiv\theta(z)$ is defined in \eqref{deftheta}. It is noted that the kernel function $N(z)$ does not have any poles or zeros in the finite complex domain in the sub-Rayleigh regime, only BPs which depend upon the values of $(h_0,m)$. In particular, we distinguish the following three cases (see Fig. B.1): 
\begin{itemize}
\item Case I. $h_0\leq 1$ and $m\leq\dfrac{\sqrt{1-\sqrt{1-h_0^2}}}{h_0}$; then the BPs are: $0$ and $\pm ib_0$. 
\item Case II. $h_0\leq 1$ and $m>\dfrac{\sqrt{1-\sqrt{1-h_0^2}}}{h_0}$; then the BPs are: $0$, $\pm ib_0$ and $\pm ib_1$. 
\item Case III. $h_0>1$; then the BPs are: $0$ and $\pm ib_0$. 
\end{itemize}

\begin{figure}[!htcb]
\centering
\includegraphics[scale=0.4]{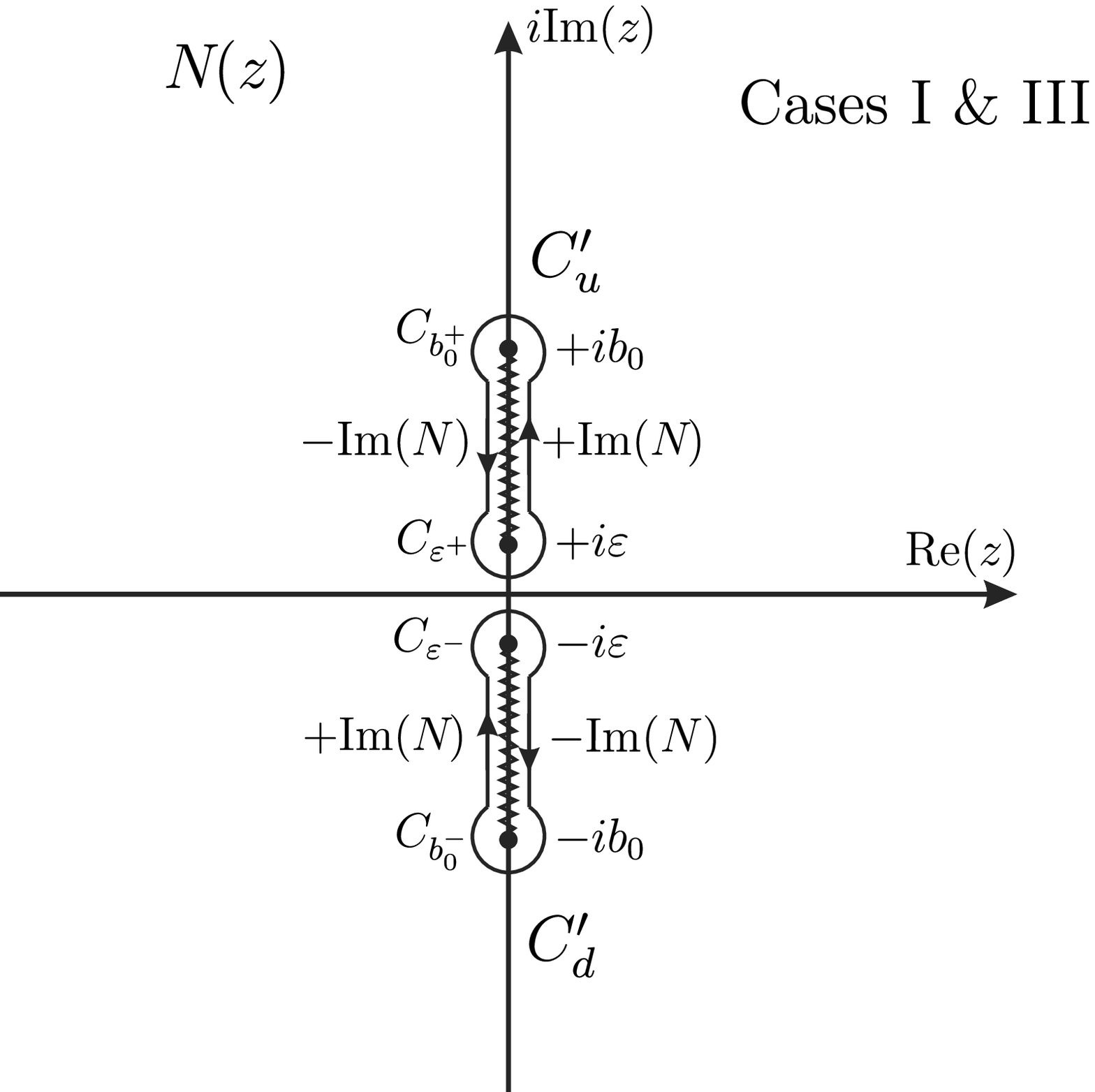}
\quad
\includegraphics[scale=0.4]{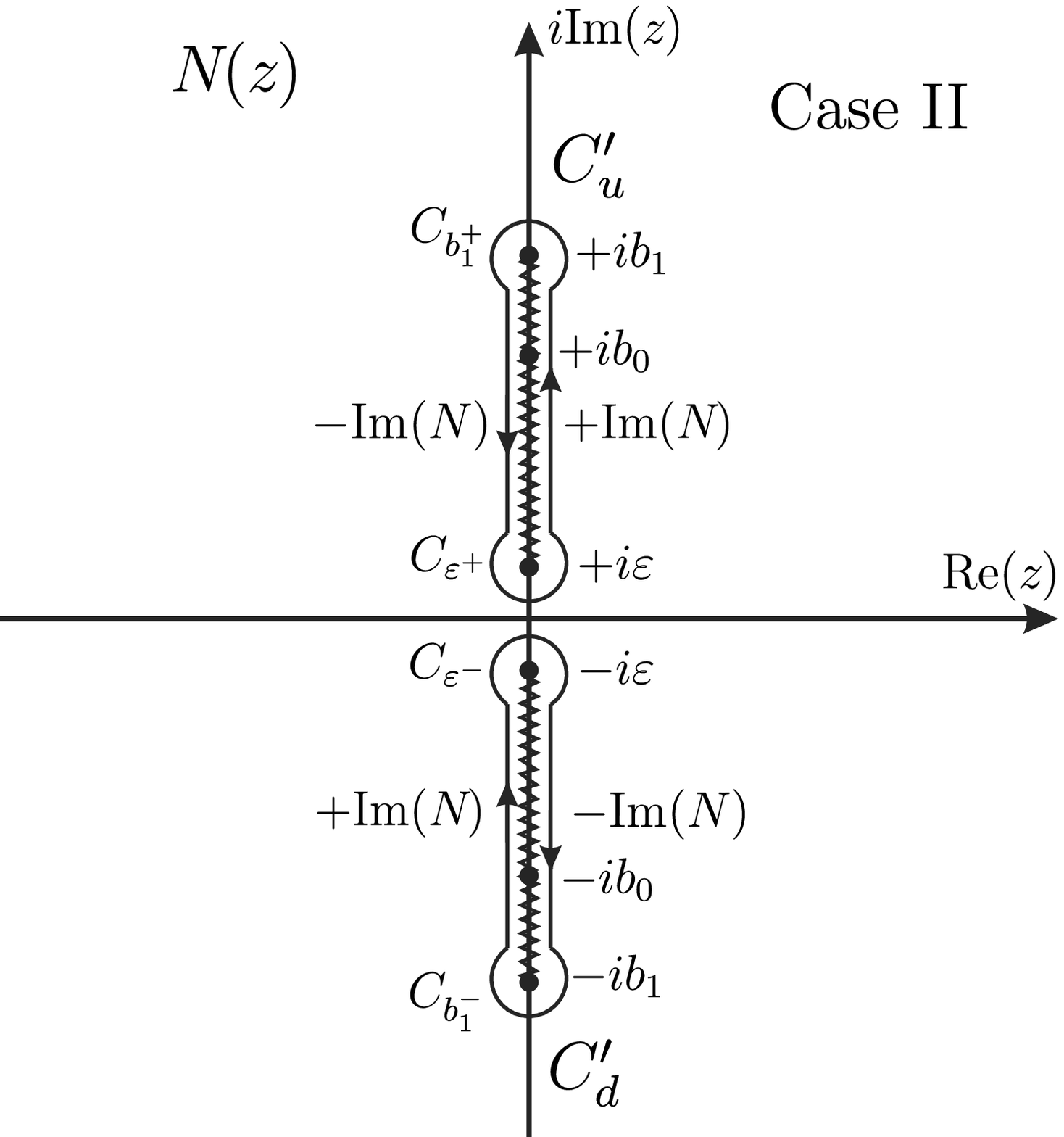}
\caption{Branch cuts and integration paths for the factorization 
of the kernel function $N(z)$. The positive imaginary parts of $N(z)$ are given by \eqref{Re&Im1} and \eqref{Re&Im2}, respectively.}
\end{figure}

\noindent
Note, that $m \leq m_R$, always. It should be remarked that for the Cases I and III, the points: $\pm ib_1$ and $\pm ib_2$ are not BPs of the kernel function $N(z)$. Indeed, although these points are BPs for the functions $\beta(z)$, $\gamma(z)$ and $\chi(z)$ as we have shown, they are removable BPs for $N(z)$. This can be readily shown by expanding $N(z)$ in series around these points
\begin{gather}
N(z)=a_0(m,h_0) \pm a_1(m,h_0) (z \pm ib_1)+O(z \pm ib_1)^2, \quad \text{for} \quad z \to \pm ib_1, \\
N(z)=a_2(m,h_0) \pm a_3(m,h_0) (z \pm ib_2)+O(z \pm ib_2)^2, \quad \text{for} \quad z \to \pm ib_2, 
\end{gather}
where $a_n$ with $n=(0,1,2,3)$ are constants that depend on the parameters $(m,h_0)$. Thus, we see that around these points the function is single-valued.  A similar situation applies for Case II where $\pm ib_2$ are removable BPs. Furthermore, for all Cases (I)-(III), when $\varepsilon<\textup{Im}(z)<b_0$ and $\textup{Re}(z)=+0$ (i.e. approaching the branch cut from the right), the real and imaginary parts of $N(z)$ take the following form
\begin{equation}
    \label{Re&Im1}
\textup{Re}\left(N(z)\right)=\frac{|z|(\theta^2+m^2-\ell^2\gamma^2)}{\ell^3(1-d)|\gamma|(\beta^2-\gamma^2)}-\frac{d}{1-d}, \qquad \textup{Im}\left(N(z)\right)=-\frac{|z|(\theta^2+m^2-\ell^2\beta^2)}{\ell^3(1-d)|\beta|(\beta^2-\gamma^2)},
\end{equation}
whereas for case II, when $b_0<\textup{Im}(z)<b_1$ and $\textup{Re}(z)=+0$, the real and imaginary parts of $N(z)$ become
\begin{equation}
    \label{Re&Im2}
\textup{Re}\left(N(z)\right)=-\frac{d}{1-d}, \qquad \textup{Im}\left(N(z)\right)=\frac{|z|(\theta^2+m^2+\ell^2\beta\gamma)}{\ell^3(1-d)\beta\gamma(\beta+\gamma)}.
\end{equation}

Next, taking into account that $N(z)\to 1$ as $|z| \to \infty$ and employing Jordan's Lemma, we evaluate the functions $N^+(z)$ and $N^-(z)$ defined in Eqs. \eqref{Nplus} and \eqref{Nminus}, by closing the original integration paths $C_u$ and $C_d$, which extend parallel to the real axis in the complex $z$-plane, with large semi-circles at infinity on the upper and lower half-planes respectively, as it is shown in Fig. 4. Then, a deformation of the integration contour along with the use of Cauchy's theorem, allows taking as equivalent integration paths the contours $C'_u$ and $C'_d$, around the pertinent branch cuts of $N(z)$ (Fig. B.1). In particular, we obtain for the function $N^-(z)$:\\
\\
\textit{Cases I and III}
\begin{equation}
\begin{aligned}
    \label{N-ac}
N^-(z)&=\exp\left\{-\frac{1}{2\pi i}\int_{C_u}\frac{\log\left[N(\zeta)\right]}{\zeta-z}\,d\zeta\right\}=\exp\left\{-\frac{1}{2\pi i}\int_{C'_u}\frac{\log\left[N(\zeta)\right]}{\zeta-z}\,d\zeta\right\}\\
&=\exp\biggl\{-\frac{1}{2\pi i}\biggl(\int_{C_{\varepsilon^+}}+\int_{+i\varepsilon}^{+ib_0}+\int_{C_{b_0^+}}+\int_{+ib_0}^{+i\varepsilon}\biggr)\frac{\log\left[N(\zeta)\right]}{\zeta-z}\,d\zeta\biggr\}\\
&=\exp\biggl\{-\frac{1}{2\pi}\biggl(\int_{+i\varepsilon}^{+ib_0}\tan^{-1}\biggl[\frac{\textup{Im}\left(N(\zeta)\right)}{\textup{Re}\left(N(\zeta)\right)}\biggr]\frac{d\zeta}{\zeta-z}\\
&\phantom{texttttttttttttttttt}+\int_{+ib_0}^{+i\varepsilon} \tan^{-1}\biggl[\frac{-\textup{Im}\left(N(\zeta)\right)}{\textup{Re}\left(N(\zeta)\right)}\biggr]\frac{d\zeta}{\zeta-z}\biggr)\biggr\}\\
&=\exp\biggl\{-\frac{1}{\pi}\int_{i\varepsilon}^{ib_0}\tan^{-1}\biggl[\frac{\textup{Im}\left(N(\zeta)\right)}{\textup{Re}\left(N(\zeta)\right)}\biggr]\frac{d\zeta}{\zeta-z}\biggr\}.					
\end{aligned}
\end{equation}
\\
\textit{Case II}
\begin{equation}
\begin{aligned}
    \label{N-b}
N^-(z)&=\exp\biggl\{-\frac{1}{2\pi i}\biggl(\int_{C_{\varepsilon^+}}+\int_{+i\varepsilon}^{+ib_1}+\int_{C_{b_1^+}}+\int_{+ib_1}^{+i\varepsilon}\biggr)\frac{\log\left[N(\zeta)\right]}{\zeta-z}\,d\zeta\biggr\}\\
&=\exp\biggl\{-\frac{1}{\pi}\int_{i\varepsilon}^{ib_1}\tan^{-1}\biggl[\frac{\textup{Im}\left(N(\zeta)\right)}{\textup{Re}\left(N(\zeta)\right)}\biggr]\frac{d\zeta}{\zeta-z}\biggr\}.
\end{aligned}
\end{equation}
where the function $\tan^{-1}(\cdot)$ is defined as
\begin{equation}
    \label{def-arctan}
		\tan^{-1}\frac{y}{x}=\Tan^{-1}\frac{y}{x}
		+\begin{cases}
           0,    &\text{for}  \quad x>0\\
           \pi,  &\text{for}  \quad x<0 \quad  \text{and} \quad  y>0\\
          -\pi,  &\text{for}  \quad x<0 \quad  \text{and} \quad y \leq 0\\
     \end{cases}
\end{equation}
where $\Tan^{-1}(\cdot)$ is the principal value of the inverse tangent with branch cuts $(-i \infty,-i]$ and $[i,i \infty)$.
Also, it is noted that the contour integrals around the branch points of $N(\zeta)$ in Eqs. \eqref{N-ac} and \eqref{N-b} are all zero. Indeed, by writing $\zeta=\zeta_k+r_0e^{i\theta}$, where $\zeta_k=\pm ik$ are the branch points of $N(\zeta)$, with $k=(\varepsilon, b_0, b_1)$, and by taking into account that $(\zeta-\zeta_k)\cdot\log N(\zeta) \to 0$ uniformly as $r_0 \to 0$, it can readily shown that
\begin{equation}
\lim_{r_0 \to 0} \int_{C_{k^{\pm }}} \frac{\log\left[N(\zeta)\right]}{\zeta-z}\,d\zeta=0 \qquad \text{with} \qquad k=(\varepsilon, b_0, b_1).
\end{equation}
Finally, by letting $\varepsilon \to 0$ and combining \eqref{N-ac} and \eqref{N-b}, yields a single formula for $N^{-}(z)$
\begin{equation}
N^{-}(z)=\exp{\left\{-\frac{1}{\pi}\left[\bigintssss_{0}^{ib_0}\!\!\tan^{-1}\!\left[\frac{\textup{Im}\left(N(\zeta)\right)}{\textup{Re}\left(N(\zeta)\right)}\right] \frac{d\zeta}{\zeta-z}+a\bigintssss_{ib_0}^{ib_1}\!\!\tan^{-1}\!\left[\frac{\textup{Im}\left(N(\zeta)\right)}{\textup{Re}\left(N(\zeta)\right)}\right] \frac{d\zeta}{\zeta-z}\right]\right\}},
\end{equation}
where the constant $a$ depends on the branches of $N(z)$ and is defined as
\begin{equation}
    \label{def-a}
a=\begin{cases}
1-\mathcal{H}\left(\sqrt{1-\sqrt{1-h_0^2}}-mh_0\right), &\text{for $h_0 \leq 1$}\\
0, &\text{for $h_0>1$}
\end{cases}
\end{equation}
with $\mathcal{H}(\phantom{x})$ being the Heaviside step function. Similarly, integrating along $C'_d$, we evaluate the function $N^+(z)$ given in \eqref{Nplus}.
%

\section{The SIF in the equilibrium case}

\noindent
In the limit $m\to 0$, we have, according to \eqref{b0} and \eqref{def-a}, that $b_0=1$ and $a=0$. In addition, the real and imaginary parts of $N(z)$ become, respectively
\begin{equation}
\textup{Re}\left(N(z)\right)=\frac{1+4(1-\nu)z^2}{3-2\nu}, \qquad \textup{Im}\left(N(z)\right)=\frac{4(1-\nu)}{3-2\nu}\,\frac{\left(-z^2\right)^{\frac{3}{2}}}{(1+z^2)^{\frac{1}{2}}},
\end{equation}
for $0 \leq \textup{Im}(z) \leq 1$ and $\textup{Re}(z)=+0$. In view of the above, the function $N^+(z)$ takes the following form
\begin{equation}
N^+(z)=\exp\biggl\{-\frac{1}{\pi}\biggl(\int_{0}^{i}\tan^{-1}\biggl[\frac{4(1-\nu)\left(-\zeta^2\right)^{\frac{3}{2}}}{\left(1+\zeta^2\right)^{\frac{1}{2}}\left(1+4(1-\nu)\zeta^2\right)}\biggr]\frac{d\zeta}{\zeta+z}\biggr\},
\end{equation}
which after changing the variable of integration from $\zeta$ to $iq$ and by taking into account the properties of the inverse tangent function, yields
\begin{equation}
\begin{split}
N^+(z)&=\exp\biggl\{\frac{1}{\pi}\biggl(\int_{0}^{1}\left(-\frac{\pi}{2}+\tan^{-1}\biggl[\frac{\left(1-q^2\right)^{\frac{1}{2}}\left(1-4(1-\nu)q^2\right)}{4(1-\nu)q^3}\biggr]\right)\frac{dq}{q+z/i}\biggr\}\\
&=\frac{1}{\left(1+i/z\right)^{\frac{1}{2}}}\exp\biggl\{\frac{1}{\pi}\biggl(\int_{0}^{1}\tan^{-1}\biggl[\frac{\left(1-q^2\right)^{\frac{1}{2}}\left(1-4(1-\nu)q^2\right)}{4(1-\nu)q^3}\biggr]\frac{dq}{q+z/i}\biggr\}.
\end{split}
\end{equation}
Finally, upon substituting $z=i\ell/L$ into the above expression, we evaluate the equilibrium SIF through the relation \eqref{KIId}, which agrees exactly with the SIF given previously by Gourgiotis et al. (2012) for the stationary mode-II crack (see Eq. (126) in the cited work, with $\tau_0 \equiv T_0/L$).

\end{appendices}

\section*{Acknowledgments}
\addcontentsline{toc}{section}{Acknowledgments}
\noindent Panos A. Gourgiotis gratefully acknowledges support from the European Union FP7 project ``Modelling and optimal design of ceramic structures with defects and imperfect interfaces'' under contract number PIAP-GA-2011-286110-INTERCER2. Andrea Piccolroaz would like to acknowledge the Italian Ministry of Education, University and Research (MIUR) for the grant FIRB 2010 Future in Research ``Structural mechanics models for renewable energy applications'' (RBFR107AKG).

\section*{References}
\addcontentsline{toc}{section}{References}
\begin{enumerate} [itemsep=1mm]

\item Achenbach JD (1973) Wave propagation in elastic solids. North-Holland, Amsterdam

\item Antipov YA (2012) Weight functions of a crack in a two-dimensional micropolar solid. Quart J Mech Appl Math 65: 239-271

\item Aravas N, Giannakopoulos AE (2009) Plane asymptotic crack-tip solutions in gradient elasticity. Int J Solids Struct 46:4478-4503

\item Atkinson C, Leppington FG (1974) Some calculations of the energy-release rate G for cracks in micropolar and couple-stress elastic media. Int J Frac 10:599-602

\item Atkinson C, Leppington FG (1977) The effect of couple stresses on the tip of a crack. Int J Solids Struct 13:1103-1122

\item Bigoni D, Drugan WJ (2007) Analytical derivation of Cosserat moduli via homogenization of heterogeneous elastic materials. ASME J Appl Mech 74:741-753

\item Cauchy AL (1851) Note sur l' equilibre et les mouvements vibratoires des corps solides. Comptes-Rendus Acad. Paris 32:323-326

\item Chang CS, Shi Q, Liao CL (2003) Elastic constants for granular materials modeled as first-order strain-gradient continua. Int J Solids Struct 40:5565-5582

\item Chen JY, Huang Y, Ortiz M (1998) Fracture analysis of cellular materials: a strain gradient model. J Mech Phys Solids 46:789-828

\item Cosserat E, Cosserat F (1909) Theorie des Corps Deformables. Hermann et Fils, Paris

\item Dal Corso F, Willis JR (2011) Stability of strain gradient plastic materials. J Mech Phys Solids 59:1251-1267

\item Engelbrecht J, Berezovski A, Pastrone F, Braun M (2005) Waves in microstructured materials and dispersion. Phil Mag 85:4127-4141

\item Fleck NA, Muller GM, Ashby MF, Hutchinson JW (1994) Strain gradient plasticity: theory and experiment. Acta Metall Mater 42:475-487

\item Fleck NA, Hutchinson JW (1997) Strain gradient plasticity. In: Hutchinson, J.W., Wu, T.Y. (Eds.), Advances in Applied Mechanics vol. 33. Academic Press, New York, 295-361

\item Fisher B (1971) The product of distributions. Q J Math 22:291-298

\item Freund LB (1972) Energy flux into the tip of an extending crack in an elastic solid. J Elast 2:341-349

\item Freund LB (1990) Dynamic fracture mechanics. Cambridge University Press, Cambridge UK

\item Gao H, Huang Y, Nix WD, Hutchinson JW (1999) Mechanism-based strain gradient plasticity - I. Theory. J Mech Phys Solids 47:1239-1263

\item Georgiadis HG (2003) The mode-III crack problem in microstructured solids governed by dipolar gradient elasticity: static and dynamic analysis. ASME J Appl Mech 70:517-530

\item Georgiadis HG, Velgaki EG (2003) High-frequency Rayleigh waves in materials with microstructure and couple-stress effects. Int J Solids Struct 40:2501-2520

\item Georgiadis HG, Vardoulakis I, Velgaki EG (2004) Dispersive Rayleigh-wave propagation in microstructured solids characterized by dipolar gradient elasticity. J Elast 74:17-45

\item Gourgiotis PA, Georgiadis HG (2007) Distributed dislocation approach for cracks in couple-stress elasticity: shear modes. Int J Fract 147:83-102

\item Gourgiotis PA, Sifnaiou MD, Georgiadis HG (2010) The problem of sharp notch in microstructured solids governed by dipolar gradient elasticity. Int J Fract 166:179-201

\item Gourgiotis PA, Georgiadis HG (2011) The problem of sharp notch in couple-stress elasticity. Int J Solids Struct 48:2630-2641.

\item Gourgiotis PA, Georgiadis HG, Sifnaiou MD (2012) Couple-stress effects for the problem of a crack under concentrated shear loading. Math Mech Solids 17:433-459

\item Gourgiotis PA, Georgiadis HG, Neocleous I (2013) On the reflection of waves in half-spaces of microstructured materials governed by dipolar gradient elasticity. Wave Motion 50:437-455

\item Graff KF, Pao YH (1967) The effects of couple-stresses on the propagation and reflection of plane waves in an elastic half-space. J Sound Vib 6:217-229

\item Grentzelou CG, Georgiadis HG (2008) Balance laws and energy release rates for cracks in dipolar gradient elasticity. Int J Solids Struct 45:551-567

\item Han SY, Narasimhan MNL, Kennedy TC (1990) Dynamic propagation of a finite crack in a micropolar elastic solid. Acta Mech 85:179-191

\item Huang Y, Zhang L, Guo TF, Hwang KC (1997) Mixed mode near tip fields for cracks in materials with strain-gradient effects. J Mech Phys Solids 45:439-465

\item Huang Y, Chen JY, Guo TF, Zhang L, Hwang KC (1999) Analytic and numerical studies on mode I and mode II fracture in elastic-plastic materials with strain gradient effects. Int J Fract 100:1-27

\item Hwang KC, Jiang H, Huang Y, Gao H, Hu N (2002) A finite deformation theory of strain gradient plasticity. J Mech Phys Solids 50:81-99

\item Itou S (1972) The effect of couple-stresses dynamic stress concentration around a crack. Int J Eng Sci 10:393-400

\item Itou S (1981) The effect of couple-stresses on the stress concentration around a moving crack. Int J Math Math Sci 4:165-180

\item Itou S (2013) Effect of couple-stresses on the mode I dynamic stress intensity factors for two equal collinear cracks in an infinite elastic medium during passage of time-harmonic stress waves. Int J Solids Struct 50:1597-1604

\item Koiter WT (1964) Couple-stresses in the theory of elasticity. Parts I and II. Proc Ned Akad Wet B67:17-44

\item Kulakhmetova SA, Saraikin VA, Slepyan LI (1984) Plane problem of a crack in a lattice. Mech Solids 19:101-108

\item Lakes RS (1983) Size effects and micromechanics of a porous solid. J Mater Sci 18:2572-258

\item Lakes RS (1993) Strongly Cosserat elastic lattice and foam materials for enhanced toughness. Cell Polym 12:17-30

\item Livne A, Bouchbinder E, Svetlizky I, Fineberg J (2010) The near-tip fields of fast cracks. Science 327:1359-1263

\item Lubarda VA, Markenskoff X, (2000) Conservation integrals in couple stress elasticity. J Mech Phys Solids 48:553-564

\item Maranganti R, Sharma P (2007a) Length scales at which classical elasticity breaks down for various materials. Phys Rev Lett 98:195504 1-4

\item Maranganti R, Sharma P (2007b) A novel atomistic approach to determine strain-gradient elasticity constants: Tabulation and comparison for various metals, semiconductors, silica, polymers and the (Ir) relevance for nanotechnologies. J Mech Phys Solids 55:1823-1852

\item Maugin GA (2010) Mechanics of generalized continua: What do we mean by that?, eds. Maugin GA and Metrikine AV. Mechanics of generalized continua. One hundred years after the Cosserats. Springer New York, 3-13

\item Mindlin RD, Tiersten HF (1962) Effects of couple-stresses in linear elasticity. Arch Ration Mech Anal 11:415-448

\item Mishuris G, Piccolroaz A, Radi E (2012) Steady-state propagation of a mode III crack in couple stress elastic materials. Int J Eng Sci 61:112-128

\item Mora R, Waas AM (2000) Mesurement of the Cosserat constant of circular-cell polycarbonate honeycomb. Philos Mag A 80:1699-1713

\item Muki R, Sternberg E (1965) The influence of couple-stresses on singular stress concentrations in elastic solids. ZAMP 16:611-618

\item Nieves MJ, Movchan AB, Jones IS, Mishuris GS (2013) Propagation of Slepyan's crack in a non-uniform elastic lattice. J Mech Phys Solids, 10.1016/j.jmps.2012.12.006

\item Noble B (1958) Methods based on the Wiener-Hopf technique. Pergamon Press, Oxford UK

\item Nowacki W (1986) Theory of asymmetric elasticity. Pergamon Press,  Oxford UK

\item Ostoja-Starzewski M, Jasiuk I (1995) Stress invariance in planar Cosserat elasticity. Proc Roy Soc Lond A 451:453-470
\item Piccolroaz A, Mishuris G, Radi E (2012) Mode III interfacial crack in the presence of couple stress elastic materials. Eng Fract Mech 80:60-71

\item Polyzos D, Fotiadis DI (2012) Derivation of Mindlin's first and second strain gradient elastic theory via simple lattice and continuum models. Int J Solids Struct 49:470-480

\item Radi E, Gei M (2004) Mode III crack growth in linear hardening materials with strain gradient effects. Int J Fract 130:765-785

\item Radi E (2008) On the effects of characteristic lengths in bending and torsion on mode III crack in couple stress elasticity. Int J Solids Struct 45:3033-3058

\item Ravi-Chandar K (2004) Dynamic fracture. Elsevier, Amsterdam

\item Rice JR (1968a)  A path independent integral and the approximate analysis of strain concentration by notches and cracks. ASME J Appl Mech 35:379-386

\item Rice JR (1968b) Mathematical analysis in the mechanics of fracture. Fracture, H Liebowitz (ed), Academic Press New York, 2:191-311

\item Roos BW (1969) Analytic functions and distributions in physics and engineering. John Wiley and Sons, New York

\item Rosakis AJ, Samudrala O, Coker D (1999) Cracks faster than the shear wave speed. Science 284:1337-1340

\item Sciarra G, Vidoli S (2012a) The role of edge forces in conservation laws and energy release rates of strain-gradient solids. Math Mech Solids 17:266-278

\item Sciarra G, Vidoli S (2012b) Asymptotic fracture modes in strain-gradient elasticity: size effects and characteristic lengths for isotropic materials. J Elast, doi:10.1007/s10659-012-9409-y

\item Sternberg E (1960) On the integration of the equations of motion in the classical theory of elasticity. Arch Ration Mech Anal 6:34-50

\item Sternberg E, Muki R (1967) The effect of couple-stresses on the stress concentration around a crack. Int J Solids Struct 3:69-95

\item Toupin RA (1962) Perfectly elastic materials with couple stresses. Arch Ration Mech Anal 11:385-414

\item Vardoulakis I, Georgiadis HG (1997) SH surface waves in a homogeneous gradient elastic half-space with surface energy. J Elast 47:147-165

\item Voigt W (1887) Theoretische Studien uber die Elasticitatsverhaltnisse der Krystalle. Abh Ges Wiss Gottingen 34:3-100

\item Willis JR (1971) Fracture mechanics of interfacial cracks. J Mech Phys Solids 19:353-368

\end{enumerate}

\end{document}